\documentclass[a4paper,11pt]{article}
\usepackage{jheppub}
\usepackage{amsmath,amssymb}
\usepackage{natbib}
\usepackage{caption}
\usepackage{subcaption}
\usepackage{graphicx}
\usepackage{hyperref}
\usepackage{braket}
\usepackage{comment}
\usepackage[normalem]{ulem}
\usepackage{multirow}
\usepackage{caption}

\newcommand{\bQ}{\boldsymbol{\mathcal{Q}}}
\newcommand{\bq}{\boldsymbol{\mathsf{Q}}}

\usepackage{float}
\usepackage{natbib}
\usepackage{hyperref}
\newcommand{\be}{\begin{equation}}
\newcommand{\ee}{\end{equation}}
\newcommand{\bse}{\begin{subequations}}
\newcommand{\ese}{\end{subequations}}
\newcommand{\ba}{\begin{eqnarray}}
\newcommand{\ea}{\end{eqnarray}}

\usepackage{color}
\usepackage[normalem]{ulem}  

\begin{document}


\title{An effective framework for strange metallic transport}



\author[a]{Beno\^{i}t Dou\c{c}ot,} 
\author[b]{Ayan Mukhopadhyay,} 
\author[c]{Giuseppe Policastro,}
\author[d]{Sutapa Samanta}
\author[e]{and Hareram Swain}
\emailAdd{doucot@lpthe.jussieu.fr, ayan.mukhopadhyay@pucv.cl, giuseppe.policastro@phys.ens.fr, samants2@wwu.edu, dhareram1993@physics.iitm.ac.in }
\affiliation[a]{Laboratoire  de  Physique  Th\'{e}orique  et  Hautes  Energies, Sorbonne Universit\'{e} and CNRS UMR 7589, 4 place Jussieu, 75252 Paris Cedex 05, France}
\affiliation[b]{Instituto de F\'{\i}sica, Pontificia Universidad Cat\'{o}lica de Valpara\'{\i}so, Avenida Universidad 330, Valpara\'{\i}so, Chile}
\affiliation[c]{Laboratoire de Physique de l'\'{E}cole Normale Supérieure, ENS, Universit\'{e} PSL, CNRS, Sorbonne Universit\'{e}, Universit\'{e} de Paris, F-75005 Paris, France}
\affiliation[d]{Department of Physics and Astronomy, Western Washington University,
516 High Street, Bellingham, Washington 98225, USA}
\affiliation[e]{Department of Physics, Indian Institute of Technology Madras, Chennai 600036, India}



\date{\today}

\abstract{Semi-holography, originally proposed as a model for conducting lattice electrons coupled to a holographic critical sector, leads to an effective theory of non-Fermi liquids with only a few relevant interactions on the Fermi surface in the large $N$ limit. A refined version of such theories has only two effective couplings, which give holographic and Fermi-liquid-like contributions to the self-energy, respectively.

We show that a low co-dimension sub-manifold exists in the space of refined semi-holographic theories in which strange metallic behavior is manifested and which can be obtained just by tuning the ratio of the two couplings. On this sub-manifold, the product of the spectral function and the temperature is approximately independent of the critical exponent, the Fermi energy, and the temperature at all frequencies and near the Fermi surface when expressed in terms of suitably scaled momentum and frequency variables. This quasi-universal behavior leads to linear-in-$T$ dc resistivity and Planckian dissipation over a large range of temperatures, and we also obtain $T^{-3}$ scaling of the Hall conductivity at higher temperatures. 

The quasi-universal spectral function also fits well with photoemission spectroscopic data without varying the critical exponent with the doping.

Combining with the results for optical conductivity, we construct a generalized version of Drude phenomenology for strange-metallic behavior, which satisfies non-trivial consistency tests. 

Finally, we discuss a possible dynamical mechanism for the fine-tuning of the ratio of the two couplings necessary to realize the strange metallic behavior in a typical state.}

\maketitle

\section{Introduction}

Strange metals form one of the most poorly understood phases of quantum matter \cite{Phillips-Hussey}. Despite a wealth of experimental studies on cuprates and heavy fermion crystalline compounds, what fundamentally characterizes the strange metallic phase, which exhibits distinctive transport properties irrespective of the material constitution, is still mysterious. Furthermore, there is sufficient experimental evidence for the absence of conventional quasi-particles in the strange metallic phase \cite{Sachdev:2011cs,Taillefer,RevModPhys.75.473,PhysRevB.78.035103,Vishik_2010,reber2015power,Lee_2018,Michon2019,RevModPhys.92.031001,Hayes2021}. Therefore, it is natural to explore how strange metallic behavior can arise within an effective field theory framework that can be applied to quantum phases without conventional quasi-particles. The goal of this paper is to take some steps in these directions within the semi-holographic effective field theory approach.

\subsection{A brief historical review and our setup}\label{Sec:Review}
\paragraph{Semi-holographic model for non-Fermi liquids:} Semi-holography was first advanced as a phenomenological tool for modelling strongly correlated electronic systems by coupling electrons on a $2$-dimensional lattice to a large $N$ and solvable infrared conformal field theory (IR-CFT) \cite{Faulkner:2010tq}. Such IR-CFTs can arise as emergent fixed points of $2+1$-dimensional holographic gauge theories at finite density \cite{Faulkner:2009wj,Iqbal:2011in}, and are described by $3+1$-dimensional near-horizon geometries of charged black holes at finite temperature \cite{Lee:2008xf,Liu:2009dm,Cubrovic:2009ye,Faulkner:2010da}. Most commonly, such an effectively $1+1$-dimensional holographic IR-CFT has a dual description in terms of the near-horizon AdS$_2$ $\times$ R$^2$ geometry of extremal black holes endowed with a constant electric field whose strength is determined by the chemical potential in the IR-CFT.  A Dirac fermion $\chi$ (or any other bulk field) living in the AdS$_2$ $\times$ R$^2$ geometry with mass $m$, momentum $\mathbf{k}$ on R$^2$,  and charge $q$ has an effective mass $m_{\rm eff}(m, \mathbf{k}, q)$ in AdS$_2$ which determines the scaling dimension $\tilde\nu(m, \mathbf{k}, q)$ of the dual operator (carrying the quantum number $\mathbf{k}$) in the $1+1$-dimensional IR-CFT. The R$^2$ factor at the boundary of AdS$_2$ is identified with the ambient space where the lattice electrons live. 

The simplest coupling of a lattice electron $\psi(\mathbf{k})$ to the IR-CFT is via linear hybridization with an IR-CFT operator $\chi(\mathbf{k})$ carrying the same momentum $\mathbf{k}$ and other relevant quantum numbers like spin. Assuming a spherical Fermi surface\footnote{The Fermi surface can be precisely defined in field theory as the locus of the poles of the two-point correlations in momentum space at zero frequency.}, and restricting the momenta to be near the Fermi surface $\vert \mathbf{k}\vert = k_F$ so that we can be concerned with only low energy excitations, we can assume $\tilde\nu(m, \mathbf{k}, q)$ to be a constant $\tilde\nu (m,q)$ since the deviation is suppressed by $(k/k_F)^2$ as far as low energy physics is concerned.\footnote{This assumption needs to be tested because if we do loop integrals, then it is not obvious that only loop momenta near the Fermi surface contribute in absence of sharply defined quasi-particles. We will demonstrate this a posteriori.} The linear coupling of $\psi$ and $\chi$ leads to a single propagating mode on the Fermi surface whose spectral function vanishes on the Fermi surface as $\omega^{2\tilde\nu}$ in the low energy limit. For later purposes, we define $\nu \equiv 2 \tilde{\nu}$.

 Incorporating other types of near-horizon geometries \cite{Kachru:2008yh,Gubser:2009qt,Hartnoll:2009ns,Charmousis:2010zz,Gouteraux:2011ce,Tarrio:2011de,Huijse:2011ef,Gursoy:2011gz,Alishahiha:2012cm,Gath:2012pg,Bhattacharya:2012zu,Donos:2014oha}, semi-holography provides a simple description of many emergent non-Fermi liquid fixed points which can be embedded into a UV-complete setup, typically a $2+1$-dimensional large $N$ strongly coupled holographic gauge theory. At the same time, it has the flexibility of being a bottom-up model that can incorporate lattice-dependent phenomenology.

\paragraph{Semi-holographic effective field theory for non-Fermi liquids:} A natural question is whether the semi-holographic models can be promoted to an effective field theory framework that can describe general non-Fermi liquid fixed points. In \cite{Mukhopadhyay:2013dqa}, it was shown that such an effective field theory can indeed be constructed in the large $N$ limit where
\begin{itemize}
    \item  the bulk gravitational action describing the holographic sector is $\mathcal{O}(N^2)$,
    \item  the couplings between the lattice electrons with holographic single trace operators are $\mathcal{O}(N)$ (and couplings to multi-trace holographic operators are sub-leading at large $N$), and
    \item the couplings between the lattice electrons are $\mathcal{O}(N^0)$.
\end{itemize}
As for instance, $\psi^3\chi$ coupling is $\mathcal{O}(N)$ while the $\psi^4$ coupling is $\mathcal{O}(N^0)$. In this large $N$ limit, the backreaction on the near-horizon geometry describing the holographic sector is suppressed. Here, we will restrict our discussion to the case where the near-horizon geometry is AdS$_2$ $\times$ R$^2$. 

It was shown that with the above large $N$ scaling of various couplings, only the linear hybridization between $\psi$ and $\chi$ is relevant in the low energy limit on the Fermi surface provided all long-range forces such as Coulomb interactions are screened when $0 <\nu \leq 1$ \cite{Mukhopadhyay:2013dqa}, and later it was shown in \cite{Doucot:2017bdm} that such a screening indeed happens when $1/2 <\nu \leq 1$. The limit $\nu\rightarrow 1$ yields a precise type of marginal Fermi liquid (which will not be relevant for this work). This implies the notion of generalized quasi-particles since the low energy behavior of the spectral function $\omega^\nu$ is not renormalized on the Fermi surface.

In \cite{Doucot:2017bdm}, it was also shown that for $1/2 <\nu \leq 1$, there are well-defined collective excitations in the continuum, which could potentially yield an interesting way to generate superconducting instability via screened Coulombic interactions. However, a similar statement can also be made about long-range interaction arising from a Yukawa coupling $\psi^2\Phi$ with a bulk scalar $\Phi$ dual to a scalar operator in the IR-CFT.

Nevertheless, the phenomenology of collective modes is dependent on the short distance cut-off because momenta far away from the Fermi surface contribute significantly in loop integrals. In particular, the density-density correlation in frequency space is always negative for frequencies less than a chosen value unless an appropriate momentum cut-off is introduced \cite{Doucot:2017bdm}. This does not invalidate the effective approach at low frequencies but implies that the location of the collective modes of the model within the particle-hole continuum at mid-infrared frequencies (obtained via the random-phase-approximation (RPA) in \cite{Doucot:2017bdm} explicitly) depend on the momentum cut-off. Similar issues arise in the computation of transport properties at finite temperatures and frequencies.

\paragraph{Refined semi-holography:} A further refined version of the semi-holographic effective theory, which was proposed in \cite{Doucot:2020fvy} and reviewed in Sec. \ref{Sec:Model} ameliorates the problems of the general semi-holographic effective theories of non-Fermi liquids described above. 

In this refined version, the fermion $\psi$ living at the boundary can be of two varieties: one is $c$, which belongs to the conduction band, and the other one is $f$, which belongs to the filled bands. Furthermore, it is assumed that all couplings that are non-linear in $c$ are suppressed just like those which are non-linear in $\chi$. Therefore, the leading couplings of $c$ with other degrees of freedom are $c\chi$  and $cf^3$ which should be  $\mathcal{O}(N)$ and $\mathcal{O}(N^0)$, respectively, as mentioned above. Here $c$ can be best understood as a baryonic field living on coincident probe flavor branes which are embedded in the gravitational background that describes the dynamics of the adjoint matter of a gauge theory with $N_c$ colors and $N_f$ flavors (see Sec. \ref{Sec:Model}). The probe-brane approximation is valid when $N_f \ll N_c$. Furthermore, if $N_f$ is also large, the couplings that are non-linear in $c$ are also suppressed. However, the latter feature is introduced here simply as an assumption in this setup, and a full justification requires specific embedding into realistic microscopic dynamics. One can think of the generalized quasi-particles arising from this theory as the degrees of freedom of the emergent lower Hubbard band. {In the large $N$ limit we consider, the refined effective theory has only two effective couplings, corresponding to a holographic and
Fermi-liquid-like contributions to the self-energy, respectively.}

The main advantage of this refined model is that the loop integrals involved in physical observables like density-density correlations and conductivities receive major contributions from loop momenta \textit{only} near the Fermi surface, as will be explicitly discussed later in the paper. Despite arbitrary large loop frequencies contributing to the loop integral, no divergence is encountered while integrating over loop frequencies also. Therefore, one can extract the physical consequences of the model, which are independent of the short distance cut-off, at least at the one-loop level.

\subsection{Statement of the main results}
\paragraph{The strange metallic sub-manifold:} In this work, we show that there is a \textit{co-dimension one sub-manifold} in the space of couplings and parameters of the refined semi-holographic non-Fermi liquid theories where model independent (quasi-universal) strange metallic behavior emerges. Our primary result obtained from the study of the refined semi-holographic effective theory, which can be shown to have only two effective dimensionless couplings, is as follows. \\

\noindent\fbox{
    \parbox{\textwidth}{By tuning {\textit{only}} the ratio of the two dimensionless couplings of the refined semi-holographic non-Fermi liquid theories, we obtain a \textit{strange-metallic sub-manifold} where the spectral function times the temperature is a \textit{quasi-universal} function of suitably scaled frequency and momentum to a very good approximation when the momentum is at or near the Fermi surface, and the temperature is above a threshold. This quasi-universal function has no explicit dependence on the Fermi momentum $k_F$, the scaling exponent $\nu$, and the temperature $T$. It also implies \textit{linear-in-$T$} resistivity and  \textit{Planckian dissipation} over a very wide range of temperatures.}
}
\\\\
The above feature of low co-dimension strange-metallic sub-manifold is realized when $0.66 \lessapprox \nu \lessapprox 0.95$ and the lower temperature threshold ($t_c$) for the quasi-universal spectral function to emerge depends on $\nu$. When $\nu$ is larger, then $t_c \approx 0.01 E_F$ and for smaller $\nu$, $t_c \approx 0.1 E_F$. Given that the one-loop integral gets a major contribution from loop momenta near the Fermi surface, we can readily show that the quasi-universal form of the spectral function implies linear-in-$T$ resistivity \cite{Chien,Hussey1,Hussey2,Phillips-Hussey}.\footnote{The coupling of the lattice electron to the holographic Dirac fermion living in the black hole background in semi-holography breaks Galilean boost invariance essentially because the holographic correlation function in the charged black hole background is not boost invariant. This results in a finite resistivity.} Thus, we are able to explain the numerical findings of \cite{Doucot:2020fvy} where the linear-in-$T$ resistivity was demonstrated by explicit evaluation of the one-loop contribution. 

We also find that the quasi-universal form of the spectral function fits very well with data obtained from angle-resolved photoemission spectroscopy (ARPES) across a range of temperatures with the Fermi energy fixed for a given doping and $\nu$ unchanged with the doping. Only one of the two couplings is changed with the doping for the fitting. This departs from previous fittings of the same experimental data \cite{Reber2019} in which the critical exponent (the analogue of $\nu$) was assumed to vary with the doping. This also marks a notable difference between our model and traditional holographic models, in which the resistivity at the critical point has a scaling behavior determined by a critical exponent, which has to be tuned in order to have linear behavior. 

Furthermore, we also obtain a novel real-time formula for the Hall conductivity at one loop and show we obtain $T^{-2}$ scaling at low temperatures and the strange metallic $T^{-3}$ scaling \cite{Chien,Phillips-Hussey} at higher temperatures for the Hall conductivity at the optimal ratio of the couplings. The $T^{-2}$ scaling follows from quasi-universality. However, the high temperature $T^{-3}$  also robustly appears when the scaling exponent $\nu$ takes values in the range where quasi-universality emerges. It follows that these scaling behaviors of the Hall conductivity appear on the strange metallic sub-manifold irrespective of the value of the scaling exponent $\nu$ as in the case of the dc conductivity.

\paragraph{Optical conductivity:} In this work, we obtain more insights from the study of optical conductivity in refined semi-holographic theories within and outside of the strange metallic sub-manifold. As mentioned above, the refined semi-holographic effective theory has two dimensionless couplings, namely $\alpha$ and $\gamma$, which give holographic and Fermi-liquid-like contributions to the self-energy respectively. Phenomenologically, we can identify the ratio of $\alpha$ to $\gamma$, which produces strange metallic behavior with optimal doping (where doping refers to the number of holes per CuO$_2$ in cuprates). Since overdoping leads to Fermi-liquid-like behavior, this should phenomenologically correspond to sub-optimal ratios of $\alpha$ to $\gamma$ as the Fermi-liquid-like self-energy term dominates over the holographic one in this case. Similarly, underdoping should be phenomenologically equivalent to super-optimal ratio of $\alpha$ to $\gamma$. Our main findings are as follows.\\

\noindent\fbox{
    \parbox{\textwidth}{Both for sub-optimal and optimal ratios of $\alpha$ to $\gamma$ (overdoping and optimal doping), the Drude formula for the real part of the optical conductivity $${\rm Re}\,\sigma(\Omega, T) = \frac{\sigma_{\rm{dc}}(T)}{1 + \Omega^2 \tau^2(T)}$$is remarkably valid for a wide range of temperatures $T\leq  0.6 E_F$ at least when $\Omega/T \leq 1$. For optimal doping, both $\tau(T)$ and $\sigma_{\rm{dc}}(T)$ scale as $T^{-\nu}$ at low temperatures and both have Planckian scaling $T^{-1}$ for higher temperatures (in the strange-metallic regime). These scaling behaviors of $\tau(T)$ and $\sigma_{\rm{dc}}(T)$ evolve to the Fermi-liquid-like $T^{-2}$ scaling in both low and high-temperature regimes as $\alpha/\gamma$ is decreased from the optimal value towards zero. \\
    
    For super-optimal ratios of $\alpha$ to $\gamma$ (underdoping), we obtain non-Drude scaling behaviors ${\rm Re}\,\sigma(\Omega, T) \sim \Omega^\kappa T^\eta$ in both low and high-temperature regimes. In the limit of large $\alpha/\gamma$, $\kappa \sim -1$ while $\eta \sim \nu/2$ and $\sim 1$ in low and high-temperature regimes, respectively.}
}

\vspace{0.5cm}

Our results reproduce some experimental features of the optical conductivity of crystalline compounds exhibiting strange metallic behavior at optimal doping.
\begin{itemize}
    \item Analysis of the experimental data of optical conductivity in cuprates in \cite{PhysRevB.106.054515} shows that the Drude formula explains about 90 \%  of the optical conductivity for optimal doping and overdoping at low temperatures (up to 290 K) over a large range of frequencies, while at high temperatures, there is a lack of sensitivity for any definite conclusions. At underdoping, the Drude-like contribution disappears. This is in agreement with our results (assuming $E_F \sim 10^4$K).
    \item The non-Drude like $\Omega^{-1}$ scaling of the real part of the optical conductivity has been reported in many underdoped samples, especially at low temperatures as in \cite{10.3389/femat.2022.934691}. This is in broad agreement with our results for large $\alpha/\gamma$, and we can quantitatively reproduce the range of temperatures and frequencies and other features reported in \cite{10.3389/femat.2022.934691} choosing the value of $\nu = 0.73$ which gives the best approximation to linear-in-$T$ dc resistivity (and assuming $E_F \sim 10^4$K).
\end{itemize}

\paragraph{Generalized Drude phenomenology and refined Planckian dissipation:} We show that the dc conductivity, Hall conductivity and optical conductivity at the optimal ratio of couplings leads to a compelling generalized version of Drude phenomenology for strange metallic behavior which satisfies some non-trivial internal consistency tests. Particularly, we are able to match the scattering time $\tau(T)$ obtained from the Drude-like behavior of the optical conductivity with the dissipation time scale $\tau_d(T)$ obtained from the combined results of the dc and Hall conductivities. The generalized version of Drude phenomenology also leads us to estimate the threshold temperature $t_c$ where quasi-universality emerges accurately while explaining why the effective carrier density is set only by the Fermi energy and is independent of the scaling exponent $\nu$ at lower temperatures where quasi-universality of the spectral function is not exhibited. 

Furthermore, the generalized version of Drude phenomenology leads to a refined picture of Planckian dissipation where the dissipation time scale is $f\times \frac{\hbar}{k_B} T^{-1}$, but with $f$ being close to 1 only at strong coupling and even as large as 25 at very weak coupling. Interestingly, the analysis of experimental data presented in heavy fermion crystalline compounds in \cite{Taupin2022AreHF} argues that the scattering time can indeed be large as $10\, T^{-1}$  in the Planckian domain. 

\subsection{Plan of the paper}

The plan for the rest of the paper is as follows. 

In Sec. \ref{Sec:Model}, we review the refined semi-holographic effective theory proposed in  \cite{Doucot:2020fvy}, including the spectral function of the propagating mode responsible for charge conduction. In Sec. \ref{Sec:Univ}, we show that these theories have a low co-dimensional sub-manifold where the spectral function times the temperature becomes an approximately model-independent (quasi-universal) function of suitably scaled frequency and momentum variables for all frequencies and near the Fermi surface. This sub-manifold is reached only by tuning the ratio of two dimensionless couplings, which can be phenomenologically related to the doping strength, and the quasi-universal spectral function fits with experimental data without varying the scaling exponent on the Fermi surface with the doping strength.  In Sec. \ref{Sec:DC}, we show that the quasi-universal spectral function implies linear-in-$T$ dc resistivity, $T^{-2}$ scaling of the Hall conductivity, and generalized Drude phenomenology leading to a refined picture of Planckian dissipation over a wide range of temperatures. We also find $T^{-3}$ Hall conductivity at higher tempratures. In Sec. \ref{Sec:OC}, we study the optical conductivity and show that (i) the Drude formula holds for overdoping and optimal doping over a wide range of temperatures and for frequencies less than the temperature but with a scattering time which becomes Planckian at optimal doping and larger temperatures, and (ii) non-Drude scaling is obtained in various ranges of frequencies and temperatures at underdoping with a generic behavior at large values of the ratio of the two dimensionless couplings. We also further validate the generalized Drude phenomenology.

In Sec. \ref{Sec:Micro}, we compare the refined semi-holographic effective theory with an effective description of black hole microstates and discuss a mechanism for the fine-tuning of the ratio of the two couplings for obtaining the strange metallic submanifold which can be \textit{dynamically} realized in a typical state.  Finally, in Sec. \ref{Sec:Disc}, we summarize the shortcomings of our model and how they can be addressed in a larger effective framework, which is also relevant for understanding aspects of typical black hole microstates.

Unless stated otherwise, we will use units in which $\hbar = k = 1$.

Note added: Preliminary versions of some of the results in Secs. \ref{Sec:Univ} and \ref{Sec:DC} of this paper have been made available by the authors in the arXiv preprint \cite{Swain:2022yez}.

\section{The refined semi-holographic effective theory}\label{Sec:Model}
\noindent
Here, we briefly review the refined semi-holographic setups proposed in \cite{Doucot:2020fvy} and elaborate on how it is indeed an effective theory from which one can reliably compute observables such as transport coefficients in the large $N$ limit. 

Building on the semi-holographic phenomenological setups of Faulkner and Polchinski \cite{Faulkner:2010tq}, it was shown in \cite{Mukhopadhyay:2013dqa} that semi-holography can provide an effective theory for a class of non-Fermi liquids in terms of \textit{generalized} quasi-particles on the Fermi surface in the large $N$ limit. In the Faulkner-Polchinski setup, the conducting electrons ($c$) living on a two-dimensional lattice hybridize linearly with fermionic operators of an IR-CFT ($\chi$). The IR-CFT is described holographically by an AdS$_2$ $\times$ R$^2$ geometry while $\chi$ is dual to a charged Dirac fermion living in the bulk of this geometry. The $c$ fermions live at the boundary of the AdS$_2$ $\times$ R$^2$. This setup was generalized in \cite{Mukhopadhyay:2013dqa} by including arbitrary couplings that have specific large $N$ scaling as described in Sec. \ref{Sec:Review} and here later. It was shown that the spectral function of the propagating fermion on the Fermi surface has a robust and simple low energy behavior given only by the linear hybridization term between $c$ and $\chi$ in the large $N$ limit, and therefore the spectral function behaves as $\omega^{\nu}$ when $\omega \sim 0$ and the momentum is on the Fermi surface ($\vert \vec{k}\vert = k_F$ ). The scaling exponent $\nu$ is essentially the effective mass of the charged Dirac fermion dual to $\chi$ living in AdS$_2$ when it has a momentum of magnitude of $k_F$ on R$^2$. Since only the linear interaction of $c$ and $\chi$ is relevant on the Fermi surface at low energies, the propagating fermion with Fermi momentum is a \textit{generalized} quasi-particle.

However, these general large $N$ semi-holographic effective theories suffer from a drawback that limits their application for the study of transport, collective behavior and possible superconducting instability as mentioned in Sec. \ref{Sec:Review}. The loop integrals do not necessarily get a dominant contribution from loop momenta near the Fermi surface. Although this does not invalidate the effective field theory, the finite temperature transport properties, the dispersion relations of the collective modes (which are sharply defined at mid-infrared frequencies within the particle-hole continuum as shown in \cite{Doucot:2017bdm}), etc., depend on the momentum cut-off and therefore can be sensitive to the ultraviolet completion of the theory.

\begin{figure}[ht]
\centering
\includegraphics[scale=0.12 ]{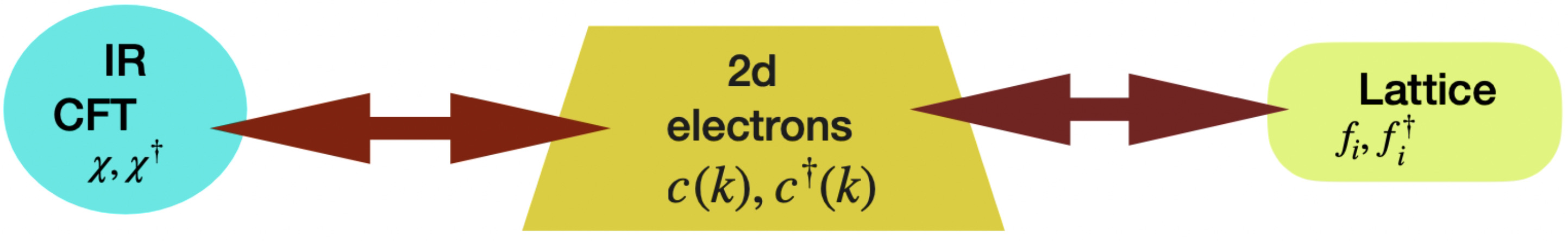}
\caption{Cartoon of the refined semi-holographic model introduced in \cite{Doucot:2020fvy}.}
\label{Fig:model-cartoon}
\end{figure}

This drawback was overcome in the refined semi-holographic setup introduced in \cite{Doucot:2020fvy}. Here, the boundary fermions are of two varieties, namely $c$ and $f$, with $c$ denoting the conducting electrons and $f$ denoting the electrons of filled bands of the lattice, which behave as Landau quasi-particles. The IR-CFT does not directly interact with the $f$-fermions, but $f$-fermions can have arbitrary self-interactions. Furthermore, the $c$-electron has interactions with both the IR-CFT operators and also with the $f$-fermions, but all these interactions should be linear in $c$, e.g. $c\chi$, $cf^3$, $cf^5$ etc. A cartoon of the model is presented in Fig. \ref{Fig:model-cartoon}. 

 In the large $N$ limit, couplings that are multi-linear in single-trace IR-CFT operators are suppressed. The large $N$ scaling of the various couplings are inherited from the general theories of \cite{Mukhopadhyay:2013dqa}, and are as follows:
 \begin{itemize}
     \item The boundary couplings $cf^3$, $cf^5$, $f^4$, etc. at the boundary are $\mathcal{O}(N^0)$.
     \item The coupling $c\chi$ is $\mathcal{O}(N)$.
     \item The gravitational action which describes the IR-CFT holographically is $\mathcal{O}(N^2)$, and therefore, the bulk vertices (which can include bulk scalars) are also $\mathcal{O}(N^2)$.
 \end{itemize}
It is easy to see that in the large $N$ limit, the backreaction on the bulk AdS$_2$ $\times$ R$^2$ geometry is suppressed.

As mentioned in Sec. \ref{Sec:Review}, the refined semi-holographic model can be best understood by thinking of the $c$-fermion as an emergent baryonic degree of freedom living on coincident $N_f$ flavor branes at the boundary of the AdS$_2$ $\times$ R$^2$ geometry. The latter bulk geometry is the holographic dual of the adjoint matter of the $2+1$-dimensional gauge theory with $N_c$ colors. (Note that both $c$ and $\chi$ are gauge-invariant operators.) In the limit in which $N_f$ and $N_c$ are both large with $N_f/N_c \ll 1$, the coincident flavor branes can be treated as probes, while the dynamics of the dual AdS$_2$ $\times$ R$^2$ bulk geometry can be described by Einstein's equations minimally coupled to matter if the adjoint matter is strongly self-interacting. Furthermore, the interactions which are multi-linear in $c$ should also be suppressed in this (large $N_f$) limit. Although we do not claim to understand how such a holographic gauge theory with $N_f$ flavors and $N_c$ colors can emerge in the infrared from the microscopic interactions in a material, we will demonstrate that we can indeed explore the phenomenology systematically in such refined semi-holographic theories. Essentially, we assume that the generalized quasi-particle of these refined semi-holographic theories with the $c$-fermion and the IR-CFT coupled to the lattice bands has the characteristics of the conducting degree of freedom of the lower Hubbard band.

It follows from our description that the Hamiltonian of the refined semi-holographic theory takes the form
\begin{align}\label{Eq:modelhamiltonian}
\hat{H} = &\sum_\mathbf{k} \epsilon(\mathbf{k}) \hat{c}^\dagger (\mathbf{k})\hat{c} (\mathbf{k}) + N\sum_\mathbf{k} \left( g \hat{c}^\dagger(\mathbf{k}) \hat\chi_{\text{CFT}} (\mathbf{k}) + c.c. +\cdots\right)+ N^2 \hat{H}_{\text{IR CFT}}\nonumber \\
&+ \sum_{i,j,k} \left(\lambda_{ijk, \mathbf{k}_1, \mathbf{k}_2, \mathbf{k}_3 }  \hat{c}^\dagger(\mathbf{k}_1) \hat{f}_i (\mathbf{k}_2) \hat{f}_j^\dagger(\mathbf{k}_3) \hat{f}_k(\mathbf{k}_1 - \mathbf{k}_2 + \mathbf{k}_3)+ c.c + \cdots\right) + \cdots ,
\end{align}
where $g$ and $\lambda$ are the two possible couplings of $c$ with $\chi$ (the IR-CFT operator) and $f$ (the Landau quasi-particles of the lattice), respectively. We have not shown the other couplings explicitly, but $\hat{H}_{\text{IR CFT}}$ includes all self-interactions of the IR-CFT and the $``\cdots"$ terms in the second line of \eqref{Eq:modelhamiltonian} denote other possible $cf^5$ couplings which are linear in $c$, and self-interactions of the $f$-fermions (and therefore include $f^4$ terms, etc.)

The propagating degree of freedom on the Fermi surface is essentially $c$ plus a $\mathcal{O}(1/N)$ contribution from $\hat{\chi}_{\rm{CFT}}$ \cite{Mukhopadhyay:2013dqa}. The self-energy of this propagating fermion on the Fermi surface can be readily computed from \eqref{Eq:modelhamiltonian} as shown in \cite{Doucot:2020fvy}. Up to two loop order, the self-energy receives two types of $\mathcal{O}(N^0)$ local (i.e. $k$-independent) contributions on the Fermi surface each with a specific temperature dependence, one of which is given by the Green’s function of the charged Dirac fermion (carrying Fermi momentum on R$^2$) in the two-dimensional AdS$_2$ black hole, and another which is Fermi liquid-like (arising from $f$ fermions in the loops). In the large $N$ limit, all other contributions are suppressed at two-loop order.  Therefore, in the large $N$ limit, the  retarded propagator takes the following form 
\begin{align}\label{Eq:propagator1}
G_R(\omega,\mathbf{k}) = &\Bigg( \omega + i \tilde{\gamma} (\omega^2 +\pi^2 T^2) +\tilde{\alpha} \mathcal{G}(\omega) - \tfrac1{2m}(\mathbf{k}^2 - k_F^2)+\mathcal{O}\left(\frac{1}{N}\right)\Bigg)^{-1}.
\end{align}
which is parametrized by two effective coupling constants denoted by $\tilde{\alpha}$ and $\tilde{\gamma}$. Here, $\tilde{\gamma} \sim \mathcal{O}(\lambda^2)$ determines contribution from the  Fermi-liquid type of self-energy, whereas $\tilde{\alpha} \sim \mathcal{O}(|g|^2)$ determines contributions from the holographic sector.\footnote{Note the result does not change even if we include $cf^5$ and other couplings which are higher orders in $f$. For the low energy phenomenology, only two effective coupling coupling constants, namely $\tilde\alpha$ and $\tilde\gamma$, are relevant.} $\mathcal{G}(\omega)$, the self-energy contribution from the holographic sector is explicitly given by (see ~\cite{Iqbal:2009fd}):
\begin{align}\label{Eq:holographic propagator}
 \mathcal{G}(\omega) = e^{i(\phi + \pi\nu/2)} (2\pi T)^\nu \frac{\Gamma\left(\tfrac12 +\tfrac{\nu}2 -i\tfrac{\omega}{2\pi T}\right)}{\Gamma\left(\tfrac12 -\tfrac{\nu}2 -i\tfrac{\omega}{2\pi T}\right)} \, .
\end{align}
Here, $\nu$ is the scaling exponent determined by the effective mass of the charged Dirac fermion carrying Fermi momentum on R$^2$ and $\phi$ is a phase determined primarily by the charge of the Dirac fermion. For the spectral function $\rho=-2\text{Im} G_R$ to be positive, $\phi$ should take value in the range $0<\phi<\pi(1-\nu)$. We set $\phi =\pi(1-\nu)/2 $ for the rest of our paper so that (only) at the Fermi momentum, the
spectral function has particle-hole symmetry, as discussed in \cite{Doucot:2020fvy}. However, we do not need significant fine-tuning of $\phi$ for our conclusions to hold.

We note that the propagator \eqref{Eq:propagator1} is valid only when the momentum is near the Fermi surface ($\vert \mathbf{k} \vert = k_F$). Otherwise, the exponent $\nu$ has $k$-dependence which is suppressed only near the Fermi surface by $(k-k_F)/k_F$. Loop calculations with the propagator \eqref{Eq:propagator1} can be justified only if the loop integrals get dominant contributions from loop momentum near the Fermi surface.

Furthermore, since the propagator \eqref{Eq:propagator1} is valid only up to two-loop order, we also need the two effective couplings to be small. To quantify this, we note that the Fermi energy $E_F =  k_F^2/2m$ sets the single intrinsic scale in the theory, and therefore we can rewrite the propagator \eqref{Eq:propagator1} in terms of dimensionless frequency $x=\omega/E_F$ and momentum $\mathbf{y} = \mathbf{k}/k_F $ variables. We also introduce the dimensionless temperature $x_T = T/E_F$. In terms of these variables, the propagator \eqref{Eq:propagator1} takes the form as follows 
\begin{align}\label{Eq:rescaledpropagator}
G_R(x,\mathbf{y}) = & E_F^{-1} \Bigg( x+i \gamma\left(x^2 +(\pi x_T)^2\right)  +\alpha e^{i(\phi +\pi\nu/2)} \nonumber\\
&\times(2\pi x_T)^\nu\frac{\Gamma\left(\tfrac12 +\tfrac{\nu}2 -i\tfrac{x}{2\pi x_T}\right)}{\Gamma\left(\tfrac12 -\tfrac{\nu}2 -i\tfrac{x}{2\pi x_T}\right)} -(\vert\mathbf{y}\vert^2 -1) \Bigg)^{-1} 
\end{align}
in which $\alpha = \tilde{\alpha}E_F^{-(1-\nu)}$ and $\gamma= \tilde{\gamma}E_F$ appear as the two dimensionless couplings of the refined semi-holographic theories. These dimensionless couplings $\alpha$ and $\gamma$ should be small for the two-loop approximation to be valid. 

The validity of the effective field theory on the Fermi surface at energies and temperatures small compared to the Fermi energy follows from the general arguments of \cite{Mukhopadhyay:2013dqa}. Although $\alpha$ and $\gamma$ are not renormalized at low frequencies and temperatures on the Fermi surface, we need to assume that the renormalization of these effective couplings is not significant if we are computing observables such as optical conductivity at finite frequencies and temperatures. This assumption can be justified from the effective field theory for sufficiently small frequencies and temperatures provided we can restrict $\alpha$ and $\gamma$ to their effective values on the Fermi surface ($\vert \mathbf{k} \vert = k_F$). For the latter to be justified, the loop integrals contributing to the observable should receive major contributions from loop momenta only near the Fermi surface, and this is indeed the case, as will be demonstrated soon.

\section{The strange metallic sub-manifold and the quasi-universal spectral function}
\label{Sec:Univ}

One of the key features of refined semi-holographic theories is that they demonstrate strange metallic linear-in-$T$ resistivity over a wide range of temperatures irrespective of the value of the scaling exponent $\nu$ provided $0.66 \lessapprox \nu\lessapprox0.95$ \cite{Doucot:2020fvy}. For each value of $\nu$ in this range, the strange metallicity appears when the ratio of the two couplings $\alpha$ to $\gamma$ is tuned to an \textit{optimal} value.  

We will show that the strange metallicity is indeed a robust feature of large $N$ refined semi-holographic theories. This follows from two key results. Firstly, the loop integrals receive major contributions from loop momenta near the Fermi surface ($\vert \mathbf{k}\vert \sim k_F$), and therefore, as discussed in the previous section, we can treat the two dimensionless couplings $\alpha$ and $\gamma$ as the leading couplings of an effective theory which determine the observables, e.g. transport coefficients. In fact, these are the only couplings that contribute to the two loop self energy in the large $N$ limit, and these couplings should be small so that we can restrict ourselves to the two loop result. Secondly, there exists a low co-dimension submanifold in the space of these refined semi-holographic theories, which we call the \textit{strange metallic submanifold} where the product of the temperature and the spectral function obtained from the retarded propagator \eqref{Eq:propagator1} can be expressed as a function that is independent of the temperature, the Fermi energy and even the scaling exponent $\nu$ to a very good approximation. This quasi-universality of the spectral function, which holds for arbitrary frequencies but only for momenta near the Fermi surface, translates to robust features such as (quasi-universal) linear-in-$T$ resistivity. The strange metallic submanifold is attained by tuning only the ratio of the two couplings for each value of $\nu$. 

Clearly, the tuning of the ratio of $\alpha$ to $\gamma$ is an additional input which is needed beyond the arguments based on the Wilsonian renormalization group which justifies the refined semi-holographic setups as low energy effective theories. We will argue in Sec. \ref{Sec:Micro} that the fine-tuning of the ratio of $\alpha$ to $\gamma$ can be achieved dynamically in typical states by embedding the simple refined semi-holographic setups into a more general effective approach, which is more realistic and in which the coupling $\alpha$ can be promoted to a field. 

In this section, we examine how the quasi-universal behavior of the refined semi-holographic theories arises on the strange metallic sub-manifold, which is obtained by the tuning of the ratio of $\alpha$ to $\gamma$. We refer to the approximate model independent behavior as quasi-universality as we cannot obtain this from the renormalization group approach alone, and an additional dynamical mechanism that can be realized in typical states is needed. In the following section, we will demonstrate that the loop integrals for the observables of interest indeed receive dominant contribution only from loop momenta near the Fermi surface and that the quasi-universal spectral function leads to strange metallic transport properties.

In order to demonstrate the quasi-universality, we need to introduce new dimensionless frequency and momentum variables, which are namely, $\tilde{x} = \omega/T =x/x_T$ and $\tilde{\mathbf{y}}$ defined via $ \vert\tilde{\mathbf{y}}\vert^2 -1 = x_T^{-1} \left(\vert\mathbf{y}\vert^2 - 1\right)$, respectively (recall $x_T = T/E_F$, $x = \omega/E_F$ and $\mathbf{y} = \mathbf{k}/k_F$). Note that $\vert\tilde{\mathbf{y}}\vert = 1$ corresponds to the Fermi surface. We introduce $F$, the dimensionless product of the temperature and the retarded propagator \eqref{Eq:rescaledpropagator}, as a function of these new dimensionless frequency and momentum variables:
\begin{align}\label{Eq:Spec-Universal}
&F(\tilde{x},\tilde{\mathbf{y}}):= T G_R(\omega,\mathbf{k}) = T G_R(\tilde{x},\tilde{\mathbf{y}}) 
\nonumber\\
&=\Big( \tilde{x}+ i \gamma x_T(\tilde{x}^2 + \pi^2) +\alpha x_T^{-1}(2\pi x_T)^\nu e^{i(\phi +\pi\nu/2)}
 \frac{\Gamma\left(\tfrac12 +\tfrac{\nu}2 -i\tfrac{\tilde{x}}{2\pi}\right)}{\Gamma\left(\tfrac12 -\tfrac{\nu}2 -i\tfrac{\tilde{x}}{2\pi }\right)} - (\vert\tilde{\mathbf{y}}\vert^2 -1) \Big)^{-1}.
\end{align}

Mathematically we define the quasi-universality as follows.  The dimensionless scaled spectral function $$ \tilde\rho(\tilde{x},\tilde{\mathbf{y}}) :=-2 {\rm Im}F(\tilde{x},\tilde{\mathbf{y}})=-2 T {\rm Im} G_R(\omega,\mathbf{k}) = T\rho(\omega,\mathbf{k})$$ as a function of $\tilde{x}$ and $\tilde{\mathbf{y}}$ is approximately independent of 
\begin{itemize}
    \item $x_T$ (and therefore of temperature $T$ and the Fermi energy $E_F$), and
    \item the scaling exponent $\nu$,
\end{itemize}
when the ratio of the two couplings $\alpha$ to $\gamma$ is tuned to the \textit{optimal} value corresponding to a given value of $\nu$ between $0.66$ and $0.95$. Furthermore, this quasi-universal form of $\tilde{\rho}$ holds
\begin{itemize}
    \item for all $\tilde{x}$, i.e. for all values of the dimensionless frequency,
    \item for $\vert\tilde{\mathbf{y}}\vert \sim 1$, i.e. for dimensionless momenta corresponding to the vicinity of the Fermi surface (the precise range of $\vert\tilde{\mathbf{y}}\vert$ will be characterized later), 
    \item for $x_T > t_c$, i.e. for temperatures above a threshold value that is determined by $\nu$ and the overall strength of the two couplings, and 
    \item for $x_T  < \tilde{t}_c < 1$, i.e. for temperatures below a threshold value $\tilde{t}_c$ that is lesser than the Fermi energy, and with $\tilde{t}_c$ determined by the overall strength of the couplings and $\nu$.
\end{itemize}
Note that the optimal value of the ratio of the couplings is determined only by $\nu$ and is therefore independent of the overall strength of the couplings, etc. In Table \ref{tab:tc vs nu}, we have listed $t_c$ and $\tilde{t}_c$ for some values of $\nu$ when $\gamma = 0.001$ (and the ratio of couplings is optimal). We will describe how $t_c$ and $\tilde{t}_c$ can be determined quantitatively later in this section.

\begin{table}[h]
    \centering
    \begin{tabular}{|c|c|c|}
    \hline
         $\nu$ & $t_c$ & $\tilde{t}_c$ \\ \hline
         0.95  &  $\sim$ 0.01  & $\sim$ 0.1 \\ \hline
         0.8   &  $\sim$ 0.05 & $\sim$ 0.22 \\ \hline
         0.7   & $\sim$ 0.17  & $\sim$ 0.5\\ \hline
    \end{tabular}
    \caption{The value of $t_c$ and $\tilde{t}_c$ for different values of $\nu$ when $\gamma = 0.001$.}
    \label{tab:tc vs nu}
\end{table}

It is really important that the quasi-universal form of the scaled spectral function $\tilde{\rho}$ holds for all frequencies because the loop integrals relevant for transport receive contributions even from the large frequency tails in the loops.\footnote{This is not true at very small temperatures. However, the contribution from large frequency tails is already significant when $T\sim 0.01 E_F$.} However, it is sufficient that quasi-universality holds only near the Fermi surface as the loop integrals receive significant contributions only from loop momenta near the Fermi surface when temperatures are not larger than the Fermi energy (we will characterize the range of values of momenta which give maximum contribution to the loop integral quantitatively in the next section). The spectral function can be modified at higher frequencies in a material, and so we expect that, realistically, quasi-universality holds up to only a cutoff frequency. The contribution to the loop integral from loop frequencies beyond the cutoff is sub-leading for temperatures smaller than the cut-off frequency. Nevertheless, this cut-off should be material dependent, and therefore it is necessary that quasi-universality holds beyond low frequencies so that strange metallic behavior is obtained robustly.

The origin of the quasi-universality is essentially the competition between the temperature dependence of the local Fermi liquid and holographic contributions to the self-energy. Clearly, a threshold temperature is needed as otherwise the Fermi liquid and holographic contributions to the self energy are simply $\omega^2$ and $\omega^\nu$ at zero temperature, respectively, and therefore these cannot compete at any arbitrary value of frequency (unless of course, both couplings vanish). We will discuss this more at the end of this section.

When the overall strength of the couplings is small (say $\gamma \sim 0.001$), the threshold temperature $t_c$ is about $0.1$ for small values of $\nu$ (e.g. $0.66$), but it can be much smaller when $\nu$ is larger (e.g. $t_c \sim 0.01$ when $\nu = 0.95$). For any $\nu$, $t_c$ reduces when the overall strength of the couplings is reduced. Our effective approach need not be valid when the external frequency and/or temperature is large compared to the Fermi energy. 

\begin{figure}[h]
\includegraphics[width= 0.5\textwidth]{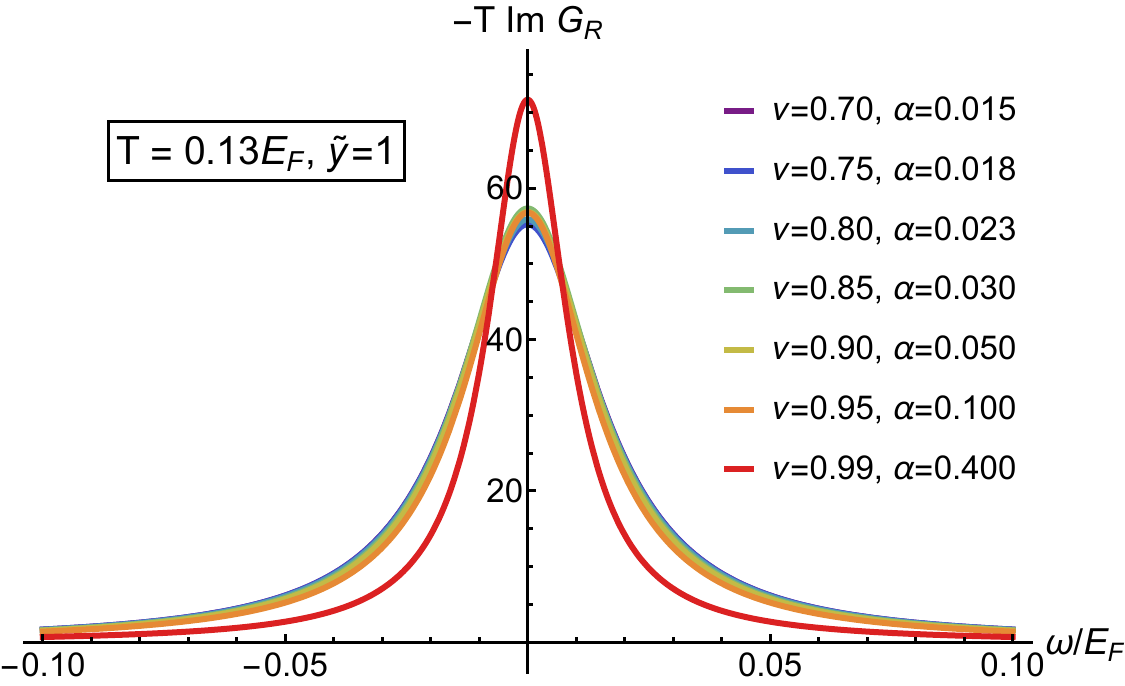}
\includegraphics[width= 0.5\textwidth]{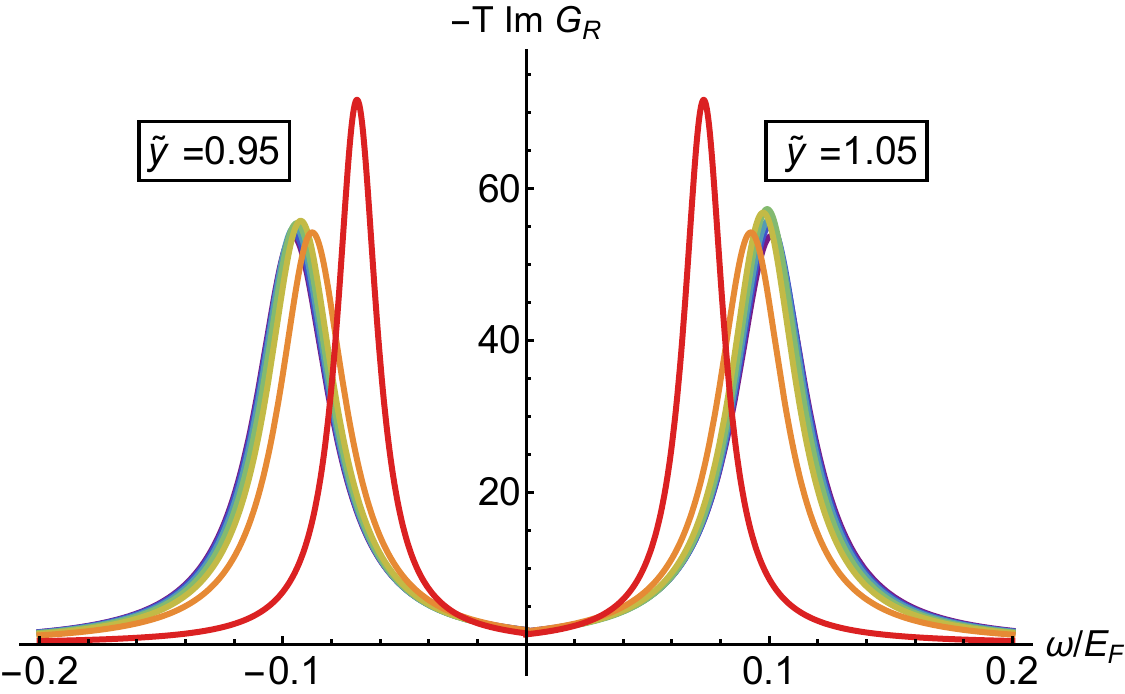}
\caption{Universality for different values of $\nu$ on the Fermi surface (left) as well as near it (right) for corresponding optimal $\alpha/\gamma$, and $\gamma = 0.001$. We have chosen $T=0.13E_F$.}\label{Fig:spectraluniversalityWnu}
\end{figure}

The range of values of the temperatures ($t_c < x_T < \tilde{t}_c$) in which quasi-universality holds shrinks if the overall strength of the couplings is increased. In fact, this range shrinks significantly beyond $\gamma = 0.01$ when our effective theory, which requires the overall strength of \textit{both} the couplings to be small, also becomes unreliable.\footnote{As evident from Fig. \ref{Fig:spectraluniversalityWnu} , the optimal ratio of $\alpha$ to $\gamma$ is typically much larger than $1$. In fact when $\nu = 0.95$, the optimal ratio is $100$, so that $\alpha \sim 1$ when $\gamma = 0.01$. (Note that in Fig. \ref{Fig:spectraluniversalityWnu}, we have set $\gamma = 0.001$, so the optimal value of $\alpha$ is $0.1$ when $\nu = 0.95$.) }

\begin{figure}[ht]
\includegraphics[width= 0.45\textwidth]{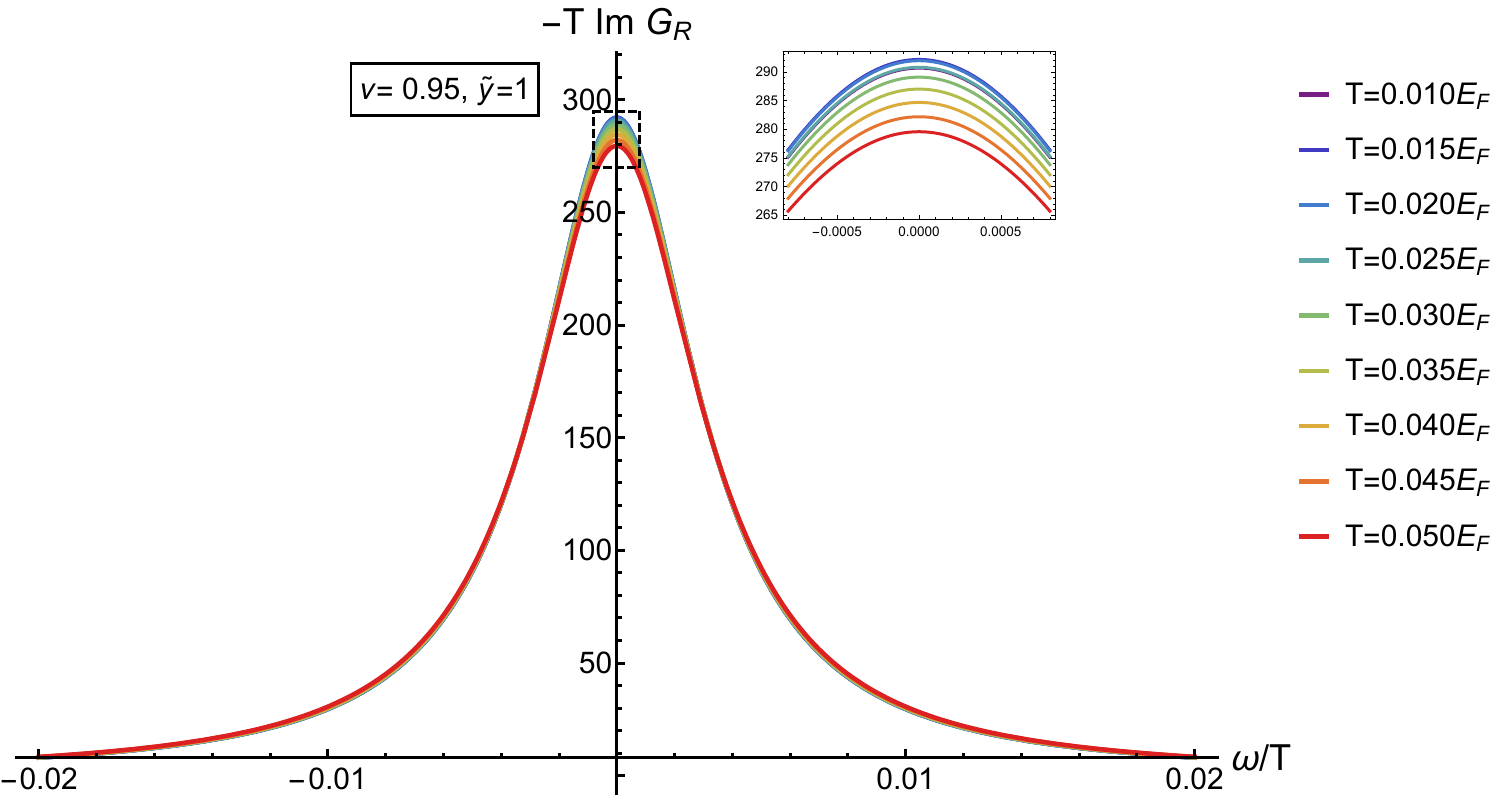}
\includegraphics[width= 0.45\textwidth]{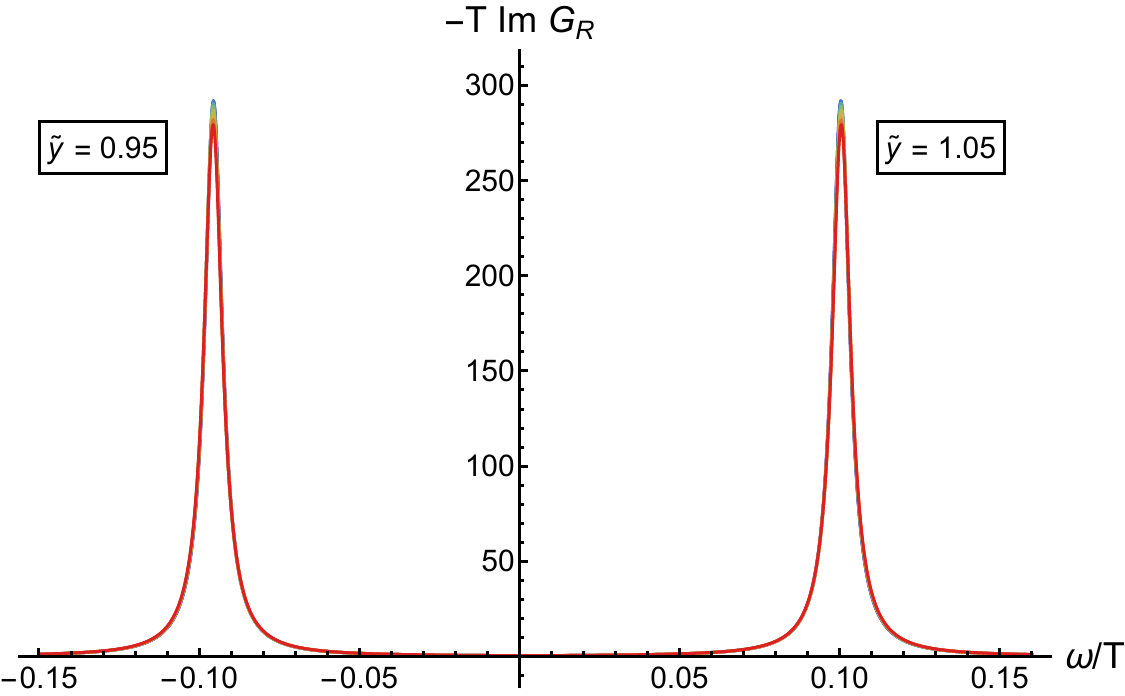}
\caption{ Plot of $(1/2)T \rho$ as a function of $\omega/T$ for various $T \geq 0.01 E_F$ at the optimal $\alpha/\gamma = 100$ for $\nu = 0.95$ on the Fermi surface (left), as well as near it (right). In all plots, $\gamma = 0.001$.}\label{Fig:spectraluniversality}
\end{figure}

Before quantifying quasi-universality appropriately, let us present some plots which demonstrate it. Fig.~\ref{Fig:spectraluniversality} shows plots of $\tilde{\rho}/2$  as a function of the dimensionless frequency $\tilde{x}$ for different fixed values of the dimensionless momentum $\vert \tilde{\mathbf{y}} \vert \sim 1$ (i.e. in the vicinity of the Fermi surface) and different temperatures at $\nu = 0.95$ for which the corresponding optimal ratio of $\alpha/\gamma$ is $100$ (we have chosen $\gamma = 0.001$). These plots show that the $\tilde{\rho}$ is indeed the same function of  $\tilde{x}$ and $\vert \tilde{\mathbf{y}}\vert$, and thus independent of the temperature for temperatures above $0.01 E_F$. Clearly, the quasi-universal form is valid for all frequencies and near the Fermi surface at the optimal ratio of the coupling. Similarly, from the plots in Fig.~\ref{Fig:spectraluniversalityWnu}, we can infer that $\tilde{\rho}$ is also independent of the scaling exponent $\nu$ at the corresponding values of the optimal ratio of the couplings both at and near the Fermi surface, and for all frequencies. The optimal ratio of $\alpha$ to $\gamma$ at various values of $\nu$ can be read off from Fig. \ref{Fig:spectraluniversalityWnu} where the value of $\gamma$ is fixed to $0.001$. 

\begin{figure}[h]
\centering
\includegraphics[width= 0.45\textwidth]{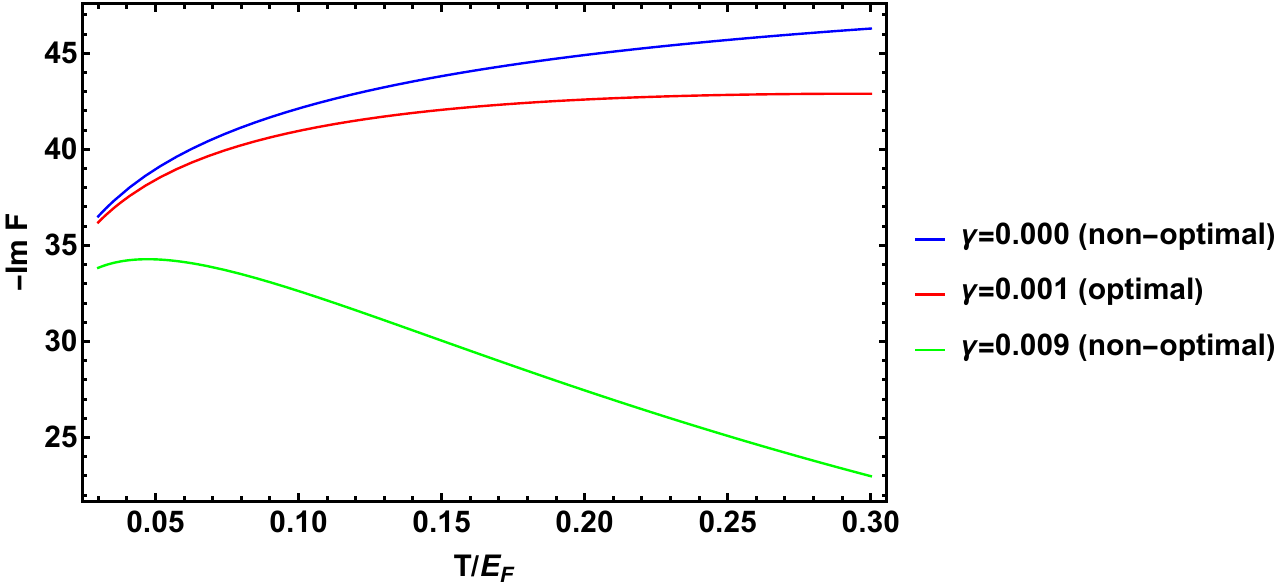}
\includegraphics[width= 0.45\textwidth]{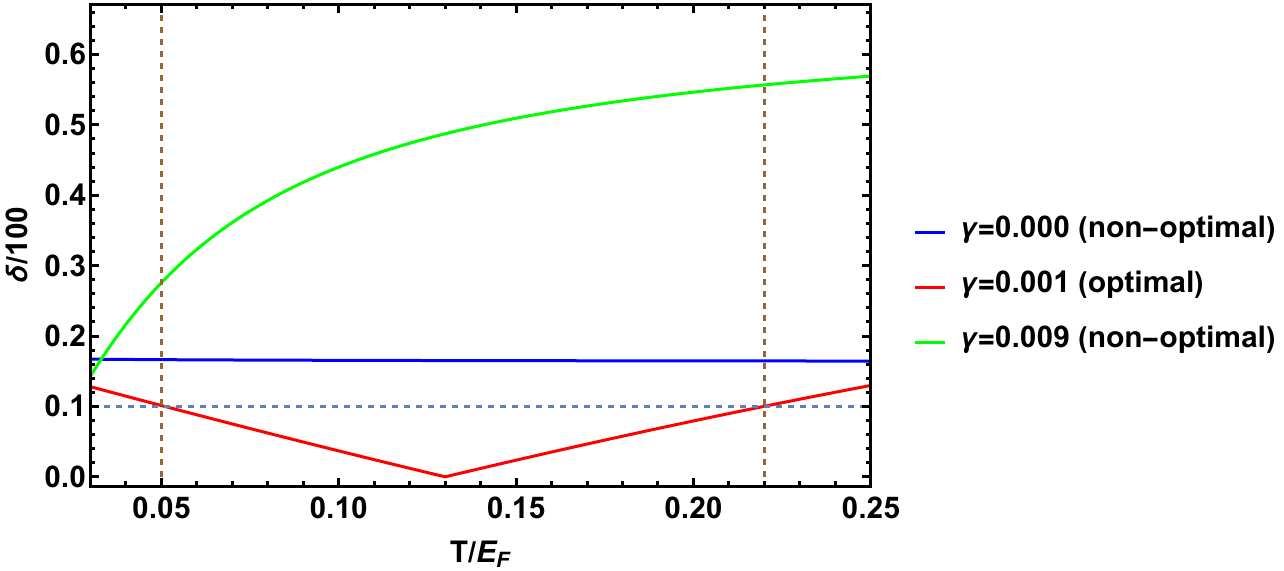}
\caption{\textbf{Left:} Plots of $- {\rm Im}F = - T {\rm Im}G_R$ as a function of $x_T = T/E_F$ at $\nu =0.8$ for $\gamma =0$ (the blue curve) and $\gamma =0.001$ (the red curve) and $\gamma = 0.009$ (the green curve) on the Fermi surface at $\alpha = 0.023$ and $\tilde{x}=\omega/T = 0.01$. The red curve is in the universal regime and is, as expected, flatter than the other curves. However, a precise characterization of universality, as explained in the text and shown explicitly in the tables, should rely on the global dependence of $\tilde{x}$ and not based on the behavior at a particular value of the $\tilde{x}$. \textbf{Right:} Plots of $\delta/100$ vs $x_T$ for the three cases demonstrating that $\delta < 10$ only for $\gamma = 0.001$ for a wide range of temperatures. Thus, $\delta <10$ is a good measure of universality and takes into account the global dependence on $\tilde{x}$. We also readily see that when $\gamma = 0.001$ and $\nu = 0.8$, $\delta < 10$ is satisfied for $t_c < x_T < \tilde{t}_c$ with $t_c\approx 0.05$ and $\tilde{t}_c \approx 0.22$.}\label{Fig:UnivComp}
\end{figure}

However, a more quantitative way of characterizing quasi-universality is required. Let's compare a point on the strange metallic submanifold, $p_s ~(\alpha=0.023,~ \gamma=0.001)$ for $\nu = 0.80$ where quasi-universality is expected, with another point $p_n ~(\alpha=0.023,~ \gamma=0)$ for the same $\nu = 0.80$ which corresponds to the simple Faulkner-Polchinski model. If we compare the temperature dependence of the spectral function at $p_s$ (the red curve in Fig. \ref{Fig:UnivComp}) with that at point $p_n$ (the blue curve in Fig. \ref{Fig:UnivComp}) for a fixed value of the dimensionless frequency $\tilde{x} = 0.01$ and on the Fermi surface, then the dependence on the temperature at $p_s$ is indeed flatter, as expected. However, the temperature dependence even at $p_s$ does not seem to be negligible at least for such a low value of the frequency, and apriori it is not clear why the strange metallic linear-in-$T$ resistivity can be found at the point $p_s$, which is in the quasi-universal regime but not at the point $p_n$ corresponding to the simple Faulkner-Polchinski model (this is indeed the case as discussed in the next section). The point is that the temperature dependence of $\tilde\rho$ has to be sufficiently small for all dimensionless frequencies and not just at a specific value of the frequency. This is crucial to reproduce linear-in-$T$ resistivity as will be explicitly shown in the next section. Therefore, we need a quantitative measure of quasi-universality on the strange metallic submanifold which takes into account all frequencies.

\begin{table}[h]
\centering
\begin{tabular}{|c|c|c||c|c|c|}
\hline
 \multicolumn{3}{|c||}{  Optimal }  &\multicolumn{3}{c|} {Non-Optimal}   \\
\hline
\hline
    $\nu$ & $\alpha$ & $\delta$ & $\nu$ & $\alpha$ & $\delta$  \\ \hline
    0.95 & 0.100 & 7.3 & 0.95 & 0.300 & 20.5 \\ \hline 
    0.90 & 0.050 & 2.3 & 0.90 & 0.200 & 16.7 \\ \hline
    0.85 & 0.030 & 0.7 & 0.87 & 0.900 & 33.2 \\ \hline
    0.75 & 0.018 & 0.4 & 0.99 & 0.300 & 35.6 \\ \hline
    0.70 & 0.015 & 0.5 & 0.85 & 0.400 & 16.2 \\ 
\hline
\end{tabular}
\caption{Left: $\delta(\vert \tilde{\mathbf{y}}\vert =1)$ for ``optimal set of values" of $\alpha$ for given choices of $\nu$ and $\gamma =0.001$. We see that $\delta(\vert \tilde{\mathbf{y}}\vert =1) <10$. Right: The same for some set of values which are ``not optimal" for which $\delta$ is significantly larger than $10$. }
\label{universalitytablefig2}
\end{table}

A suitable measure of quasi-universality which correlates strongly with linear-in-$T$ resistivity is as follows. Let us first define a standard spectral function at a point on the strange metallic submanifold:

\begin{equation}\label{Eq:standard}
    f_s(\tilde{x}, \tilde{\mathbf{y}}) := \tilde\rho(\tilde{x}, \tilde{\mathbf{y}})[\nu = 0.8, \alpha = 0.023, \gamma = 0.001, x_T = 0.13].
\end{equation}

The measure of quasi-universality is $\delta$, which takes into account the modulus of the deviation of $\tilde{\rho}$ from $f_s$ at all frequencies, and is simply\footnote{At high frequencies, the spectral function can be modified in the material, and we have also restricted the self-energy corrections only to two loops. So the spectral sum rule is not necessarily satisfied for $f_s$.}
\begin{equation}\label{Eq:delta}
    \frac{\delta(\vert \tilde{\mathbf{y}}\vert, x_T)}{100} = \frac{\int_{-\infty}^\infty \vert\tilde{\rho}(\tilde{x}, \tilde{\mathbf{y}}, x_T)- f_s(\tilde{x}, \tilde{\mathbf{y}})  \vert {\rm d}\tilde x}{\int_{-\infty}^\infty f_s (\tilde{x}, \tilde{\mathbf{y}}){\rm d}\tilde x}.
\end{equation}
On the strange metallic submanifold, we require that $\delta(\vert \tilde{\mathbf{y}}\vert, x_T) < 10$ for a range of $\vert\mathbf{y}\vert$ near the Fermi surface ($\vert \tilde{\mathbf{y}}\vert =1$) such that the loop integrals get maximum contribution from loop momenta taking values within this range. For brevity, we will refer to $\delta(\vert \tilde{\mathbf{y}}\vert =1, x_T)$ as $\delta(x_T)$. As shown in the right figure of Fig. \ref{Fig:UnivComp},  $\delta(x_T) < 10$ is realized for $0.05< x_T < 0.22$ when $\gamma = 0.001$ and $\nu = 0.80$ at the optimal ratio of coupling (corresponding to $\alpha = 0.023$). Thus we can determine that $t_c = 0.05$ and $\tilde{t}_c = 0.22$ when $\gamma = 0.001$ and $\nu = 0.80$. Away from the optimal ratio of the couplings (i.e. the strange metallic manifold), the range of temperatures where $\delta (x_T) < 10$ is realized is either vanishing or narrow. Similarly, $\delta (x_T) < 10$ is realized for a sufficiently wide range of temperatures only if the range of $\nu$ is between 0.66 and 0.95. Thus, the extent of the strange metallic sub-manifold can be estimated from $\delta(x_T)$. The choice of the standard curve \eqref{Eq:standard} gives the maximum range of temperatures for which $\delta (x_T) < 10$ is realized, and the $t_c$ thus obtained coincides with the lower temperature threshold for linear-in-$T$ resistivity to a very good approximation.

\begin{table}[h]
\centering
\begin{tabular}{|c|c|c||c|c|c|}
\hline
 \multicolumn{3}{|c||}{  Optimal }  &\multicolumn{3}{c|} {Non-Optimal}   \\
\hline
\hline
    $\nu$ & $\gamma$ & $\delta$ & $\nu$ & $\gamma$ & $\delta$  \\ \hline
    0.95 & 0.0015 & 6.1 & 0.95 & 0.0030 & 22.7 \\ \hline 
    0.85 & 0.0012 & 2.7 & 0.85 & 0.0030 & 23.8 \\ \hline
    0.75 & 0.0011 & 1.7 & 0.75 & 0.0025 & 19.6 \\ \hline
    0.90 & 0.0014 & 5.2 & 0.90 & 0.0020 & 13.0 \\ \hline
\end{tabular}
\caption{$\delta(\vert \tilde{\mathbf{y}}\vert =1)$ for  optimal and non-optimal values of $\gamma$ for corresponding values of $\nu$ with $\alpha =0.023$ and $T/E_F = 0.13$ fixed. We see that $\delta(\vert \tilde{\mathbf{y}}\vert =1)< 10$ for optimal cases, while for non-optimal cases $\delta$ is much larger than $10$. This also provides evidence that the universality is determined only by the ratio of $\alpha/\gamma$.}
\label{non-optimal gamma}
\end{table}

In Table ~\ref{universalitytablefig2}, we report the values of $\delta$ when we move away from the strange metallic submanifold by varying the value of $\alpha$ for various fixed values of $\nu$ and $\gamma=0.001$ at $x_T =0.13$. In Table ~\ref{non-optimal gamma}, we do a similar analysis when $\gamma$ is varied from the optimal value at various fixed values of $\nu$ while keeping $\alpha =0.023$ and $x_T =T/E_F = 0.13$. Both the tables clearly indicate that the value of $\delta$ is significantly smaller than $10$ for all points on the strange metallic submanifold, whereas the points outside of the manifold exhibit values of $\delta$ much larger than $10$. These tables together suggest that the quasi-universal regime is determined by the ratio of the couplings rather than their individual values.  The optimal ratio corresponds to the smallest value of $\delta$ at a fixed value of $\nu$. 

 Furthermore, we observe in Table ~\ref{non-optimal gamma} that a small variation in $\gamma$ can result in a significant change in $\delta$, exhibiting a significant departure from quasi-universality. Therefore, even though $\gamma$ is small and $\alpha$ is typically larger at the optimal values of the ratio of the two couplings, the effect of $\gamma$ is indeed significant at finite temperatures where quasi-universality holds.

Crucially, we need to find the range of $\vert \tilde{\mathbf{y}}\vert$ over which $\delta(\vert \tilde{\mathbf{y}}\vert) < 10$. We find that $\delta(\vert \tilde{\mathbf{y}}\vert) \lessapprox 10$ for $0.25\lessapprox \vert\tilde{\mathbf{y}}\vert \lessapprox 1.75$ over the entire range of $\nu$ from $0.66$ to $0.95$ and within most of the corresponding range of temperatures $t_c < x_T  < \tilde{t}_c < 1$. As discussed in the next section, this explains the $\nu$-independent linear-in-$T$ resistivity from the semi-holographic effective theory over a wide range of temperatures.

An estimate of how much fine-tuning of the two couplings is needed to produce such a quasi-universality can be obtained from studying the extent of variation of the couplings leads to the disappearance of the feature of linear-in-$T$ resistivity in the quasi-universal regime. From  the results of \cite{Doucot:2020fvy} (see Fig. \ref{Fig:slope dc optimal} which is reproduced from \cite{Doucot:2020fvy} particularly), we may infer that the required fine-tuning is about $10$ \%.

\paragraph{Comparing with ARPES data:}  We expect that there is a range of doping where strange metallicity holds at finite temperatures. Since $\alpha$ and $\gamma$ are the coefficients of the holographic and Fermi-liquid-like contributions to the self-energy, respectively, we expect that significant overdoping which leads to appearance of Fermi liquid behavior corresponds to values of $\alpha/\gamma$ much smaller than the optimal value at a fixed value of $\nu$, and similarly significant underdoping where pseudo-gap and other phases emerge corresponds to $\alpha/\gamma$ much larger than the optimal value. 

\begin{figure}[h]
\includegraphics[width=1.0\textwidth]{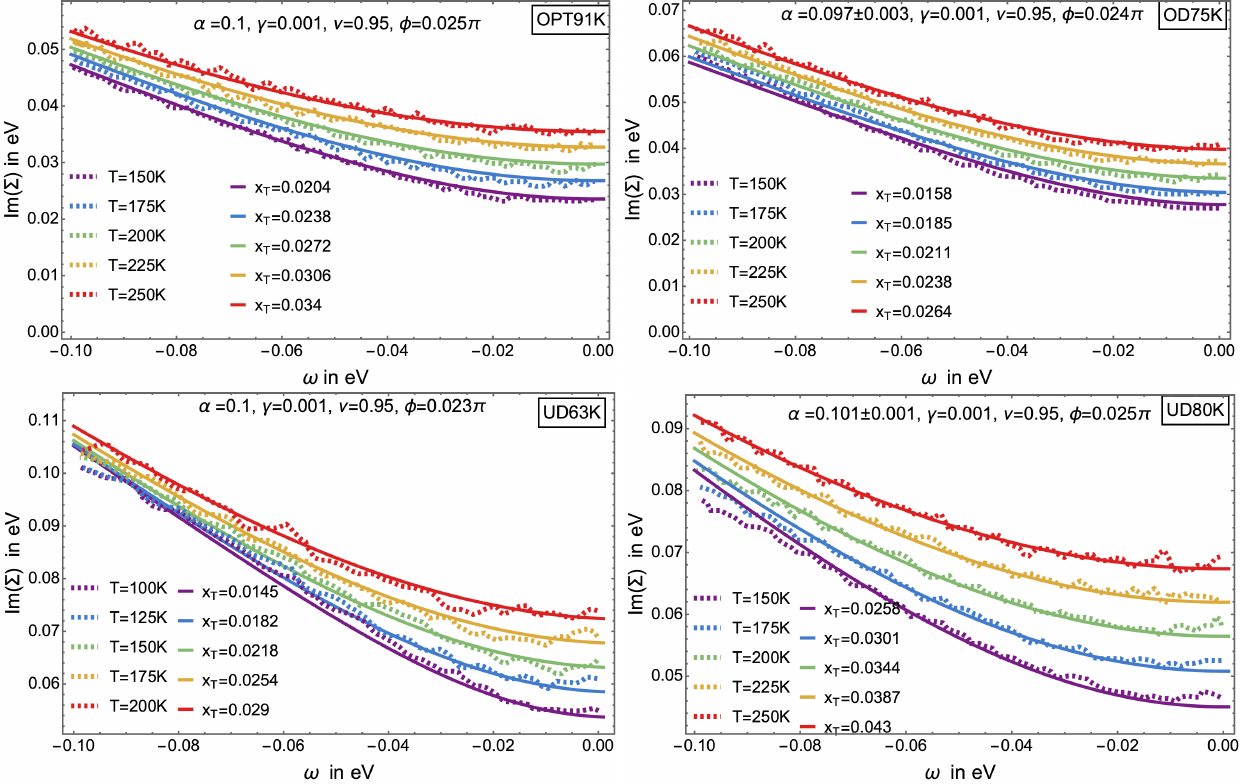}
\caption{Fit of the imaginary part of self-energy obtained from Eq. \eqref{Eq:propagator1} with the nodal ARPES data of Bi$_2$Sr$_2$CaCu$_2$O$_{8+\delta}$ reported in  \cite{Reber2019}. The labels OPT, OD and UD denote optimal/over/under-doping, and the following numbers denote the critical for the superconducting transition in K. The coupling $\gamma$ and the exponent $\nu$ are not varied with the doping (and the temperature). The best fit values of $\alpha$ are indicated above for each doping. Note $\alpha$ decreases mildly with overdoping and increases slightly with under-doping but remarkably always close to the value where quasi-universality is exhibited.}\label{selfenergy}
\end{figure}

As shown in Fig.~\ref{selfenergy}, our spectral function fits well with the \textit{nodal} ARPES data for the imaginary part of the self-energy for different underdoped, overdoped and optimally doped samples of Bi$_2$Sr$_2$CaCu$_2$O$_{8+\delta}$ at various temperatures (where strange metallicity is exhibited) as reported in \cite{Reber2019}.\footnote{The data have been extracted from the paper using WebPlot Digitizer tool available at \url{https://automeris.io/WebPlotDigitizer/}.} Crucially, our fits work well
\begin{itemize}
    \item keeping the scaling exponent $\nu = 0.95$ fixed as the doping (and also the temperature) is varied, 
    \item the Fermi energy ($\sim 5500-7500$K) fixed with the temperature at any value of the doping, and
    \item also keeping $\gamma = 0.001$ fixed for all dopings (and temperatures).
\end{itemize}

The exponent $\nu = 0.95$ is remarkably stable across doping, as evident from Fig.~\ref{selfenergy}. Note in Fig.~\ref{selfenergy} that we fit at temperature $100K$ also. Only when $\nu = 0.95$ the lower threshold temperature where quasi-universality emerges is as low as $x_T = 0.01$ and $E_F$ is not expected to $10^4K$. This actually limits our choice of $\nu$ used in the fitting, and so we have chosen $\nu = 0.95$.  Our fits with adjusted R-square value above 0.9 in Fig.~\ref{selfenergy} have been obtained by varying only $\alpha$ and $E_F$ with the doping (but of course not with the temperature) and keeping $\gamma = 0.001$ fixed. Remarkably, we indeed find that the range of $\alpha$ obtained by fitting corresponds is very close to the optimal value where quasi-universality is exhibited (for all dopings). Furthermore, we note from  Fig.~\ref{selfenergy} that $\alpha$ decreases mildly with overdoping and increases mildly with underdoping, as expected. 

From our discussion in Sec. \ref{Sec:Micro} where we promote $\alpha$ to an effective field while embedding the refined semi-holographic model in a more general effective approach to explain how the fine-tuning of the ratio of couplings can be achieved dynamically in a typical state, it can be expected that the doping affects only $\alpha$ and not $\gamma$. Therefore, the fits confirm the intuitive understanding of the microscopic origin of our model. However, note that the data is only for lower frequencies and temperatures where varying $\gamma$ gives sub-dominant effects.

In \cite{Reber2019}, good fits with the same data were obtained by varying multiple exponents with the doping using a different phenomenological ansatz which was improved in \cite{Smit} from other theoretical considerations. However, our fits work well with more constraints on parameters obtained from the physical considerations of our effective approach, and especially without varying the scaling exponent $\nu$. 

All our fits have adjusted R-squared value above 0.9. However, we are only using a digitized version of the published data (as the raw data is not accessible to us) for fitting and we have used noise filtering before fitting. Therefore, our fitting can have potential systematic errors which do not allow us to compare with competitive models reliably. Regardless, it is significant that we have been able to fit the nodal ARPES data with the quasi-universal spectral function at multiple temperatures and dopings, changing essentially only $\alpha$ and $E_F$ with the doping, and with $\nu$ kept fixed to a value chosen from physical considerations.

\begin{figure}[h]
\centering
\includegraphics[width= 0.45\textwidth]{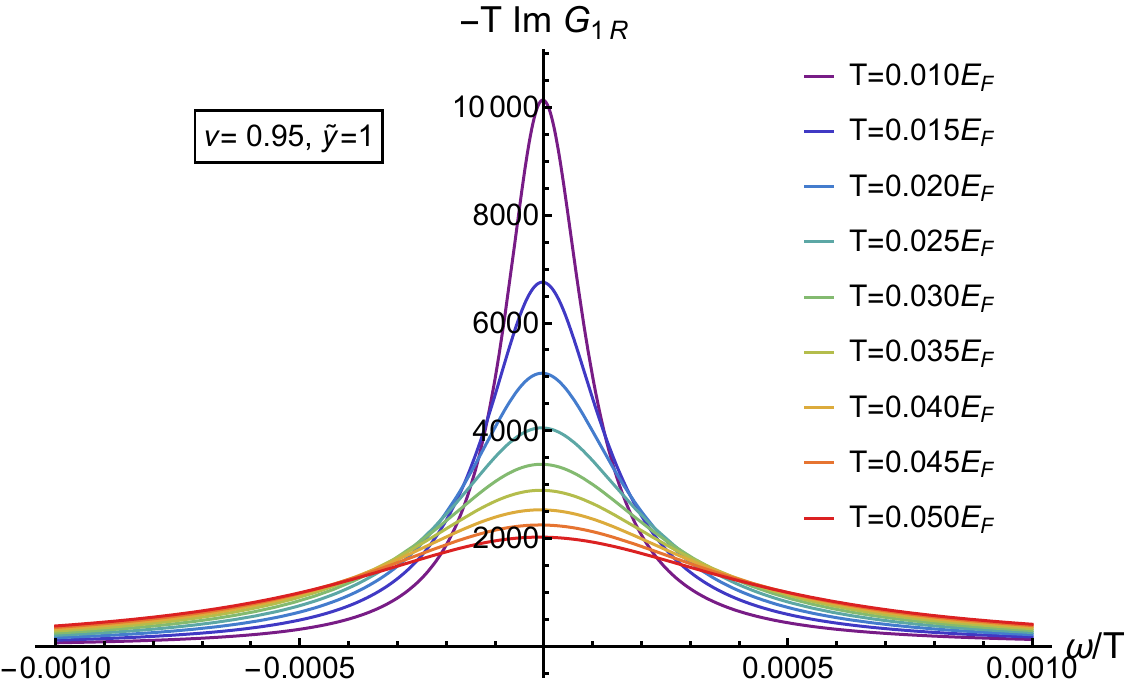}
\includegraphics[width= 0.45\textwidth]{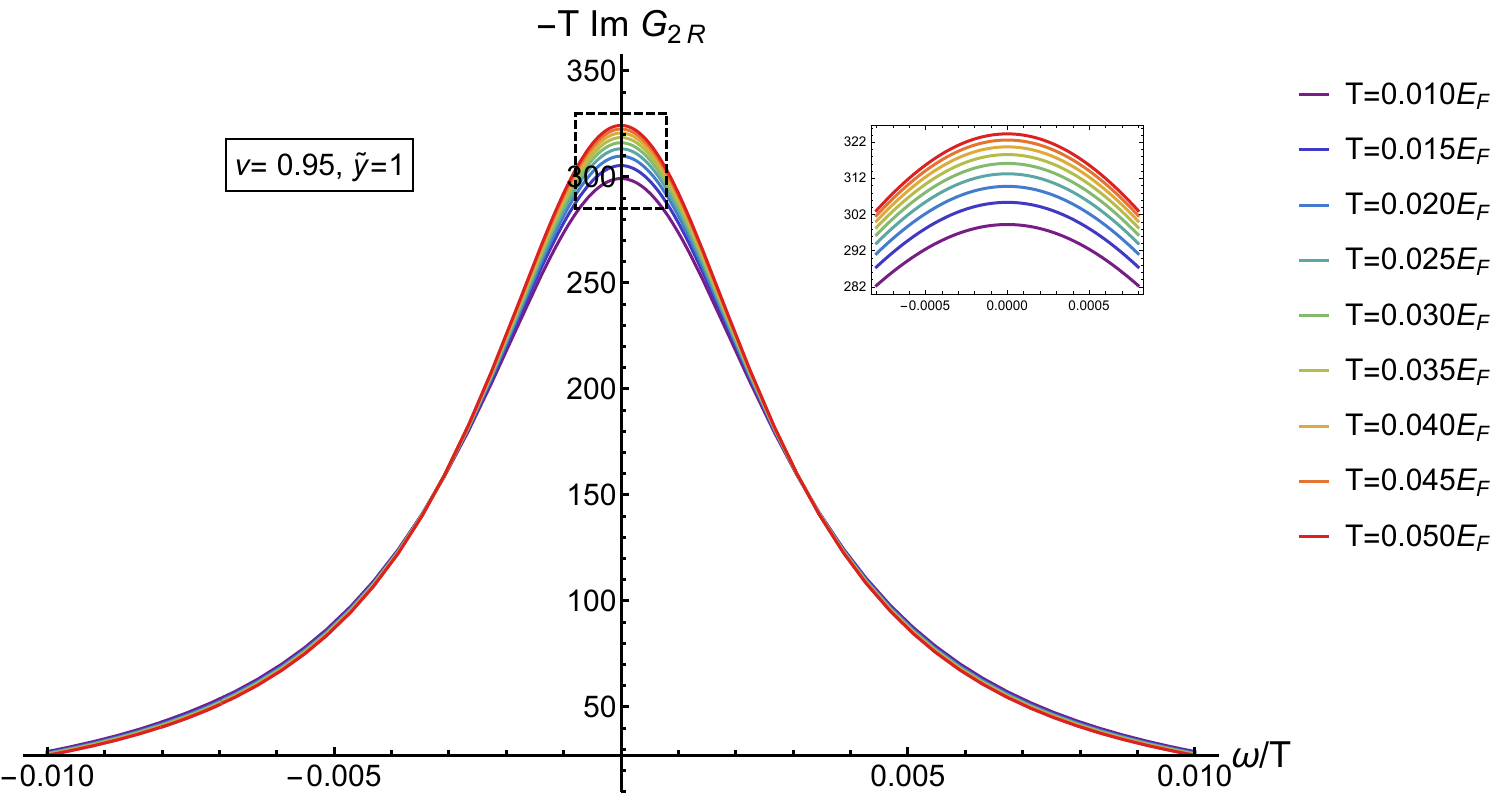}
\caption{\textbf{Left --} Plot of -$T$Im$G_{1R}$ as a function of $\tilde{x}$ on the Fermi surface with same values of parameters as in Fig.~\ref{Fig:spectraluniversality}.  $G_{1R}$ is the propagator in which the holographic contribution to the self energy is replaced by its zero temperature form. \textbf{Right --} Plot of -$T$Im$G_{2R}$ as a function of $\tilde{x}$ on the Fermi surface  with same values of parameters as in Fig.~\ref{Fig:spectraluniversality}. $G_{2R}$ is the propagator in which 
the Fermi liquid contribution to self energy is replaced by its zero temperature form. Comparison with the inset in Fig.~\ref{Fig:spectraluniversality} and numerical tests indicate deviation from quasi-universality which turns out to be indeed significant leading to loss of strange metallic properties.}
\label{Fig:Temp dependence oh each sector}
\end{figure}

\paragraph{Comments on the importance of the temperature dependence:} A pertinent question is which special features of the retarded propagator \eqref{Eq:Spec-Universal} of our model obtained by retaining only two loop contributions to the self-energy are responsible for quasi-universality. We have verified that the temperature dependence in both the holographic and Fermi liquid contributions to the self-energy contributions is crucial. Let us construct a new propagator $G_{1R}$ in which we replace the holographic contribution to the self energy by its zero temperature form (which is simply proportional to $\omega^\nu$). As shown in the left plot of Fig. \ref{Fig:Temp dependence oh each sector}, in this case unlike the original propagator $-T {\rm Im}G_{1R}$ does not demonstrate quasi-universality even on the Fermi surface. In this figure, we have retained the same ratio of couplings which gives rise to quasi-universality (at $\nu = 0.95$) but we do not recover quasi-universality even after varying the ratio of couplings away from this value. Similarly, we can check that quasi-universality is not recovered with a new propagator $G_{2R}$ in which the temperature dependent part of the Fermi-liquid self energy (i.e. the $\pi^2 T^2$ term) is dropped. As shown in the right plot of Fig. \ref{Fig:Temp dependence oh each sector}, here the loss of quasi-universality seems less catastrophic, but upon closer inspection and also checking with a numerical criterion (e.g. by evaluating $\delta$ and checking if it is larger than our prescribed value) we find that the deviation from quasi-universality is significant. Indeed we also checked that the linear-in-$T$ resistivity is not recovered in this case as well.

We therefore learn that a microscopic derivation of the temperature dependence of the spectral function has been crucial for us to obtain a quasi-universal behavior of the spectral function on a sub-manifold in the space of couplings. Furthermore, we also learn how the temperature dependence of the holographic and Fermi liquid contributions to the self-energy is crucial for these to compete and realize quasi-universality as mentioned before. Although we have not precisely determined which form of temperature dependence of self-energy leads to quasi-universality more generally, it is worth emphasizing that such a temperature dependence should be derived from an underlying effective theory as in our model.

\section{DC conductivity, Hall conductivity, and Planckian dissipation}\label{Sec:DC}

\subsection{Linear-in-$T$ resistivity and the Hall conductivity}
With the underlying assumption that the charge conduction is primarily due to the propagating degrees of freedom on the Fermi surface, we can readily derive that the dc conductivity takes the form
\begin{align}\label{Eq:dcconductivity}
\hspace{-0.27cm}\sigma_{\text{dc}} = \frac{e^2}{m^2}\int \frac{d\omega}{2\pi} \int \frac{d^2 k}{4\pi^2} k^2 \rho(\omega, \mathbf{k}, T)^2 \left(-\dfrac{\partial n_F(\omega/T)}{\partial \omega}\right).
\end{align}
at one loop and in terms of the spectral function $\rho$ obtained from the propagator in Eq. \eqref{Eq:propagator1}. Above $n_F(\omega/T) = (e^{\omega/T}+ 1)^{-1}$ is the Fermi-Dirac distribution. Assuming that both the couplings $\alpha$ and $\gamma$ are small, we can ignore the higher loop corrections, including corrections to the electromagnetic vertex.

We can demonstrate that the linear-in-$T$ dc resistivity should follow from the quasi-universality of the spectral function discussed in the previous section provided the integration over the momenta in Eq. \eqref{Eq:dcconductivity} receives dominant contribution only from the vicinity of the Fermi surface. To see this we can do a change of variables  $(\tilde{x},\tilde{\mathbf{y}})$ from $(\omega,k)$ and replace  $\rho(\omega, \mathbf{k}, T)$ by $-2 T^{-1}{\rm Im} F(\tilde{x},\tilde{\mathbf{y}})= T^{-1} \tilde{\rho}(\tilde{x},\tilde{\mathbf{y}})$ in \eqref{Eq:dcconductivity} following Eq. \eqref{Eq:Spec-Universal}. Note that ${\rm d}\omega = T{\rm d}\tilde{x}$ and ${\rm d}k \propto T{\rm d}\vert \tilde{y}\vert$. With the further assumption that the integration over the loop momenta receives a major contribution from the vicinity of the Fermi surface ($1- y_c \leq\vert \tilde{\mathbf{y}} \vert \leq 1 + y_c$), the one loop formula \eqref{Eq:dcconductivity} for the dc conductivity reduces to
\begin{align}\label{Eq:dc2}
   & \sigma_{\text{dc}} = \frac{e^2}{m^2}\int_{-\infty}^\infty \frac{T{\rm d}\tilde{x}}{2\pi} \int_{1-y_c}^{1+y_c} \frac{2\pi \vert \tilde{\mathbf{y}} \vert{\rm d} \vert \tilde{\mathbf{y}} \vert }{4\pi^2} k_F^4 \frac{T}{E_F}(1+ \mathcal{O}(x_T \vert \tilde{\mathbf{y}}\vert)) \nonumber\\&\qquad\qquad \left(-T^{-1} \tilde{\rho}(\tilde{x},\tilde{\mathbf{y}})\right)^2 \left(-T^{-1}\dfrac{\partial n_F(\tilde{x})}{\partial \tilde{x}}\right)\nonumber\\
   &=\frac{2}{ \pi^2}e^2 E_F T^{-1}  \int_{-\infty}^\infty {\rm d}\tilde{x}\int_{1-y_c}^{1+y_c} {\rm d} \vert \tilde{\mathbf{y}} \vert \,\,\vert \tilde{\mathbf{y}}\vert(1+ \mathcal{O}(x_T \vert \tilde{\mathbf{y}}\vert))  \tilde{\rho}(\tilde{x},\tilde{\mathbf{y}})^2 \left(-\dfrac{\partial n_F(\tilde{x})}{\partial \tilde{x}}\right)
\end{align}
where we have used $E_F = k_F^2/(2m)$. For a simpler accounting of $T$-dependence in \eqref{Eq:dcconductivity}, note that ${\rm d\omega}\propto$ $T$ and ${\rm d^2}k \propto$ $T$ (near the Fermi surface), while  $\partial/\partial\omega \propto$ $T^{-1}$, $\rho\propto$ $T^{-1}$ and $k^2 \sim k_F^2$ (near the Fermi surface). Therefore, we have an overall $T^{-1}$ dependence above.

Quasi-universality of the spectral function implies that $\tilde{\rho}$ is approximately independent of the temperature and even the exponent $\nu$ at the optimal value of the ratio of $\alpha$ to $\gamma$ over the corresponding range of temperatures $t_c \leq x_T \leq  \tilde{t}_c <1$ for all dimensionless frequencies $\tilde{x}$ and for dimensionless momenta $\vert\tilde{\mathbf{y}}\vert$ near the Fermi surface i.e. for $1-\epsilon <\vert\tilde{\mathbf{y}}\vert < 1 +\epsilon$ with $\epsilon < 1$. If quasi-universality  holds for $1-y_c < \vert\tilde{\mathbf{y}}\vert < 1+y_c$ with $y_c \geq \epsilon$ and the $\mathcal{O}(x_T\vert \tilde{y}\vert)$ can be ignored, then it follows from Eq. \eqref{Eq:dc2} that the dc conductivity scales as $T^{-1}$  and is independent of the exponent $\nu$ over the wide range of temperatures $t_c \leq x_T \leq \tilde{t}_c < 1$ to a very good approximation. Thus, the quasi-universality of the spectral function implies the strange metallic linear-in-$T$ resistivity\footnote{Resistivity is the inverse of dc conductivity.} if the loop integral gets a dominant contribution from loop momenta near the Fermi surface. Since quasi-universality emerges only when the exponent $\nu$ assumes values in the range $0.66\lessapprox\nu\lessapprox0.95$, the linear-in-$T$ resistivity should be exhibited only in this range of $\nu$ at the corresponding optimal ratio of $\alpha$ to $\gamma$ (for temperatures $t_c \leq x_T \leq \tilde{t}_c < 1$), and the resistivity should also be independent of $\nu$ within the same range of $\nu$. It turns out that the $\mathcal{O}(x_T\vert \tilde{y}\vert)$ terms can indeed be ignored as the maximum temperature $\tilde{t}_c E_F$ where quasi-universality holds satisfies $\tilde{t}_c <1$.

As discussed in the previous section, we have used the requirement $\delta (\vert\tilde{\mathbf{y}}\vert, x_T) < 10$ to characterize quasi-universality and have found that it holds when $0.25<\vert\tilde{\mathbf{y}}\vert<1.75$ (implying that $y_c \sim 0.75$) for the range of $\nu$ from 0.66 to 0.95 and for most of the entire corresponding range of temperatures $t_c \leq x_T \leq \tilde{t}_c <1 $. We also find that for these ranges of $\nu$ and temperatures, the integral \eqref{Eq:dcconductivity} gets 90 \% contribution from $0.3<\vert\tilde{\mathbf{y}}\vert<1.7$ implying that $\epsilon \sim 0.7$. We readily observe that $y_c > \epsilon$ so that we should indeed expect linear-in-$T$ resistivity to hold at least in the range of temperatures given by $t_c \leq x_T \leq \tilde{t}_c <1 $ where quasi-universality holds.

\begin{figure}[ht]
\centering
  \includegraphics[width= 0.6\textwidth]{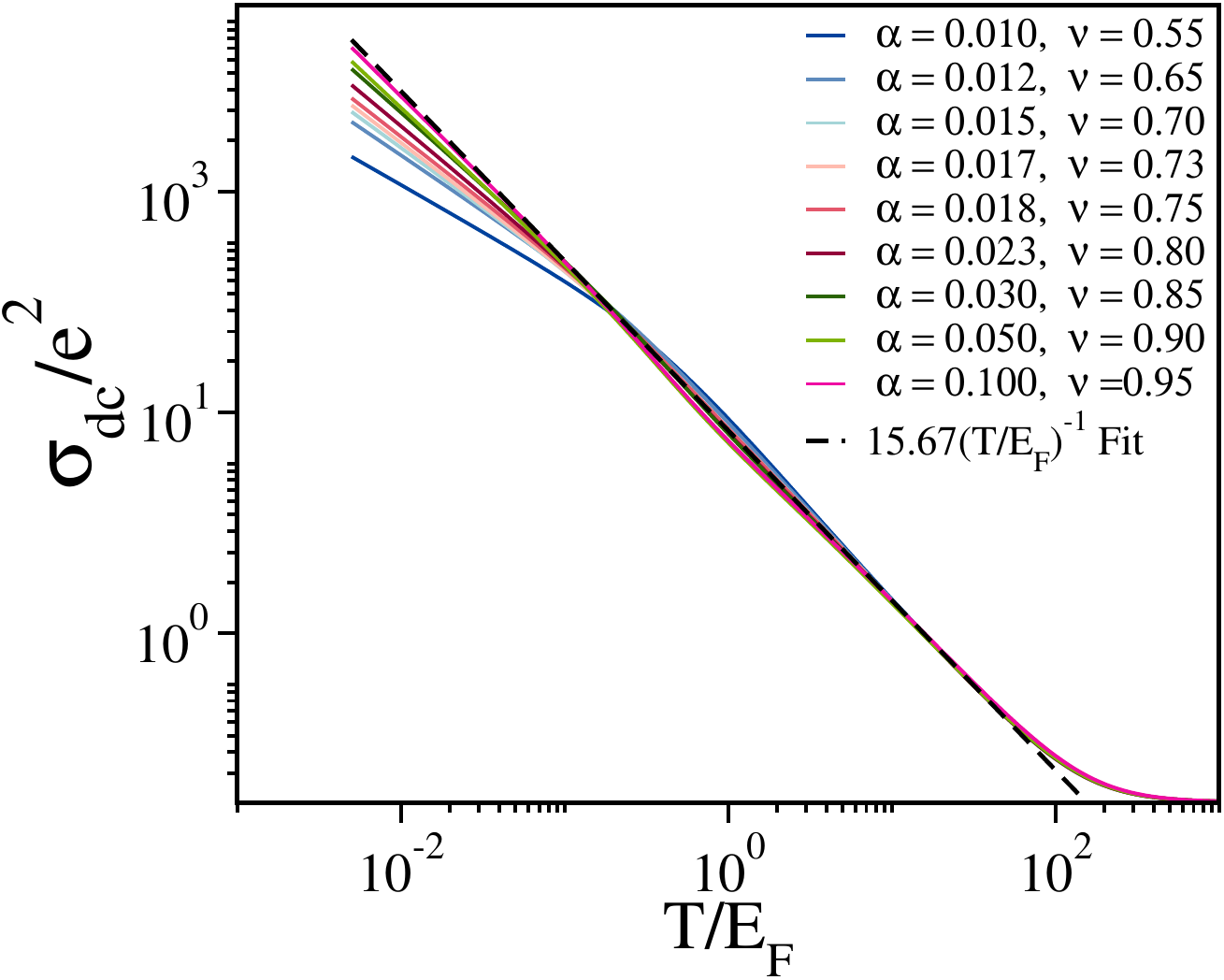}
\caption{Plot of $\sigma_{\rm{dc}}/e^2$ vs $T/E_F$ at optimal values of $\alpha$ corresponding to various values of $\nu$ with $\gamma = 0.001$. The $T^{-1}$ scaling holds in the range of temperatures in which quasi-universality of the spectral function holds. Also, $\sigma_{\rm{dc}}$ is independent of $\nu$ in this regime, as expected from quasi-universality. The only exception is $\nu = 0.55$, which is outside of the domain of values of $\nu$ where quasi-universality appears (more in text), where the $T^{-1}$ scaling appears only for $T\geq E_F$.}
\label{Fig:DCconductivity}
\end{figure}

Our expectations are indeed borne out by actual numerical computations of the dc conductivity obtained via the exact integral \eqref{Eq:dcconductivity}. As shown in Fig. \ref{Fig:DCconductivity}, the dc conductivity fits very well with
\begin{equation}\label{Eq:dcfit}
    \sigma_{\rm{dc}} = {{(15.67 \pm 0.22)}}\ e^2 \left(\frac{T}{E_F}\right)^{-1}
\end{equation}
when $\gamma = 0.001$, $\alpha$ takes optimal values corresponding to $\nu$ in the range 0.66 to 0.95 corresponding to each $\nu$. Remarkably, the linear-in-$T$ resistivity actually holds for the range of temperatures given by $t_c < x_T < t_u$ with $t_u \gg \tilde{t}_c > 1$. Although we can accurately explain the lower threshold temperature $t_c E_F$ for the validity of linear-in-$T$ resistivity from quasi-universality, it turns out that the scaling continues to hold up to much higher temperatures beyond what we can expect from quasi-universality.  Crucially, $ \sigma_{\rm{dc}}$ is indeed independent of $\nu$ in the range of $\nu$ and corresponding ranges of temperatures expected from the validity of quasi-universality. As mentioned in the previous section, the lower threshold temperature $t_c$ is larger when $\nu$ is smaller. However, the range of temperatures for which the linear-in-$T$ resistivity is exhibited for $\nu$ between 0.66 and 0.95 overlap at $0.1 \lessapprox x_T \lessapprox 10$. 

\begin{figure}[ht]
    \centering
          \includegraphics[width=0.9 \textwidth]{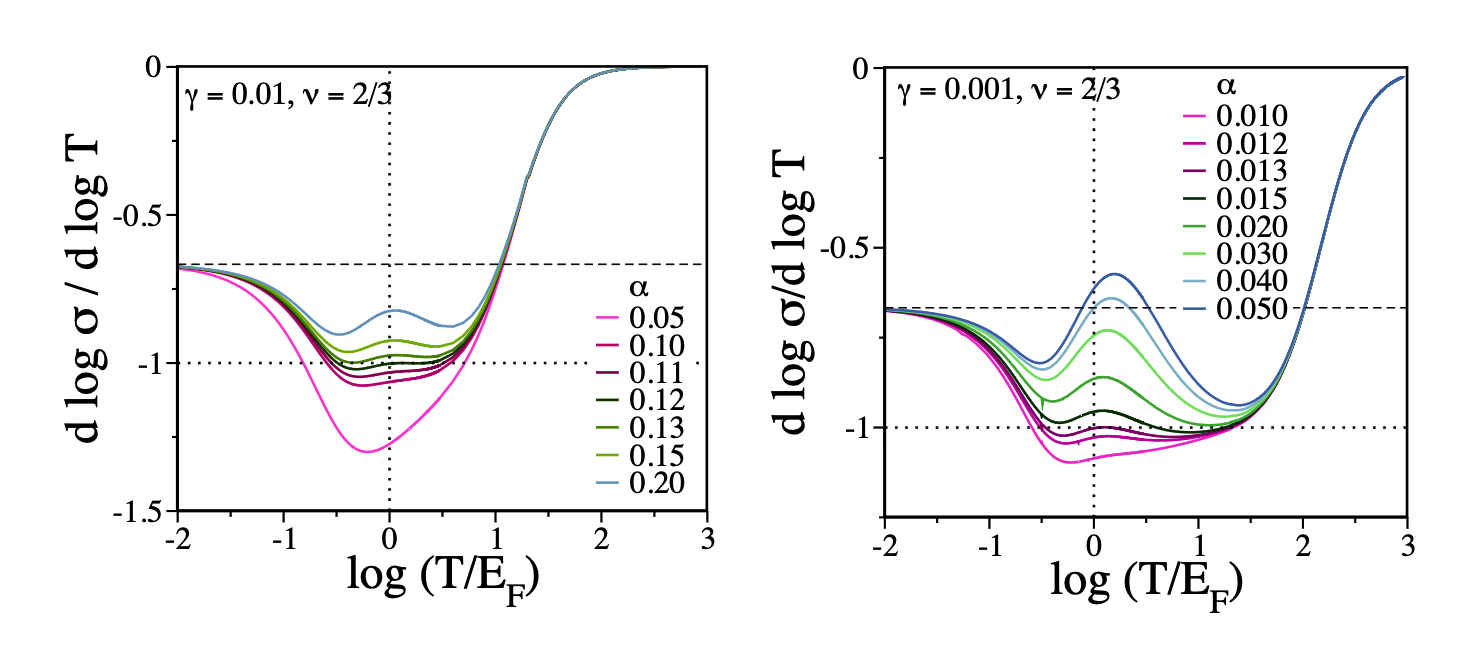}
         \caption{The temperature scaling of the dc conductivity for fixed $\nu=0.66$ is shown for different values of $\alpha$. The left plot is for $\gamma=0.01$ and the plot on the right is for $\gamma=0.001$. The upper horizontal dashed lines are at $-0.66$. It indicates that the $\sigma_{\rm{dc}} \sim$ $T^{-\nu}$ at very low temperatures. We note that the scaling $-1$ appears for a range of temperatures at $\alpha = 0.013$ as expected from the appearance of quasi-universality of the spectral function. This figure is borrowed from \cite{Doucot:2020fvy}.}
         \label{Fig:slope plot dc fixednu}
\end{figure}

\begin{figure}[ht]
    \centering
    \includegraphics[scale=0.3 ]{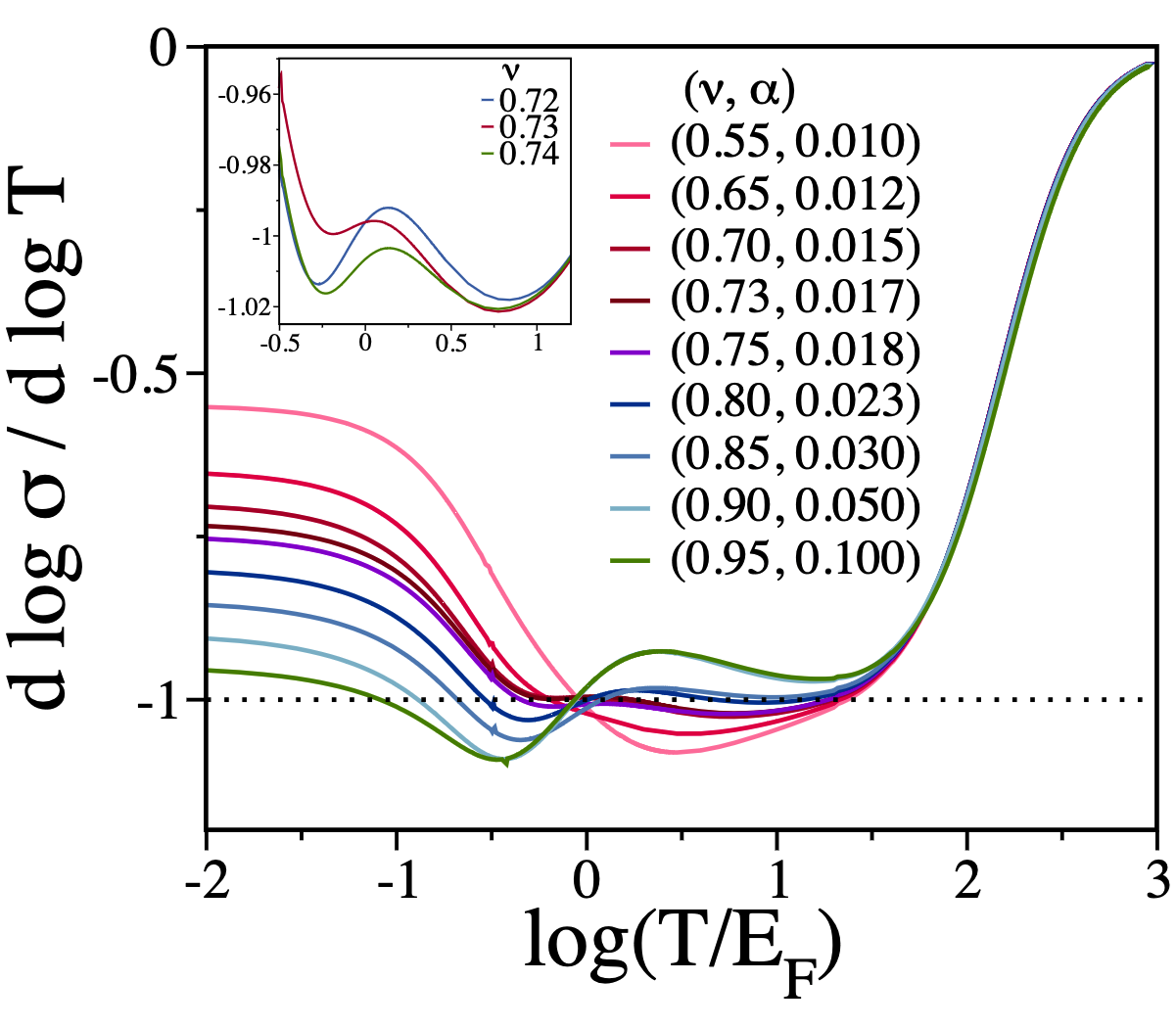}
    \caption{The plot of scaling exponent of the dc conductivity against temperature for different sets of optimal values of $\alpha$ corresponding to different values of $\nu$ when $\gamma=0.001$. Note that the scaling is almost exactly $-1$ when $T\sim E_F$. The inset plot shows that the best linear-in-$T$ scaling is obtained for $\nu = 0.73, \alpha = 0.017$. This figure is borrowed from \cite{Doucot:2020fvy}.}
    \label{Fig:slope dc optimal}
\end{figure}

In fact, the computation of the dc conductivity gives further insights into quasi-universality. If we observe the plots in Fig. \ref{Fig:slope plot dc fixednu} reproduced from \cite{Doucot:2020fvy} where the ${\rm d}\log \sigma_{\rm{dc}}/ {\rm d}\log T$ has been plotted as a function of $x_T$ at $\nu = 0.66$ for different values of $\alpha$ with $\gamma$ fixed at 0.001 (on the left) and 0.01 (on the right), we note that in both cases there is a range of temperatures (where $t_c$ correlates well with emergence of quasi-universality) where the $\frac{{\rm d}\log \sigma_{\rm{dc}}} {{\rm d}\log T }\sim -1$ when the ratio $\alpha/\gamma$ takes the optimal value 13. Interestingly, at this optimal ratio,
\begin{equation}\label{Eq:SlopeEx}
   \bigg| \frac{{\rm d}\log \sigma_{\rm{dc}}}{{\rm d}\log T}  + 1 \bigg| \quad {\rm is \,\, minimum \,\, at}\,\, x_T \sim 1
\end{equation}
for both values of $\gamma$. Although the lower threshold temperature where quasi-universality holds increases as the overall strength of the couplings increases, the scaling of $\sigma_{\rm{dc}}$ with $T$ is the closest to $T^{-1}$ at $T \sim E_F$. In fact, we also note from Fig. \ref{Fig:slope dc optimal} (reproduced from \cite{Doucot:2020fvy}) that \eqref{Eq:SlopeEx} is valid for all $\nu$ between 0.66 and 0.95 at the corresponding optimal ratio of the couplings.

We also observe from Fig. \ref{Fig:slope dc optimal} that \eqref{Eq:SlopeEx} holds for a certain ratio of couplings for $T\geq E_F$ even when $\nu = 0.55$ where quasi-universality (the temperature independence of $\tilde\rho$) is not exhibited (see also Fig. \ref{Fig:DCconductivity} for the plot of the dc conductivity in this case). However, based on the validity of \eqref{Eq:SlopeEx} at the optimal ratio of couplings, we can argue that quasi-universality is a feature which is driven primarily by the dynamics at the scale $E_F$ where the Fermi-liquid and holographic contributions to the self-energy balance out in loop integrals to produce a new emergent scaling at $T \sim E_F$. Only when the range of $\nu$ is between 0.66 and 0.95, this feature permeates over a larger range of temperatures, including lower temperatures ($T\ll E_F$) to a very good approximation. This manifests through the approximate temperature and $\nu$ independence of $\tilde\rho$. Essentially \eqref{Eq:SlopeEx} gives an important clue to understand why the linear-in-$T$ resistivity can persist up to very high temperatures $t_u E_f$ beyond the validity of quasi-universality. As our effective theory cannot be trusted for $x_T > 1$, we do not pursue this further here and leave a rigorous analytic understanding of this phenomenon to the future.

From Fig. \ref{Fig:slope plot dc fixednu} and Fig. \ref{Fig:slope dc optimal}, we note that $\sigma_{\rm{dc}}$ scales as $T^{-\nu}$ for small temperatures. This is expected because, at small temperatures and frequencies, the holographic contribution $\sim\omega^\nu$ in the self-energy dominates over the Fermi liquid contribution. We also emphasize that the lower threshold temperature $t_c E_F$ where linear-in-$T$ resistivity emerges coincides with the emergence of quasi-universality as determined by the criterion $\delta(x_T) < 10$ mentioned in the previous section. It is to be noted also that $T^{-1}$ is the only other emergent scaling which is stable over a large range of temperatures due to the emergence of quasi-universality of the spectral function. The study of the dc conductivity itself tells us that \textit{there is a unique submanifold in the space of couplings and parameters of the semi-holographic effective theory where such quasi-universal scaling behavior is exhibited}. We also emphasize that such a universal scaling behavior is not realized in the limit $\gamma \rightarrow 0$ with fixed $\alpha$, i.e. for the simple Polchinski-Faulkner model. The fact that quasi-universality occurs at a specific ratio of couplings at any $\nu$ implies that we cannot achieve linear-in-$T$ resistivity by keeping $\alpha$ finite and taking $\gamma$ to zero, and this is borne out by the explicit computations. Even though $\alpha/\gamma$ varies from 13 to 100 as $\nu$ varies from 0.66 to 0.95, the effect of the Fermi liquid contribution which is proportional to $\gamma$ is significant in the range of temperatures where quasi-universality emerges.

When semi-holography is embedded into a more general approach, as discussed in Sec. \ref{Sec:Micro}, it is possible that another scale replaces $E_F$ as the ultraviolet scale. In fact, we will discuss in Sec. \ref{Sec:Disc} that the general approach should also be able to predict the shape of the Fermi surface for a given lattice.

With the assumptions that the couplings $\alpha$ and $\gamma$ are small and that only the propagating degrees of freedom near the Fermi surface contribute to electric conduction, we can derive the following form of the Hall conductivity $\sigma_{\rm H}$ in the presence of weak magnetic fields (with $\omega_c = eB/m$ being the cyclotron frequency):
\begin{align}\label{Eq:hallconductivity}
&\sigma_{\rm H} =\sigma_{xy}= {2\omega_c \frac{e^2}{m}} \int \frac{d\omega}{2\pi} \int \frac{d^2 k}{4\pi^2} k_x \rho(\omega,\mathbf{k},T) \dfrac{\partial n_F(\omega, T)}{\partial \omega}\nonumber\\
&\hspace{3cm}\times\left( \dfrac{\partial}{\partial k_x} \text{Re } G_R (\omega,\mathbf{k}, T) \right) \ .
\end{align}
We have derived the above real-time expression for the hall conductivity from the one-loop form obtained in ~\cite{Fukuyama1969,Itoh_1985} in the Matsubara (imaginary) time formalism. Our real time expression \eqref{Eq:hallconductivity} reproduces the result for the  Hall conductivity of the Fermi liquid as shown in Appendix \ref{Sec:AppendixA}. 

Using the same change of variables which we have done earlier in the context of the dc conductivity, we readily see that the one loop integral \eqref{Eq:hallconductivity} giving the Hall conductivity can be approximated by
{\begin{align}
    &\sigma_{\rm H} \sim {2\omega_c \frac{e^2}{m}} \int_{-\infty}^{\infty} \frac{T d\tilde{x}}{2\pi} \int_{1-y_c}^{1+y_c} \frac{d\theta d\vert \tilde{y}\vert}{4\pi^2}\ k_F^2\ x_T \vert \tilde{y}\vert\ k_F \vert y \vert(1+ \mathcal{O}(x_T\vert \tilde{y}\vert^2)) \cos{\theta}\ \left(2\ T^{-1} {\rm{Im}}F(\tilde{x},\tilde{y})\right)\nonumber\\
    &\hspace{3cm}\times
    \left(T^{-1}\dfrac{\partial n_F(\tilde{x})}{\partial \tilde{x}}\right) \left(k_F^{-1}\ T^{-1}\ \cos{\theta} \frac{\partial}{\partial \vert y \vert }  {\rm{Re}} F(\tilde{x},\tilde{y})\right) \nonumber\\
    &\hspace{0.5cm}\sim -\omega_c e^2 \frac{1}{2\pi^2} E_F T^{-2} \int_{-\infty}^{\infty} d\tilde{x} \int_{1-y_c}^{1+y_c} d\vert \tilde{y}\vert\,\,(1+ \mathcal{O}(x_T\vert \tilde{y}\vert^2)) \nonumber\\ &
    \hspace{3cm} 
    \times \tilde{\rho}(\tilde{x},\tilde{y})
    \dfrac{\partial n_F(\tilde{x})}{\partial \tilde{x}}\ \frac{\partial}{\partial \vert \tilde{y} \vert}{\rm{Re}} F(\tilde{x},\tilde{y}).\label{Eq:Hallapprox}
\end{align}
}

From the above expressions, we can argue that the Hall conductivity should scale as $T^{-2}$ and is also independent of $\nu$ to a good approximation in the range of $\nu$ and the corresponding ranges of temperatures where quasi-universality holds.\footnote{Note that ${\rm Re}F$ and $\tilde\rho = - 2{\rm Im}F$ are related by Kramers-Kronig relation as $F$ is just the  retarded propagator times the temperature. Although the Kramers-Kronig relation involves integration over all frequencies, we recall that quasi-universality holds at all frequencies. Therefore, the quasi-universality of $\tilde\rho$ also implies the same for ${\rm Re}F$.} As in the case of the dc conductivity, this argument is valid provided quasi-universality holds for the range of momenta $1-y_c < \vert\tilde{{y}}\vert < 1 + y_c$ where the integral gets its maximum contribution and the $ \mathcal{O}(x_T\vert \tilde{y}\vert^2)$ corrections are small.

\begin{figure}[ht]
\centering
  \includegraphics[width= 0.6\textwidth]{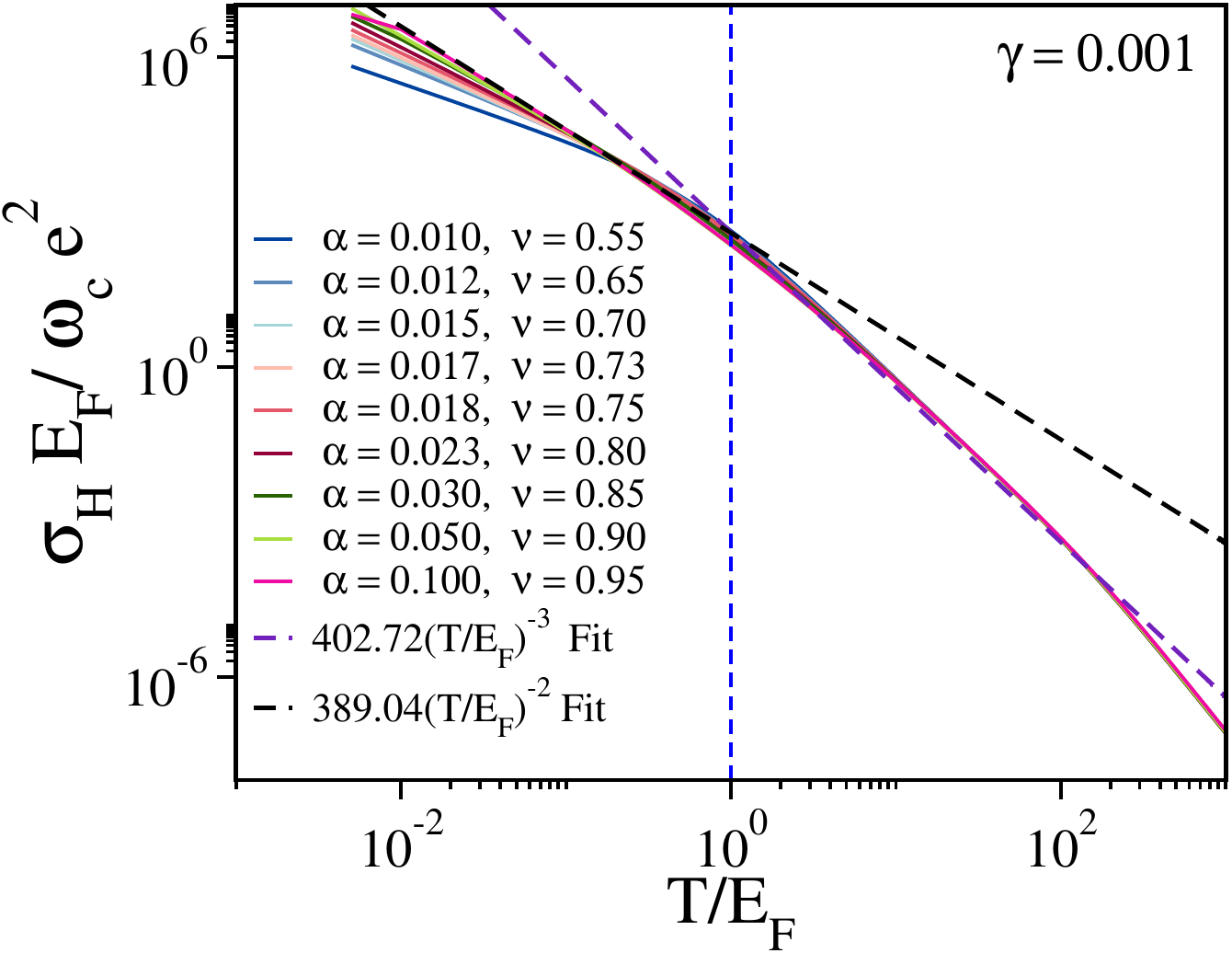}
\caption{Plot of $\sigma_{\rm H} E_F/(\omega_c e^2)$ as a function of $T/E_F$ is shown above at optimal values of $\alpha$ corresponding to different values of $\nu$ setting $\gamma = 0.001$. Except for the case of $\nu =0.55$ (which is outside of the domain of quasi-universality), we see that $\sigma_{\rm H} \sim T^{-2}$ as argued from \eqref{Eq:Hallapprox} for $T\leq E_F$ and in the range where quasi-universality appears. For $T\geq E_F$, we find a $T^{-3}$ scaling. Also note that in both scaling regimes, $\sigma_{\rm H}$ is independent of $\nu$ as expected from quasi-universality.}
\label{Fig:Hall conductivity}
\end{figure}

Computing the exact integral \eqref{Eq:hallconductivity}, we find that indeed $\sigma_{\rm H}$ scales as $T^{-2}$ for $T\leq E_F$ for optimal ratios of the couplings and in the range of temperatures corresponding to the values of $\nu$ where quasi-universality appears as can be expected from \eqref{Eq:Hallapprox}. See the plot \ref{Fig:Hall conductivity} where this has been demonstrated at $\gamma = 0.001$. As in the case of the dc conductivity, although the lower threshold temperature where the $T^{-2}$ scaling emerges for the Hall conductivity coincides with $t_c E_F$ where quasi-universality also emerges, the scaling continues to hold for temperatures above $\tilde{t}_c E_F$ (the maximum temperature where quasi-universal behavior of the spectral function is exhibited) but not above $E_F$.  Explicitly,
\begin{align}\label{Eq:Hallfit1}
    \sigma_{\rm{H}} = {{(389.04 \pm 34.23)}} \frac{\omega_c e^2}{E_F}\left(\frac{T}{E_F}\right)^{-2}
\end{align}
to a good approximation in this scaling regime when $t_c E_F \leq T \leq E_F$ and $\gamma = 0.001$ and thus independent of $\nu$. (At very low energies, $\sigma_{\rm H}\sim$ $T^{-2\nu}$ as the holographic self-energy contribution dominates.) Remarkably, as we can readily see from Fig. \ref{Fig:Hall conductivity}, we obtain a scaling $T^{-3}$ (as observed in strange metals \cite{PhysRevB.68.094502}) for $T \geq E_F$. Furthermore, the Hall conductivity is also independent of $\nu$ in this scaling regime, and is given by 
\begin{align}\label{Eq:Hallfit2}
   \sigma_{\rm{H}} = {{(402.72 \pm 7.35)}} \frac{\omega_c e^2}{E_F}\left(\frac{T}{E_F}\right)^{-3}
\end{align}
when $\gamma = 0.001$. This scaling regime is not explained by the approximation \eqref{Eq:Hallapprox} because it turns out that the $\mathcal{O}(x_T\vert\tilde{y}\vert^2)$ correction in the integrand is significant. Interestingly, this $T^{-3}$ scaling regime of the Hall conductivity emerges sharply at $T\sim E_F$ as evident from \eqref{Eq:Hallfit1} and \eqref{Eq:Hallfit2}. Note that the Hall conductivity is itself not discontinuous at $T\sim E_F$ although the scaling with the temperature changes sharply at $T\sim E_F$. Furthermore, the independence of the Hall conductivity from $\nu$ in the $T^{-3}$ scaling regime follows from the independence of the Hall conductivity in the quasi-universal regime given that there is no discontinuity and that the $T^{-2}$ scaling transits to $T^{-3}$ scaling at $T\sim E_F$.

The $T^{-3}$ scaling regime of the Hall conductivity, which appears at $T\geq E_F$ is not within the range of the validity of our simple effective field theory.  Nevertheless, it is encouraging for the purpose of comparisons with materials exhibiting strange metallicity. The $T^{-1}$ scaling of the dc conductivity and the $T^{-2}$ scaling of the Hall conductivity at low temperatures is also obtained in the marginal Fermi liquid phenomenological approach \cite{PhysRevB.68.094502}. In this approach, the $T^{-3}$ scaling of the Hall conductivity is expected to emerge only when the Fermi surface has portions of equal areas with positive and negative curvatures but not for a spherical Fermi surface \cite{PhysRevB.68.094502}. In our model, a non-spherical Fermi surface can lead to the $T^{-3}$ scaling of the Hall conductivity emerging at lower temperatures. Even with a spherical Fermi surface, our effective theory can lead to a $T^{-3}$ scaling of the Hall conductivity at reasonable temperatures if $E_F$ is replaced by a more suitable ultraviolet cut-off. This can happen when we embed our effective theory in a more general effective framework. We will discuss these issues more in Sec. \ref{Sec:Disc}.

\paragraph{Summary:} Let the ratio of couplings takes the optimal values corresponding to each value of $\nu$ between 0.66 and 0.95. We recall that the quasi-universality of the spectral function emerges for temperatures $t_c E_F \leq T \leq \tilde{t}_c E_F < E_F$, where $t_c$ and $\tilde{t}_c$ depends on $\nu$ and the overall strength of the couplings (and can be as small as 0.01 when $\nu$ is 0.95 and the couplings are small).  We obtain three scaling regimes as summarized below. 
\begin{itemize}
\item\textbf{Very low temperatures:} For temperatures $0< T \ll t_c E_F$, we obtain that
\begin{equation}\label{Eq:h1h20}
    \sigma_{\rm dc}(T) = \tilde{h}_1  e^2 \left(\frac{T}{E_F}\right)^{-\nu} ,\qquad\sigma_{\rm{H}} =  \tilde{h}_2 \frac{\omega_c e^2}{E_F}\left(\frac{T}{E_F}\right)^{-2\nu}
\end{equation}
to a good approximation. Above $\tilde{h}_1$ and $\tilde{h}_2$ are pure numbers which depend both on the overall strength of the coupling and on the scaling exponent $\nu$.

\item\textbf{Quasi-universal regime and temperatures lesser than the Fermi energy:} For temperatures $t_c E_F \leq T \leq E_F$, we obtain
    \begin{equation}\label{Eq:h1h21}
    \sigma_{\rm dc}(T) = h_1  e^2 \left(\frac{T}{E_F}\right)^{-1} ,\qquad\sigma_{\rm{H}} =  h_2 \frac{\omega_c e^2}{E_F}\left(\frac{T}{E_F}\right)^{-2}
\end{equation}
to a good approximation. Above ${h_1}$ and ${h_2}$ are pure numbers that depend only on the overall strength of the coupling and are therefore \textit{independent} of the scaling exponent $\nu$. The latter follows from the validity of quasi-universality.

\item\textbf{Temperatures greater than the Fermi energy:} For temperatures $E_F \leq T < t_u E_F$ (with $t_u \approx 10$ at weak coupling), we obtain
    \begin{equation}\label{Eq:h1h22}
    \sigma_{\rm dc}(T) = h_1  e^2 \left(\frac{T}{E_F}\right)^{-1} ,\qquad\sigma_{\rm{H}} =  h_2 \frac{\omega_c e^2}{E_F}\left(\frac{T}{E_F}\right)^{-3}
\end{equation}
to a good approximation. Above ${h_1}$ and ${h_2}$ are the same pure numbers as those appearing in \eqref{Eq:h1h21} to a good approximation. This is expected from the validity of quasi-universality and that the scaling of the Hall conductivity changes at $T \sim E_F$.
\end{itemize}

When $\gamma = 0.001$, $h_1 = 15.67\pm 0.22$ and $h_2 = 395.88\pm 34.23$ as already mentioned (see Eqs. \eqref{Eq:dcfit}, \eqref{Eq:Hallfit1} and \eqref{Eq:Hallfit2}).  For the same value of $\gamma$ and $\nu =0.8$, $\tilde{h}_1 = 21.30 \pm 0.02$, $\tilde{h}_2 = 879.48 \pm 4.05$ when $\alpha$ takes the optimal value $0.023$.

\subsection{Generalized Drude phenomenology and a refined picture of Planckian dissipation} 
The dc conductivity and the Hall conductivity obtained from the one loop integrals \eqref{Eq:dcconductivity} and \eqref{Eq:hallconductivity}, respectively, with the spectral function derived from the Fermi liquid form of the retarded propagator
\begin{align}\label{Eq:FLG}
G_{R} (\omega,\mathbf{k}) = \left( a^{-1} \omega - b^{-1}\epsilon_{k} + i \tau^{-1} \right)^{-1},
\end{align}
reproduces results similar to the Drude model (see Appendix \ref{Sec:AppendixA} for derivations). Explicitly, these results are
 \begin{equation}\label{Eq:Drude1}
\sigma_{\rm dc} = \frac{n e^2}{m} b\,\tau,\qquad \sigma_{\rm H} = \frac{ne^2}{m}\omega_c\tau^2.
\end{equation}
Here, $\tau$ is the scattering time of the electrons moving in a lattice with impurities, and N is the carrier density. Also, $b$ is the renormalization of the density of states at the Fermi surface, and $b/a$ is the renormalization factor of the specific heat. Defining $\overline{n} := n/m$, we can rewrite \eqref{Eq:Drude1} as
\begin{equation}\label{Eq:Drude12}
\sigma_{\rm dc} = \overline{n} e^2 b\,\tau,\qquad \sigma_{\rm H} = \overline{n} e^2\omega_c\tau^2.
\end{equation}
For the two-dimensional Fermi liquid, $\overline{n} = E_F \pi^{-1}$. 

Quite remarkably, it turns out that a generalized version of Drude phenomenology can capture many aspects of the results of the semi-holographic effective theory. Postponing a full justification and a non-trivial consistency check to the next section, we can merely try to see if the following generalized Drude formulae
\begin{equation}\label{Eq:Drude2}
\sigma_{\rm dc} = \overline{n}(T)\, e^2\, \tau_{\rm d}(T),\qquad \sigma_{\rm H} = \overline{n}(T)\, e^2\omega_c\,\tau_{\rm d}(T)^2,
\end{equation}
where we have introduced temperature dependent dissipation time scale $\tau_{\rm d}$ and carrier density $\overline{n}$, which can make sense for the dc conductivity and the Hall conductivity. We will find that the non-trivial consistency tests can be satisfied by setting $b=1$.\footnote{In the future, we would like to compute the specific heat for a better understanding of the emergence of the generalized Drude phenomenology.} Assuming the validity of \eqref{Eq:Drude2} and $b=1$, we obtain that  
\begin{equation}\label{Eq:Drude3}
\tau_{\rm d}(T) = \frac{\sigma_{\rm H}(T)}{\omega_c\sigma_{\rm dc}(T)}, \qquad \overline{n}(T)= \frac{\sigma_{\rm dc}(T)^2\omega_c}{e^2\sigma_{\rm H}(T)}.
\end{equation}

In order to focus on the strange metal-like phase, we should let the ratio of the couplings take optimal values corresponding to each value of $\nu$ between 0.66 and 0.95. Using \eqref{Eq:h1h20} and \eqref{Eq:Drude3}, we obtain   
\begin{equation}\label{Eq:taun0}
    \tau_{\rm d}  = \frac{\tilde{h}_2}{\tilde{h}_1} E_F^{-1} \left(\frac{T}{E_F}\right)^{-\nu},\qquad \overline{n}  =  \frac{\tilde{h}_1^2}{\tilde{h}_2} E_F 
\end{equation}
for very low temperatures $0 < T\leq t_c E_F$.  Using \eqref{Eq:h1h21} and \eqref{Eq:Drude3}, we obtain 
\begin{equation}\label{Eq:taun1}
    \tau_{\rm d}  = \frac{{h}_2}{{h}_1} T^{-1},\qquad \overline{n}  =  \frac{{h}_1^2}{{h}_2} E_F.
\end{equation}
for temperatures $t_c E_F \leq T \leq E_F$ including the range of temperatures where quasi-universality holds. This is the Planckian dissipation regime where the scattering time scales as T$^{-1}$. Finally, assuming \eqref{Eq:h1h22} and \eqref{Eq:Drude3} we obtain that
\begin{equation}\label{Eq:taun2}
    \tau_{\rm d}  = \frac{h_2}{h_1} E_F T^{-2},\qquad \overline{n}  =  \frac{h_1^2}{h_2} T .
\end{equation}
for temperatures $E_F \leq x_T \leq t_u E_F$ greater than the Fermi energy. As mentioned before, we obtain the same $h_1$ and $h_2$, which are pure numbers determined solely by the overall strength of the couplings (and are thus independent of $\nu$) in both cases \eqref{Eq:taun1} and \eqref{Eq:taun2}. However, $\tilde{h}_1$ and $\tilde{h}_2$ also depend on $\nu$.

Postponing non-trivial consistency checks of the assumptions leading to the definitions of $\tau_{\rm d}$ and $\overline{n}$ to the next section, here we can do some simple consistency checks and see their usefulness. Firstly, we note that the change of the temperature scaling of $\tau_{\rm d}$ from sub-Planckian regime \eqref{Eq:taun0} at very low temperatures to the Planckian regime \eqref{Eq:taun1} where quasi-universality is valid, gives us a remarkably accurate way of computing the lower threshold temperature $t_c$ where quasi-universality holds. Since there is no discontinuity of $\tau_d$ as a function of the temperature, the two formulae \eqref{Eq:taun0} and \eqref{Eq:taun1} should agree when $T = t_c E_F$ assuming that the scaling changes sharply here. Therefore, we can estimate $t_c$ using
\begin{equation}\label{Eq:tc}
    t_c \sim \left(\frac{\tilde{h}_2}{\tilde{h}_1}\times \frac{h_1}{h_2}\right)^{-\frac{1}{1-\nu}}
\end{equation}
As reported before, when $\gamma =0.001$, $\nu = 0.8$ and $\alpha$ takes the optimal value 0.023, $h_1 = 15.67\pm 0.22$ and $h_2 = 395.88\pm 34.23$; while $\tilde{h}_1 = 21.30 \pm 0.02$, $\tilde{h}_2 = 879.48 \pm 4.05$. From the above, we can then readily estimate that $t_c \sim 0.08 \pm 0.02$ when $\nu = 0.8$ and $\gamma = 0.001$. Indeed it turns out that we obtain $t_c \sim 0.05\pm 0.015$ for $\nu = 0.8$ by using the criterion $\delta < 10$ mentioned in the previous section (with the uncertainty obtained by varying the choice of $x_T$ in the standard curve $f_s$).

Secondly, we note from \eqref{Eq:taun0} and \eqref{Eq:taun1} that $\overline{n}$ is temperature independent for both low temperatures and in the Planckian regime where quasi-universality holds. Since the conductivities do not have any discontinuity, $\overline{n}$ cannot change sharply at $T\sim t_c E_F (\ll E_F)$ as it is independent of the temperature to a good approximation for both the low temperature regime and the Planckian regime below and above $t_c E_F$, respectively. Therefore, comparing $\overline{n}$ in \eqref{Eq:taun0} and \eqref{Eq:taun1}, we can conclude that
\begin{equation}\label{Eq:ratiocheck}
    \frac{\tilde{h}_1^2}{\tilde{h}_2} \sim \frac{h_1^2}{h_2}
\end{equation} 
should hold. It then follows that $\tilde{h}_1^2/\tilde{h}_2$ should also be approximately independent of $\nu$ although individually $\tilde{h}_1$ and $\tilde{h}_2$ do depend on $\nu$. In fact, for the case $\nu =0.8$ and $\gamma =0.001$, we find that  $\tilde{h}_1^2/\tilde{h}_2\sim 0.62\pm 0.06$, while $h_1^2/h_2 \sim 0.52\pm 0.05$ with the values of $\tilde{h}_1$, $\tilde{h}_2$, $h_1$ and $h_2$ quoted above. 

Furthermore, given that the scaling behaviors transit at $T\sim E_F$ sharply without discontinuity in the conductivities and that the conductivities are independent of $\nu$ in the quasi-universal regime, we can expect that $\tau_d$ should be approximately independent of $\nu$ also for temperatures above $E_F$ as indeed implied by \eqref{Eq:taun2}.

\begin{figure}[h]
 \includegraphics[width=1.0\textwidth]{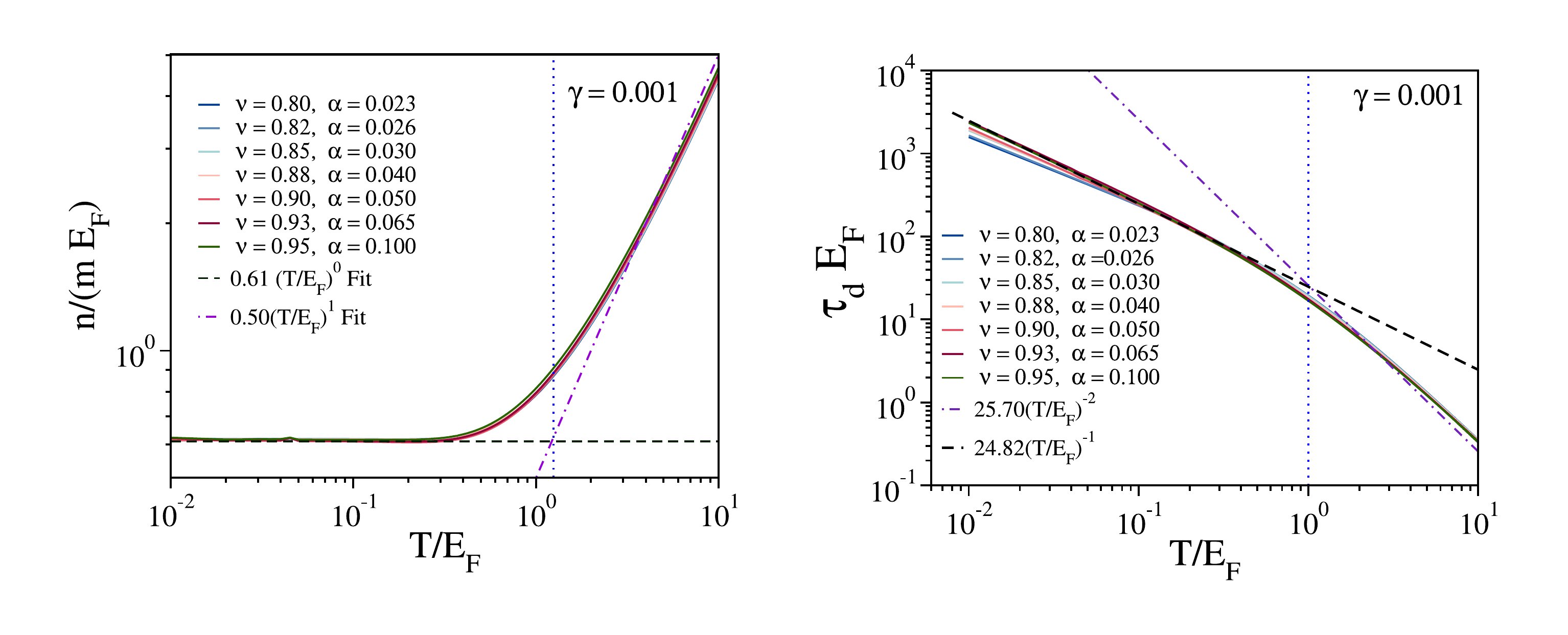}
\caption{The plots of effective $ n/(m E_F)$ (left) and $\tau_{\rm d} E_F$ (right) as functions of $T/E_F$ are shown for various values of $\nu$ (at $\gamma =0.001$) and the corresponding optimal ratios $\alpha/\gamma$. 
}\label{Fig:CDandST}
\end{figure}

The third consistency check is as follows. We note that the behavior of $\tau_{\rm d}$ and $\overline{n}$ given by \eqref{Eq:taun0}, \eqref{Eq:taun1} and \eqref{Eq:taun2} have been obtained from the approximate behaviors of the dc conductivity and Hall conductivity in the corresponding regimes which are given by \eqref{Eq:h1h20}, \eqref{Eq:h1h21} and \eqref{Eq:h1h22}, respectively. We should check that $t_c$ estimated from \eqref{Eq:tc}, the relation \eqref{Eq:ratiocheck} implying that $\overline{n}$ is temperature independent for $T \lessapprox E_F$, and also that $\overline{n}$ is independent of $\nu$ for temperatures $0< T \leq E_F$ (or at least for $0 < T \leq \tilde{t}_c E_F$ where we can explain the scalings using analytic arguments presented before) should hold if we use the exact definitions of $\tau_{\rm d}$ and $\overline{n}$ given by \eqref{Eq:Drude3}, i.e. if we use the exact integrals \eqref{Eq:dcconductivity} and \eqref{Eq:hallconductivity} giving the dc conductivity and Hall conductivity at one loop, respectively, instead of the approximations involving simple scalings w.r.t. the temperature.  

In Fig. \ref{Fig:CDandST}, the dimensionless dissipation time $\tau_{\rm d}(T)E_F$ and the dimensionless carrier density $\overline{n}(T)/E_F$ have been plotted as a function of $T/E_F$ using the definitions \eqref{Eq:Drude3} and the exact one-loop dc conductivity and Hall conductivity given by Eqs. \eqref{Eq:dcconductivity} and \eqref{Eq:hallconductivity}, respectively. We readily observe that $\overline{n}$ (i) is approximately independent of $\nu$, (ii) approximately independent of the temperature for $T < 0.3 E_F$ (also $0.3 < \tilde{t}_c$ for typical values of $\nu$ in the range where quasi-universality holds), and (iii) increases linearly with the temperature for $T> E_F$. We also observe that $\tau_{\rm d}$ (i) is independent of $\nu$ for $T > t_c E_F$ as expected from quasi-universality, and (ii) behaves as \eqref{Eq:taun0}, \eqref{Eq:taun1} and \eqref{Eq:taun2} in the respective regimes. It also turns out that $t_c E_F$ given by \eqref{Eq:tc} gives a very good estimate of the temperature at which sub-Planckian scaling transits to Planckian scaling of $\tau_{\rm d}$.

Note also the fits in Fig. \ref{Fig:CDandST} give $\overline{n}\sim (0.5 \pm 0.1) E_F$ for $T< E_F$ and $(0.61 \pm 0.1) T$ for $T\geq E_F$ which agree well with the estimates given earlier. Similarly, the fits for $\tau_{\rm d}(T) E_F$ in Fig. \ref{Fig:CDandST} also agree with the estimates given earlier in different regimes.

Let us focus on the Planckian regime, which can be defined as the regime of temperatures where
\begin{equation}
    \tau_{\rm d} \sim f T^{-1}, \qquad \overline{n} \sim \tilde{f} E_F.
\end{equation}
In this regime where temperatures are not very high, our effective field theory can be trusted if both couplings are small. Comparing the above equation with \eqref{Eq:taun1} we obtain
\begin{equation}
    f = \frac{h_2}{h_1}, \qquad \tilde{f}  = \frac{h_1^2}{h_2},
\end{equation}
which are then independent of $\nu$ but do depend on the overall strength of the couplings. In Table \ref{Table:Planckian}, we have reported the values of $f$ and $\tilde{f}$. We note that $f$ is large ($\sim 25$) when the couplings of small ($\gamma = 0.001$), while it is closer to unity ($\sim 2.5$) when the couplings are larger ($\gamma = 0.01$). Note that we cannot trust the semi-holographic effective field theory for $\gamma$ larger than 0.01 as the optimal values of $\alpha$ exceed unity, particularly when $\nu$ is close to 0.95. Also, for larger couplings, $t_c$ increases and the overall range of temperatures where Planckian scaling holds shrink.

\begin{table}[h]
\begin{center}
\begin{tabular}{ |c|c|c|c| c |} 
 \hline
 $\gamma$ & $h_1$ & $h_2$ & $f$ & $\tilde{f}$\\
 \hline
 0.001 & $15.67\pm 0.22$ & $395.88\pm 34.23$   & $25.26 \pm 2.4$  & $0.62 \pm 0.06$ \\ 
 \hline
  0.01 & $\sim$ 2  & $\sim$ 5 & $\sim$ 2.5 & $\sim$ 0.8 \\ 
 \hline
\end{tabular}
\end{center}
\caption{Table of values for the Planckian regime at weak and strong coupling}
\label{Table:Planckian}
\end{table}

We also observe from Table \ref{Table:Planckian}, $\tilde{f}$ remarkably does not also vary much with the overall strength of the coupling, suggesting that the effective carrier density is set simply by the Fermi energy in a model-independent way (recall that $\overline{n}$ is almost independent of $\nu$ from low to high temperatures).

Our effective approach thus leads to a refined picture of Planckian dissipation where $f$ can be much larger than unity and implies a temperature independent carrier density. This has been argued to be the case in \cite{Taupin2022AreHF} for many strange-metallic heavy metal crystalline compounds from the analysis of experimental data.

\section{Optical conductivity}\label{Sec:OC}

\noindent
The optical conductivity can be readily extracted from the semi-holographic effective theory with the assumption that the charge conduction is due to the propagating degrees of freedom near the Fermi surface. When the couplings are small, we can rely on the one loop result. The real part of the optical conductivity is
\begin{align}\label{Eq:realpartsigma}
   \text{Re } \sigma (\Omega, T)= & \frac{e^2}{m^2}\int\frac{d\omega}{2\pi}\int\frac{d^2k}{4\pi^2} k^2 \rho(\omega+\Omega/2,\mathbf{k},T)\rho(\omega-\Omega/2,\mathbf{k},T)\nonumber\\    & \times \left(-\frac{n_F(\omega+\Omega/2,T)-n_F(\omega-\Omega/2,T)}{\Omega}\right)\ ,
\end{align}
where $\rho(\omega,\mathbf{k},T)=-2\text{Im}G_{R} (\omega,\mathbf{k},T)$ is the spectral function and $ n_F(\omega, T)=\tfrac{1}{1+e^{\omega/T}} $  is the Fermi-Dirac distribution and $\Omega$ is the external frequency. The dc conductivity \eqref{Eq:dcconductivity} is simply the zero frequency limit of the optical conductivity \eqref{Eq:realpartsigma}. As in the case of the dc conductivity, we can check that the integral \eqref{Eq:realpartsigma} gets most of its contribution from the loop momenta near the Fermi surface at least when the external frequency $\Omega$ is not much large compared to the temperature. It is evident that the real part of the optical conductivity is an even function of $\Omega$. Therefore, we will consider $\Omega \geq 0$ without loss of generality.

At this stage, it is useful to recall that we can phenomenologically map the regime of the optimal ratio of couplings $\alpha$ to $\gamma$ to the regime of near-optimal doping in cuprates where strange metallic characteristics are exhibited. When $\alpha/\gamma$ is smaller than the optimal value, the Fermi liquid-like term in the self-energy is dominant and this phenomenologically corresponds to the overdoped regime in cuprates. Similarly, the underdoped regime can be mapped to the regime where $\alpha/\gamma$ is larger than the optimal value.

 Here, we study the optical conductivity in the semi-holographic effective theory in all these three regimes. We find that the Drude-like behavior is present in a wide range of frequencies for both overdoping and optimal doping when the temperature is below 0.6$E_F$. However, the scattering time $\tau$ scales differently from $T^{-2}$ unless we are in the extremely overdoped region where we recover Fermi liquid behavior completely. 
 
 At optimal doping, the scattering time is sub-Planckian, i.e. $\tau \sim T^{-\nu}$ when the temperatures are low; and is Planckian $\tau \sim T^{-1}$ when the temperatures correspond to the quasi-universal regime (and are not larger than 0.6 $E_F$ so that the scattering time itself can be defined clearly). Note that in both regimes the scaling of $\tau$ with the temperature is like the dissipation time $\tau_{\rm d}$ which we introduced in the previous section via a generalized Drude phenomenology, and combining the results of dc and Hall conductivities. Quite non-trivially, we will be able to extend the generalized Drude phenomenology further (to finite frequencies) in the case of optimal doping relating the scattering time $\tau$ to the dissipation time $\tau_{\rm d}$ both at low temperatures and at temperatures corresponding to the Planckian dissipation regime. 

 At under-doping, we find new emergent scaling regimes both with respect to temperature and external frequency that are different from the Drude-like behavior.

In all cases, we can qualitatively and quantitatively match with experimental results.

\subsection {Optical conductivity and the generalized Drude phenomenology at the optimal ratio of couplings}

\noindent
In the Drude model, which is a simplified classical model for the behavior of electrons in a metal, the conduction band electrons are treated as a gas of free particles. These electrons undergo diffusive motion, characterized by their scattering and relaxation processes within the metals. The relaxation time ($\tau$), represents the average time between successive collisions or scattering events with the lattice impurities experienced by the electrons. The optical conductivity, i.e. the conductivity in the presence of an electric field at finite frequency, is predicted by the Drude model to have the form
\begin{equation}\label{Eq:DrudeOpConductivity}
    \sigma(\Omega) = \frac{ne^2\tau}{m} \frac{1}{1-i \Omega\tau}=  \frac{\sigma_{\rm dc}}{1-i \Omega\tau}\, ,
\end{equation}
with its real part given by 
\begin{equation}\label{Eq:realDrudeOpConductivity}
    \text{Re } \sigma(\Omega) =\frac{ne^2\tau}{m}\frac{1}{1+  \tau^2 \Omega^2} = \frac{\sigma_{\rm dc}}{1+  \tau^2 \Omega^2}\, ,
\end{equation}
where $n$ is the electron number density, $e$ is the charge of electron and $m$ is the (effective) electron mass. Note that when $\Omega$ vanishes, the real part of the optical conductivity is just the dc conductivity $\sigma_{\rm dc}$ as in \eqref{Eq:Drude1} (with $b=1$).

Without loss of generality, we can always write the optical conductivity in the form
\begin{equation}\label{Eq:DrudeOp-1}
    \sigma(\Omega, T) = \frac{\sigma_{\rm dc}(T)}{1-i \Omega\tau_{\rm eff}(\Omega, T)}\, ,
\end{equation}
where the above should be understood as the definition of the frequency and temperature dependent effective scattering time $\tau_{\rm eff}$. We readily see  that
\begin{equation}\label{Eq:tau-eff}
    \tau_{\rm eff}(\Omega, T) = \frac{1}{\Omega}\sqrt{\frac{\sigma_{\rm dc}}{\text{Re}\,\sigma(\Omega, T)}-1}.
\end{equation}
We say that the optical conductivity is \textit{Drude-like} for a regime of frequencies $\Omega$ if and only if the following conditions are satisfied:
\begin{itemize}
\item There exists a $\tau(T)$  such that
    \begin{equation}
    \frac{\tau_{\rm eff}(\Omega, T)- \tau(T)}{\tau(T)} < 0.1,
\end{equation}
i.e. the effective scattering time is almost frequency independent in the regime of frequencies where Drude-like behavior holds. It is expected that this range of frequencies overlaps with $0\leq \Omega \leq T$. (In practice, we can choose $\tau(T)$ to be the value of $\tau_{\rm eff}(\Omega, T)$ for any $\Omega < T$ at a given $T$ in the region where  $\tau_{\rm eff}(\Omega, T)$ flattens out as a function of $\Omega$. The error involved in the choice of $\Omega$ is insignificant.)
 \item The limit 
\begin{equation}
    \lim_{\Omega\rightarrow 0}\tau_{\rm eff}(\Omega, T) = \tilde{\tau}(T)
\end{equation}
exists so that at small frequencies, the optical conductivity behaves as
\begin{equation}\label{Eq:RessmallOmega}
{\rm Re}\,\sigma(\Omega, T) = \sigma_{\rm dc}\left(1 - {\Omega^2 }{\tilde{\tau}(T)^2}+ \mathcal{O}(\Omega^3)\right).
\end{equation}
\end{itemize}
Clearly, the above two conditions also imply that
\begin{equation}\label{Eq:Crit-3}
    \frac{\tilde{\tau}(T)- \tau(T)}{\tau(T)} < 0.1.
\end{equation}

Of course, the Drude-like behavior can be expected to hold only for a range of temperatures. These properties ensure that the optical conductivity is essentially due to a relaxation process that is governed by a scattering time $\tau(T)$, which can have a non-trivial temperature dependence. Furthermore, this relaxation time is close to $\tilde\tau(T)$, which is the time scale for dissipation in the small frequency limit. In order to see if Drude-like behavior emerges in the semi-holographic effective theory, we need to compute the real part of the optical conductivity, ${\rm Re}\,\sigma$ at one loop using \eqref{Eq:realpartsigma}, extract the effective scattering time $\tau_{\rm eff}$ using \eqref{Eq:tau-eff}, and then verify if the latter satisfies the three requirements mentioned above. Here, we examine this at the optimal ratio of the couplings.

\begin{figure}[ht]
    \centering
    \includegraphics[scale=0.4 ]{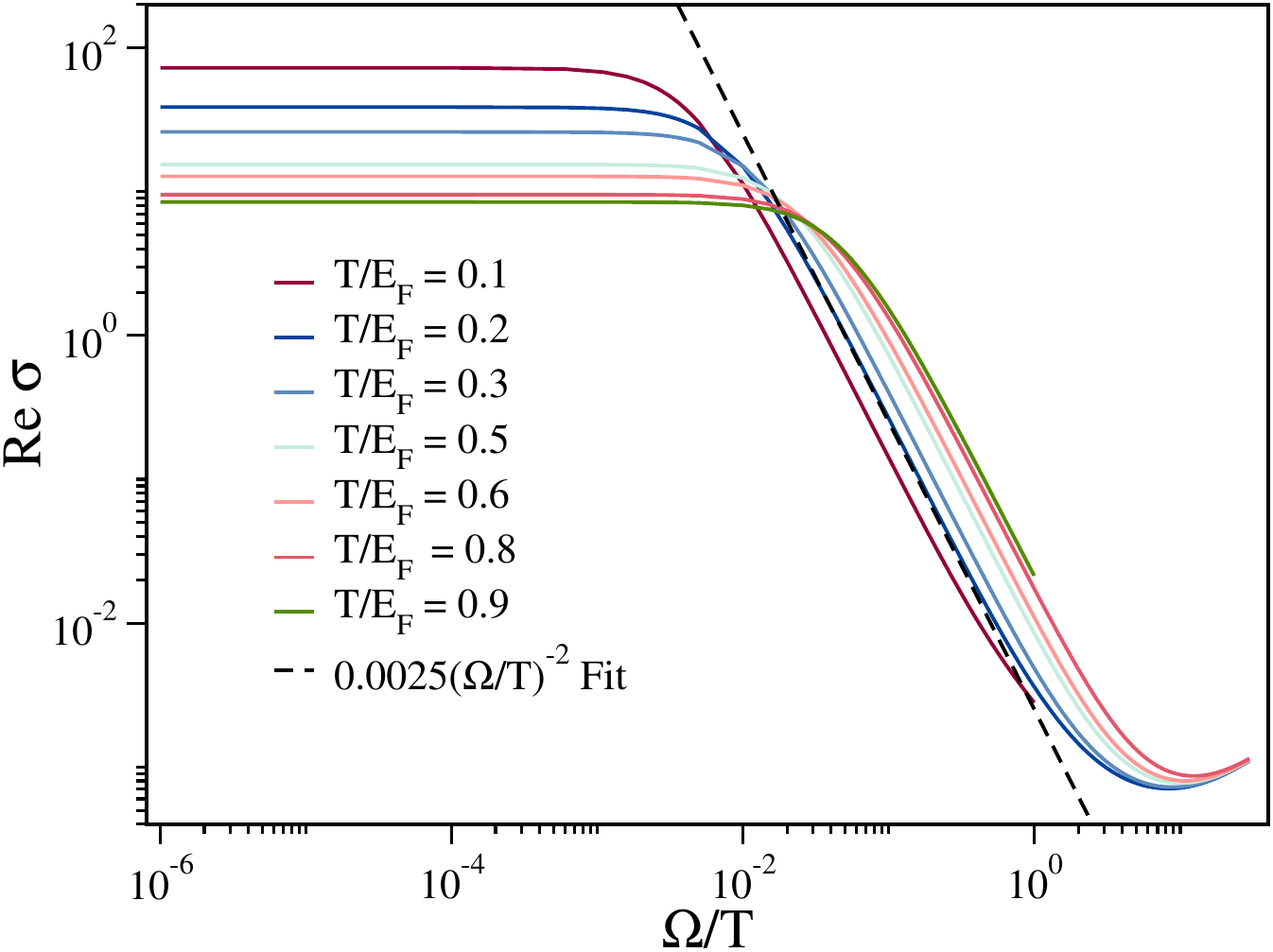}
    \caption{ The colored solid lines are log-log plots of $\text{Re}\sigma(\Omega)$ vs $\Omega/T$ for different temperatures $T$. We have set $\nu = 0.80$ and chosen one set of the optimal values of the coupling constants $\alpha=0.023, ~ \gamma=0.001$. The dashed black line is the straight line with slope $-2$.}
    \label{loglogplotrealsigmaop}
\end{figure}

An instance of the behavior of the real part of optical conductivity obtained from \eqref{Eq:realpartsigma} as a function of frequency at various fixed temperatures at the optimal ratio of couplings is shown in Fig.~\ref{loglogplotrealsigmaop} for $\nu=0.80$ and $\gamma=0.001$ (the optimal value of $\alpha$ is the 0.023). The plot shows the behavior for various temperatures $0.1E_F < T < E_F$, but it is qualitatively the same also at very low temperatures such as $T = 0.0004 E_F$. Each curve in Fig.~\ref{loglogplotrealsigmaop} has three regimes. First, a plateau part at very low frequencies, where the real part of optical conductivity is almost independent of the frequency and approaches the dc limit. The second regime is the  $\Omega^{-2}$ fall off in the intermediate frequencies up to $E_F$, which suggests a Drude-like behavior, and third, a non-trivial tail at high frequencies greater than $E_F$, which clearly deviates from Drude-like behavior. 

A similar kind of Drude response with a frequency-independent plateau followed by a $\Omega^{-2}$ falloff has been reported in optimally doped and overdoped samples of $\rm{Bi_{2-x}Pb_{x}Sr_{2-y}}$  $\rm{La_{y}CuO_{6+ \delta}}$ according to the analysis presented in \cite{PhysRevB.106.054515}. Although the data reveal the existence of $\Omega^{-2}$ fall-off, it is only visible for a range of frequencies (at lower temperatures) before crossing over to a conformal tail with a slower fall-off. In our simplistic effective theory, the slower than $\Omega^{-2}$ fall off is observed for $\Omega > E_F$ as evident from Fig.~\ref{loglogplotrealsigmaop}. We can recall our discussion in the previous section that $E_F$ is the only intrinsic energy scale in the model (where as for instance the quasi-universality of the spectral function becomes nearly exact), and a more general version, as discussed in the next section, can replace $E_F$ with a lower energy scale. 

All our criteria for Drude-like behavior are satisfied for frequencies less than the temperature ($0\leq \Omega \leq T$) and only for temperatures less than 0.6$E_F$. This is in agreement with the experimental analysis in \cite{PhysRevB.106.054515} according to which the Drude-like response accounts for 90 \% of the optical conductivity at low temperatures for frequencies below a cutoff (the data is noisy at higher temperatures and also at higher frequencies for any definite conclusion).\footnote{Emergence of Drude-like behavior of the optical conductivity at very low temperatures has been discussed before in \cite{Kiritsis:2015yna} in the context of holographic models.}

\begin{figure}[ht]
    \centering
    \includegraphics[scale=0.4]{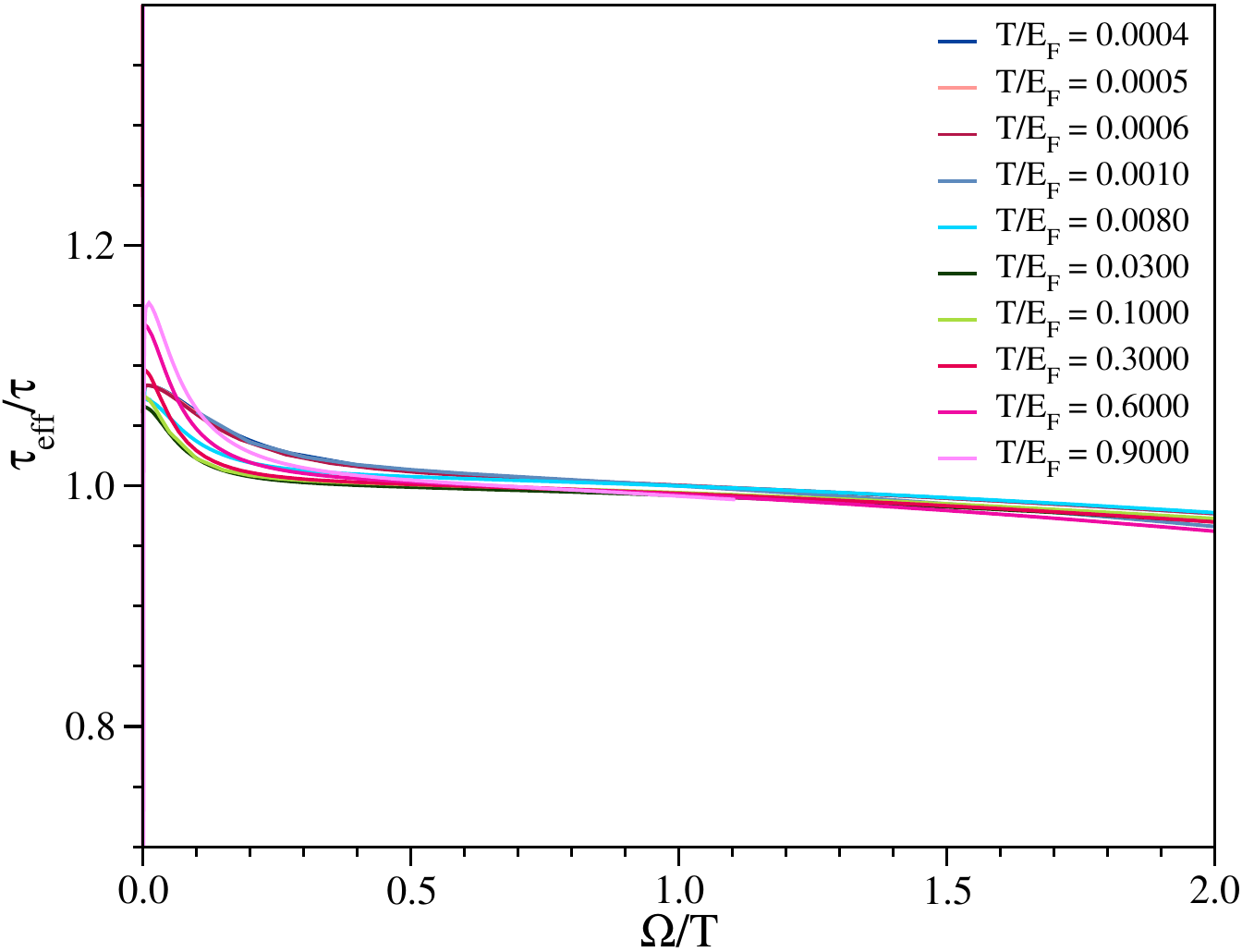}
    \caption{ $\tau_{\text{eff}}(\Omega,T)/ \tau(T) $ plot for different temperatures $T$ at optimal value ($\nu=0.8$, $\alpha=0.023$ and $\gamma=0.001$). We note that $\tau_{\text{eff}}(\Omega,T)/ \tau(T)$  is very close to unity for temperatures below 0.6$E_F$ and for frequencies $0 <\Omega \leq T$. }
    \label{Fig:taueffbytauplot}
\end{figure}

As evident from Fig. \ref{Fig:taueffbytauplot}, the effective scattering time $\tau_{\rm{eff}}(\Omega, T)$ remains almost frequency independent over the range of frequencies $0\leq \Omega \leq  T$ and approaches (an almost constant) value $\tau(T)$ for $\Omega > 0.1T$ at the optimal ratio of coupling for a wide range of temperatures.  For temperatures in the range $0.0004T$ to $0.6T$, we observe that all the criteria for Drude-like conductivity are satisfied. The strongest deviation of $\tau_{\rm{eff}}(\Omega, T)$ from the (almost) constant value $\tau(T)$ in the range of frequencies $0\leq \Omega \leq  T$ occurs for low frequencies $\Omega < 0.1T$. However, $\tau_{\rm{eff}}(\Omega, T)$ reaches a value $\tilde{\tau}(T)$ as $\Omega\rightarrow 0$ and the criterion \eqref{Eq:Crit-3} is satisfied as $\tilde{\tau}(T)$ does not deviate more than 10 \% from $\tau(T)$ for temperatures below $0.6 T$.  Although Fig. ~\ref{Fig:taueffbytauplot} shows the results for $\nu = 0.8$, all the criteria for generalized Drude phenomenology are satisfied for all $\nu$ between 0.66 and 0.95 at low frequencies ($\Omega < T$) and temperatures below $0.6E_F$ when the ratio of the couplings is optimal.

\subsubsection{A non-trivial check of generalized Drude phenomenology}

The question which naturally arises from the Drude-like behavior of the optical conductivity at the optimal ratio of couplings is whether this can justify the generalized Drude phenomenology for the dc conductivity and the Hall resistivity encapsulated in Eq. \eqref{Eq:Drude2}. Indeed if this is the case, then the dissipation time $\tau_{\rm d}(T)$, which we obtain from Eq. \eqref{Eq:Drude3} (as a consequence of Eq. \eqref{Eq:Drude2}), should agree with the $\Omega\rightarrow 0$ limit of the effective scattering time $\tau_{\rm eff}(\Omega,T)$ which we extract from the optical conductivity using the latter's Drude-like behavior, i.e. we should have
\begin{equation}\label{Eq:Crit-4}
    \tau_{\rm d}(T) = \lim_{\Omega\rightarrow 0} \tau_{\rm eff}(\Omega,T) = \tilde{\tau}(T).
\end{equation}
This is a very non-trivial requirement because the computation of $\tau_{\rm d}$ requires Hall conductivity whose evaluation is independent of that of the optical conductivity, and furthermore $\tilde{\tau}(T)$ itself comes not from the dc limit directly but from the sub-leading quadratic dependence of ${\rm Re}\,\sigma$ on the frequency for small values of the frequency as shown in Eq. \eqref{Eq:RessmallOmega}. The only unifying feature of the two is the assumption of Drude-like behavior of the Hall conductivity and the dc conductivity, and also that of the optical conductivity as a function of the external frequency.
\begin{table}[ht]
\begin{center}
\begin{tabular}{ |c|c|c|} 
 \hline
 $T/E_F$ & $\tau_{\rm d} E_F$ & $\tilde{\tau}E_F = \lim_{\Omega \to 0}\tau_{\rm{eff}}E_F$   \\
 \hline
 0.9 & 24.8 - 30.4 & 23.4   \\ 
 \hline
  0.6 & 37.2 - 45.53  & 39.87  \\ 
 \hline
 0.3 & 74.4 - 91.06  & 87.38  \\ 
 \hline
 0.1 & 223 - 273  & 245  \\ 
 \hline
 0.03 & 678 - 686 & 673 \\
 \hline
 0.008 & 1954 - 1975 & 1979 \\
 \hline
 0.001 & 10313 - 10426 & 10043 \\
 \hline
 0.0006 & 15520 - 15690 & 15621 \\
 \hline
 0.0005 & 17957 - 18154 & 18120 \\
 \hline
 0.0004 & 21647 - 21702 & 21734 \\
 \hline
\end{tabular}
\end{center}
\caption{Table of list of values of $\tau_{\rm d}$ and the zero frequency limit (dc limit) of $\tau_{\rm{eff}}$ for different temperatures in the Planckian regime for $\nu = 0.80$, $\alpha = 0.023$ and $\gamma = 0.001$ (the ratio of couplings is optimal). For a fixed temperature, the dc limit of the $\tau_{\rm{eff}}(\Omega)$ matches with the value of $\tau_{\rm d}$ for temperature below 0.6$E_F$.}
\label{Table:zero frequency limit taueff and taud}
\end{table}

As shown in Table \ref{Table:zero frequency limit taueff and taud}, we find that indeed our requirement \eqref{Eq:Crit-4} for the agreement between $\tau_{\rm d}(T)$ and $\tilde{\tau}(T)$ is satisfied in our semi-holographic effective theory at the optimal ratio of coupling for temperatures less than 0.6$E_F$ both in sub-Planckian and Planckian regimes. We recall that the criterion \eqref{Eq:Crit-3} for the Drude-like behavior requiring $\tilde{\tau}(T)$ to be close to $\tau(T)$ is also satisfied precisely in this range of temperatures. Therefore, we do obtain a very non-trivial justification for the use of generalized Drude phenomenology in our semi-holographic effective theory at the optimal ratio of couplings for temperatures below 0.6$E_F$.

We recall that we have discussed another crucial check of Drude phenomenology in the previous section, namely that the effective carrier density $\overline{n}(T)$ is independent of $\nu$ and also independent of the temperature for temperatures below 0.3$E_F$. This is also impressive as it involves very low temperatures where quasi-universality does not apply, and thus the feature that $\overline{n}(T)$ is independent of $\nu$ (along with the temperature) is not at all obvious. Together with the non-trivial agreement between the dissipation time $\tau_{\rm d}(T)$ and the zero frequency limit of the effective scattering time $\tau_{\rm eff}(\Omega,T)$ obtained from the Drude-like behavior of the optical conductivity, this provides a strong case for Drude-phenomenology with a temperature independent carrier density set just by the Fermi energy (and almost independent of the parameters of the model), and a relaxation/dissipation time, which as discussed below/before has Planckian scaling above temperatures where quasi-universality emerges and sub-Planckian scaling for lower temperatures. The full picture of generalized Drude phenomenology thus works well for temperatures below 0.3$E_F$ and for frequencies lesser than the temperature.

\subsubsection{Transition from sub-Planckian to Planckian scaling of the scattering time}\label{Sec:SubPlanck}

Given the agreement $\tilde{\tau}$ and $\tau_{\rm d}$ when the Drude-like behavior of the optical conductivity holds and that $\tilde{\tau}$ is close to the relaxation time $\tau$, we can readily expect that $\tau$ should scale as $T^{-\nu}$ at very low temperatures and as $T^{-1}$ in the regime of temperatures where quasi-universality holds following these trends for $\tau_{\rm d}$ as discussed in the previous section when the ratio of the couplings is optimal. We examine this separately for both these sub-Planckian and Planckian regimes.

\begin{figure}[ht]
    \centering
          \includegraphics[width=\textwidth]{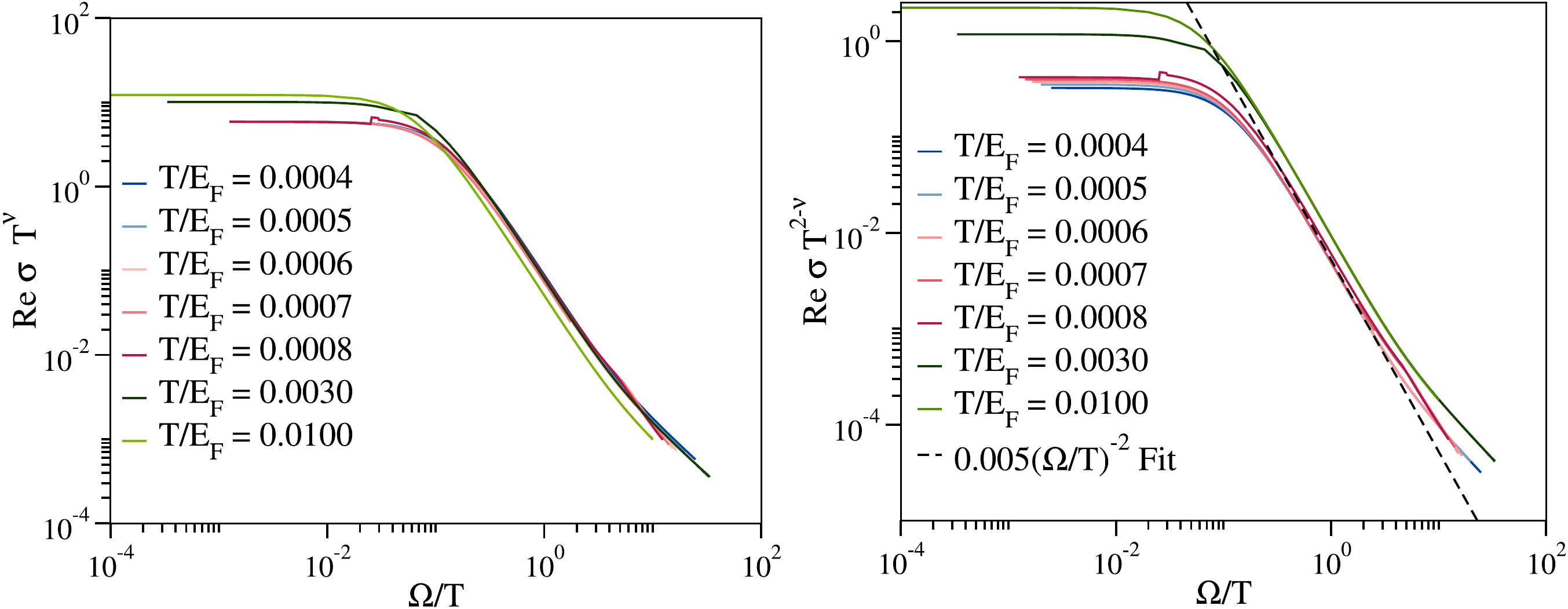}
         \caption{ \textbf{Left plot}: All the curves for $T<0.001E_F$ asymptotically collapse to a constant value at extremely low frequency, suggesting the real part of optical conductivity scales like $T^{-\nu}$. \textbf{Right Plot:} All the curves for $T<0.001E_F$ collapse to a $\Omega^{-2}$ line at intermediate frequencies following \eqref{Eq:Pred}. Both plots are at the optimal ratio of the couplings with $\nu = 0.8$, $\alpha = 0.023$, and $\gamma =0.001$.}
        \label{Fig:Lowtempfit}
\end{figure}

\vspace{0.3cm}
\noindent
\textbf{\uline{Low temperature sub-Planckian regime ( $T\lessapprox 0.001E_F$)}:} 
We readily recall from \eqref{Eq:h1h20}  that $\sigma_{\rm dc} = C T^{-\nu}$ with  $C = \tilde{h}_1 e^2 E_F^{-\nu}$ at very low temperatures. If $\tau_{\rm eff}$ is indeed independent of $\Omega$ and scales as $T^{-\nu}$, then we expect from \eqref{Eq:DrudeOp-1} that the real part of the optical conductivity should fit well with 
\be\label{Eq:optical conductivity at low temp}
\text{Re}\, (\sigma) = \frac{D T^{-\nu}}{1+D \Omega^2 T^{-2\nu}}
\ee
where $C$ and $D$ are constants, which are independent of the temperature but depend on the Fermi energy. This implies that the low frequency plateau of the 
$\text{Re} \sigma$ should scale as $T^{-\nu}$, and at higher frequencies we should have
\begin{equation}\label{Eq:Pred}
    \text{Re}\, (\sigma) T^{2-\nu} \sim k \frac{T^2}{\Omega^2},
\end{equation}
with $k$ constant and independent of the temperature. Both of these expectations are borne out in Fig.~\ref{Fig:Lowtempfit} for temperatures below 0.001$E_F$.

\vspace{0.3cm}

\begin{figure}[ht]
    \centering
          \includegraphics[width=0.6\textwidth]{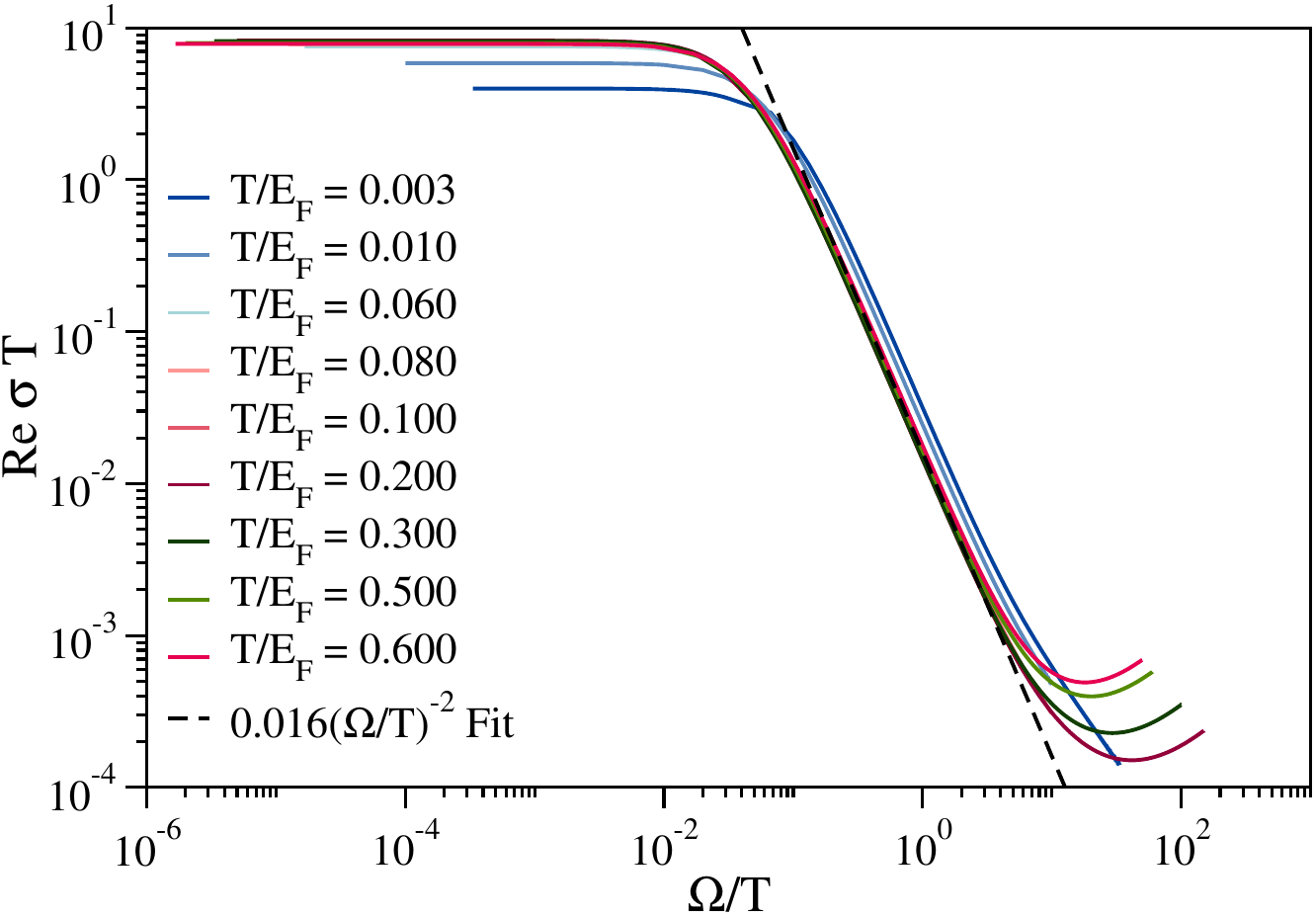}
         \caption{ Plot at optimal ratio of coupling with $\nu = 0.8$, $\alpha = 0.023$ and $\gamma =0.001$ showing that ${\rm Re}(\sigma)T$ is a function only of $\Omega/T$ for temperatures above $t_c E_F$ (at $\nu = 0.8$, $t_c = 0.05$) following \eqref{Eq:optical conductivity at high temp 2}. All curves in this plot collapse for temperatures above 0.05$E_F$.  
         }
         \label{Fig:HighTempfit}
\end{figure}

\noindent
\textbf{\uline{Higher temperature scaling ($t_c E_F \leq T \leq 0.6E_F$)}:}
In the regime where quasi-universality holds we obtain the $T^{-1}$ scaling of $\sigma_{\rm dc}$ and $\tau_{\rm d}$. We therefore expect from \eqref{Eq:DrudeOp-1} that ${\rm Re}\,\sigma$ should follow
\be\label{Eq:optical conductivity at high temp}
\text{Re}\,(\sigma) = \frac{\tilde{C} T^{-1}}{1+\tilde{D} \Omega^2 T^{-2}}
\ee
where $\tilde{C}$ and $\tilde{D}$ are also constants that are independent of temperature in the quasi-universal regime, i.e.
\be\label{Eq:optical conductivity at high temp 2}
\text{Re}\,(\sigma) T = f\left(\Omega/T\right) = \frac{\tilde{C}}{1+\tilde{D} \Omega^2 T^{-2}}.
\ee
We readily see in Fig. \ref{Fig:HighTempfit} that the above is indeed very well satisfied for temperatures above $t_cE_F$ where quasi-universality holds and below 0.6$E_F$ where the Drude-like is realized.
\vspace{0.3cm} 

\noindent
\textbf{\uline{Temperature dependence of the scattering time ($\tau$)}:} The analysis of the optical conductivity at lower and higher temperatures above mimics the analyses of experimental data. We see two different types of fits, namely \eqref{Eq:optical conductivity at low temp} for temperatures below 0.001$E_F$ and \eqref{Eq:optical conductivity at high temp} for temperatures above $t_cE_F$. This analysis is not satisfying as it does not allow us to interpolate between the temperatures 0.001$E_F$ and $t_cE_F$.

\begin{figure}[ht]
    \centering
    \includegraphics[width=0.6 \textwidth]{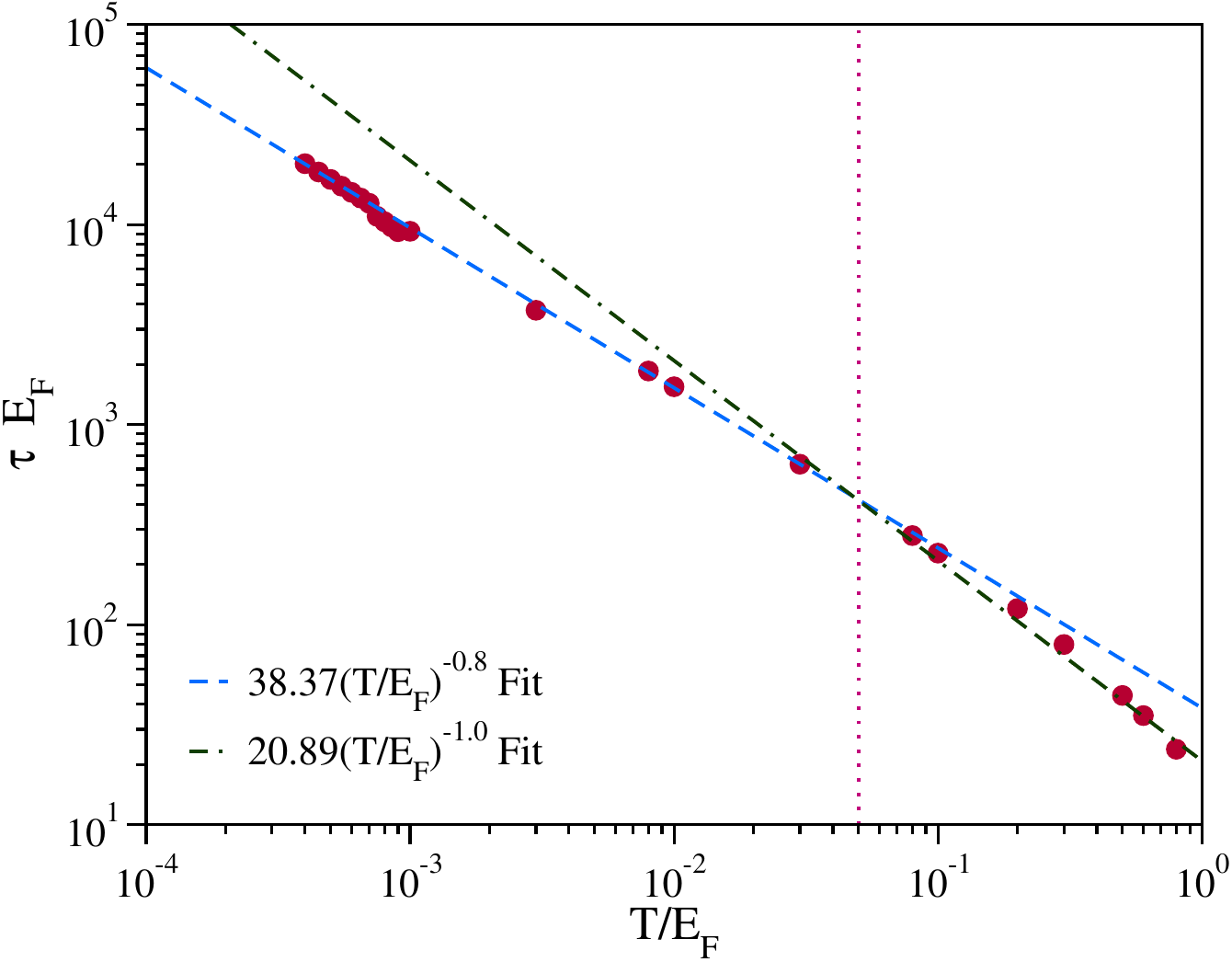}
    \caption{ The log-log plot of $\tau$ vs $T/E_F$ at optimal ratio of couplings for $\alpha = 0.023$, $\nu=0.80$, $\gamma=0.001$ The blue dashed line is the $T^{-\nu}$ fit for lower temperatures while the black one is for the high-temperature $T^{-1}$ fit. The transition between these two scaling regimes occurs sharply at $T = t_c E_F = 0.05 E_F$.
    }
    \label{temptaulogplot2fit}
\end{figure}

To get a deeper insight, we can readily extract the scattering time $\tau(T)$ from the real part of the optical conductivity and study its temperature dependence. As shown in Fig.~ \ref{temptaulogplot2fit}, we find that
\begin{equation}
    \tau(T) = a_1 T^{-\nu} \theta(t_c E_F- T) + a_2 T^{-1} \theta(T-t_c E_F)
\end{equation}
to a very good approximation, implying a sharp transition at $t_c E_F$ from sub-Planckian to Planckian behavior. Above, $a_1$ and $a_2$ are temperature independent constants. 

This sharp transition of the scattering time from sub-Planckian to Planckian scaling is quite remarkable and very non-trivially corroborates the result of our previous section that similarly in the case of $\tau_{\rm d}$ we have been able to estimate $t_c$ itself accurately assuming a sharp transition from sub-Planckian to Planckian scaling of $\tau_{\rm d}$ (c.f. Eq. \eqref{Eq:tc}). The non-triviality of the feature is that $\tau_{\rm d}$ is apriori a different time scale that requires us to know the Hall conductivity although $\tau_{\rm d}$ is close to $\tau_{\rm eff}$. 

To summarize, we have a comprehensive picture of Drude phenomenology for temperatures below 0.3$E_F$ with a scattering time $\tau(T)$ governing the Drude-like behavior of the optical conductivity and a dissipation time $\tau_{\rm d}(T)$ governing the response in the zero frequency limit. Both of these time scales are very close to each other and show a sharp transition from sub-Planckian to Planckian scaling almost exactly at the temperature above which quasi-universality of the spectral function is exhibited.

\subsection{Optical conductivity in the underdoped and overdoped regimes}

The behavior of the optical conductivity in the underdoped and overdoped regimes can be examined by considering $\alpha/\gamma$ larger and smaller than the optimal value, respectively. As in Sec. \ref{Sec:SubPlanck}, we find scaling regimes with the real part of the optical conductivity behaving as $\Omega^\kappa T^\eta$ emerge at lower and higher temperatures.\footnote{See \cite{Donos:2014uba} for holographic models with such power law behavior at very low temperatures.} The scaling with the external frequency $\Omega$ is different from Drude-like behavior at underdoping but has the $\Omega^{-2}$ behavior whenever $\Omega/T$ is sufficiently large at overdoping. The scaling with the temperature is non-trivial in both the doping regimes and it becomes Fermi-liquid-like only at extreme overdoping (i.e. in the limit $\alpha/\gamma \rightarrow 0$).

A summary of different emergent scaling behaviors both at underdoping and overdoping for various regimes of $\Omega/T$ and corresponding ranges of temperatures (less than the Fermi energy) for $\nu = 0.8$ is provided in Table \ref{Table:non-optimal}. 

\vspace{0.3cm}
\noindent
\textbf{\uline{The underdoped regime}:} We find significant departures from Drude-like behavior in the underdoped regime where $\alpha/\gamma$ is larger than the optimal value. There are, however, some generic features of $\Omega^\kappa T^\eta$ scaling of the real part of the optical conductivity which emerge at both low and high temperatures. As we increase $\alpha/\gamma$ from the optimal value, the scaling exponent $\kappa$ associated with the frequency increases monotonically from $-2$ and saturates to $-1$ in the limit  $\alpha/\gamma \rightarrow\infty$ both at small and larger temperatures (see Table \ref{Table:non-optimal}). Interestingly, the $\Omega^{-1}$ scaling of the real part of the optical conductivity has been reported for underdoped samples of the material $\rm{YbRh}_2\rm{Si}_2$ (where strange metallicity is observed at optimal doping) at low temperatures in \cite{10.3389/femat.2022.934691}. In fact, we are able to reproduce the results of \cite{10.3389/femat.2022.934691} in the reported temperature regimes choosing $\nu$ to be 0.73 and $E_F\sim 10^4$K. Interestingly, 0.73 corresponds to the value of $\nu$ for which we also obtain the best approximation to the linear-in-resistivity (see inset plot of Fig. \ref{Fig:slope dc optimal}). 

As shown in Table \ref{Table:non-optimal}, the scaling exponent $\eta$ associated with the temperature decreases monotonically from $\nu$ at optimal doping to
$\nu/2$ as $\alpha/\gamma$ is increased from the optimal value towards infinity at low temperatures. At higher temperatures, $\eta$ shows non-monotonic behavior but saturates to 1 in the limit $\alpha/\gamma\rightarrow\infty$ as shown in Table \ref{Table:non-optimal}. 

\vspace{0.3cm}
\noindent

\textbf{\uline{The overdoped regime}:}  The Drude-like $\Omega^{-2}$ tail is ubiquitously present at temperatures less than $E_F$ whenever $\alpha/\gamma $ is smaller than the optimal value. However, the scaling exponent $\eta$ associated with the temperature becomes $2$ as in the Fermi liquid only in the limit $\alpha/\gamma \rightarrow 0$ as shown in Table \ref{Table:non-optimal}. In fact, $\kappa$ increases monotonically from 1 at higher temperatures and from $\nu$ at lower temperatures as we decrease $\alpha/\gamma$ from the optimal value (recall the results of Sec. \ref{Sec:SubPlanck} for optimal doping) towards zero and eventually saturates to 2 while reproducing Fermi liquid behavior.

\begin{table}[ht]
\centering
\begin{tabular}{|c||c|c||c|c|}
\hline
 \multicolumn{1}{|c||}{  Value of $\alpha$ }  &\multicolumn{2}{|c||}{  High $T$ ($ 0.05E_F < T < E_F$)}  &\multicolumn{2}{c|} {Low $T$ ($T \sim < 0.001E_F$)}   \\
\hline
   & $\text{Re}\sigma$ scaling & Range of $\Omega/T$ & $\text{Re}\sigma$ scaling & Range of $\Omega/T$ \\
\hline
    0.0001 & $\Omega^{-2}T^2$& 0.06 - 3.16& $\Omega^{-2}T^2$& 0.003 - 5.0  \\ \hline 
    0.002 & $\Omega^{-2}T^{1.7}$ & 0.1 - 3.00& $\Omega^{-2}T^{\nu}$ & 0.05 - 2.52 \\ \hline
    0.023 & $\Omega^{-2}T^1$  & 0.15 - 2.5&$\Omega^{-2}T^{\nu}$ & 0.25 - 3.5 \\ \hline
    1 & $\Omega^{-1.45}T^{0.75}$ & 1.26 - 5.00& $\Omega^{-1.3}T^{\tfrac{\nu}{2}}$ & 1.32 - 8.00 \\ \hline
    10 & $\Omega^{-1}T^{0.85}$  & 1.32 - 6.30 & $\Omega^{-1.2}T^{\tfrac{\nu}{2}}$ & 1.8 - 12.6 \\ \hline
    100 & $\Omega^{-1}T^{1}$  &1.47 - 5.50& $\Omega^{-1.2}T^{\tfrac{\nu}{2}}$ &1.4 - 13.2 \\ \hline
\end{tabular}
\caption{Scaling behavior of the real part of optical conductivity for different values of $\alpha$ for fixed $\nu=0.80$ and $\gamma=0.001$ both at higher and lower temperatures. The scaling behaviors obtained at the optimal ratio of couplings ($\alpha = 0.023$) is as reported in Sec. \ref{Sec:SubPlanck} at both lower and higher temperatures.}
\label{Table:non-optimal}
\end{table}

\section{Black hole complementarity as a mechanism for fine-tuning}\label{Sec:Micro}

Microscopically, the refined semi-holographic model can be understood as conducting electrons coupled to a lattice of Sachdev-Yi-Kitaev (SYK) like quantum dots \cite{Sachdev:1992fk,Kitaev:2017awl} in close similarity to the approaches taken in \cite{PhysRevLett.119.216601,PhysRevX.8.021049,PhysRevX.8.031024,PhysRevLett.121.187001,Cha_2020,PhysRevLett.122.186601,PhysRevLett.123.066601,PhysRevResearch.2.033434,Patel:2022gdh}. The locality of the interaction of electrons with the quantum dots in real space would naturally lead to momentum independent self-energy contribution. Each quantum dot can be described holographically by a AdS$_2$ geometry (more precisely by Jackiw-Teitelboim gravity \cite{Almheiri:2014cka,Maldacena:2016hyu,Engelsoy:2016xyb,Maldacena:2016upp,Joshi:2019wgi}) in the large N limit, and therefore we can also think of each point of the lattice as the radial cutoff of an emergent AdS$_2$ space.  Remarkably, this setup is very similar to the black hole microstate model proposed in \cite{Kibe:2020gkx,Kibe:2023ixa}
to capture how the black hole complementarity principle \cite{PhysRevD.48.3743,PhysRevD.50.2700,Harlow:2014yka,Raju:2020smc,Kibe:2021gtw} (stated below) can emerge from microscopic dynamics along with the relaxation and absorption properties of the black hole in a local semi-classical approximation. We want to point out that the results of the latter works suggest that embedding the refined semi-holographic model into such an effective theory describing black hole microstates can lead to a \textit{dynamical} mechanism for the fine-tuning of the ratio of the two couplings needed to attain strange metallic behavior.

\begin{figure}[ht]
    \centering
    \includegraphics[width=0.6 \textwidth]{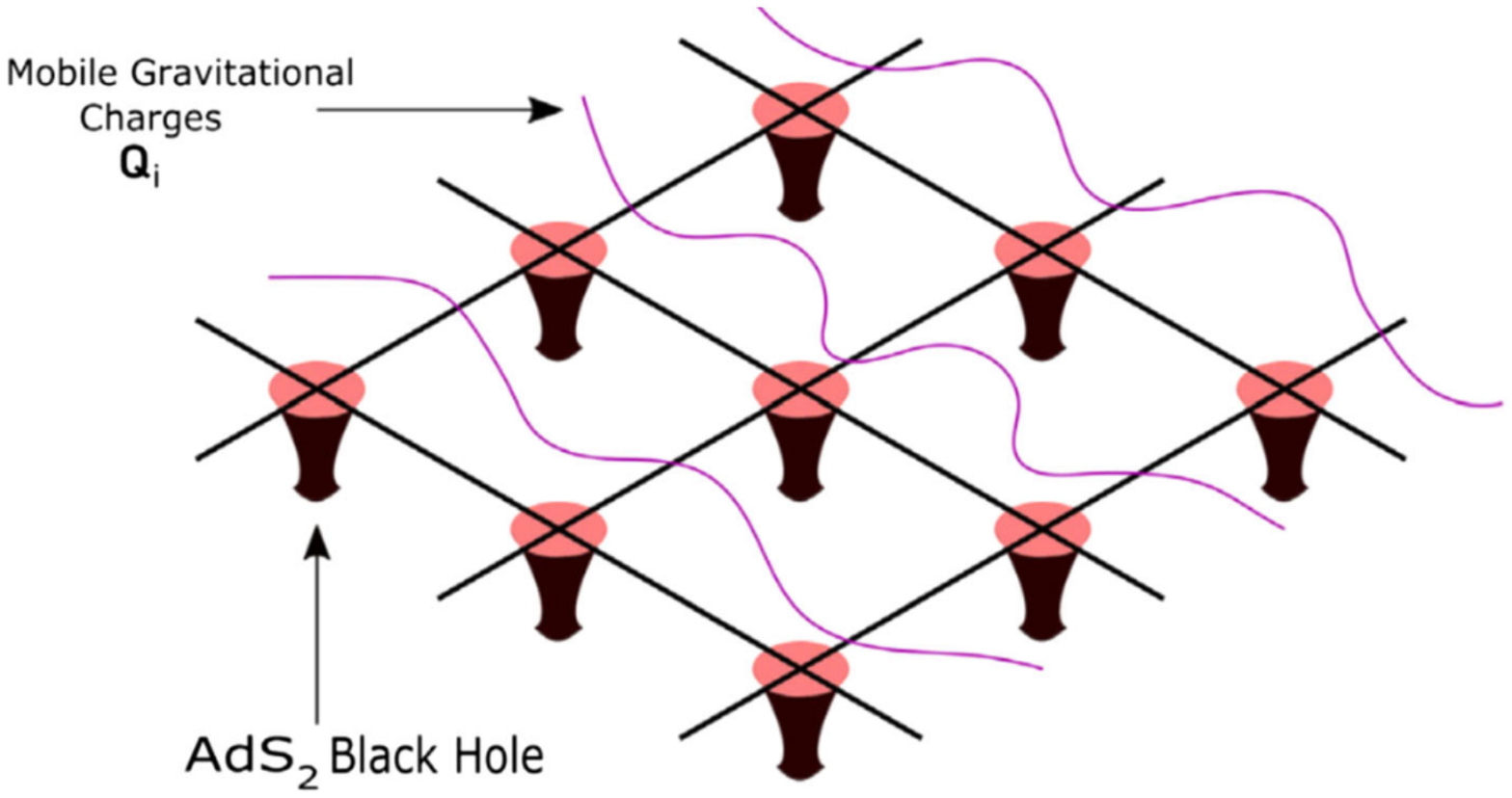}
    \caption{A schematic representation of the black hole microstate model in which the two dimensional horizon is modelled as a lattice of quantum dots, each dual to an AdS$_2$ black hole. Gravitational hair which propagates on the lattice couple these throats. Fig. from \cite{Kibe:2020gkx}.}
    \label{Fig:Microstate}
\end{figure}

The black hole microstate model which captures the fragmentation instability of the near-horizon geometry of a near-extremal black hole  in its simplest form is simply a lattice of AdS$_2$ black holes with mobile gravitational hair which can propagate on the lattice \cite{Kibe:2020gkx,Kibe:2023ixa}. It has the following degrees of freedom with $i$ labelling the lattice sites: (i) $\bQ_i$, the $SL(2,R)$ charges of the AdS$_2$ throats, and (ii) $\bq_i$, the mobile gravitational hair. From the charges $\bQ_i$, one can obtain the boundary time reparametrization mode $t_i(u)$ of each AdS$_2$ throat, i.e. the dual SYK-type quantum dot. The interactions preserve only a global $SL(2,R)$ symmetry which is the isometry of the original near-horizon geometry of the near-extremal four-dimensional black hole. See Fig. \ref{Fig:Microstate} for an illustration. In order to implement Hawking radiation of the black hole, one further assumes additional semi-classical matter in each throat, which can escape to an attached bath due to transparent boundary conditions. The model can be solved numerically in the large $N$ limit where the semi-classical approximation is valid locally in each throat.

To make contact with the strange metal models, we need these additional ingredients: (i) $c_i$, conducting lattice fermions, (ii) $\chi_i$, a Dirac bulk fermion field living in each AdS$_2$ throat, (iii) $\phi_i$, a mobile scalar field on the lattice, and (iv) $\Phi_i$, a scalar bulk field living in each AdS$_2$ throat. Note both $\chi_i$ and $\Phi_i$ are localized at the quantum dots (dual to the AdS$_2$ throats) and these fields can have different masses in different throats. The additional interactions are (i) Yukawa couplings of the form $c_i\chi_i\phi_i$, which hybridizes $c_i$ and $\phi_i$, (ii) a linear coupling $\phi_i\Phi_i$ leading to hybridization of the lattice and bulk scalar fields. These couplings are $\mathcal{O}(N)$. We should also add the $\mathcal{O}(N^0)$ couplings of the $c_i$ fermions to filled lattice band electrons $f_i$, which can act as baths for the full system. 

In order to see how such an enhanced model can naturally lead to dynamical fine-tuning towards strange metallic behavior, we need to first recall the dynamical properties of the simple version of the microstate model established in \cite{Kibe:2020gkx,Kibe:2023ixa}. It is useful to first see the properties of the model by switching off the Hawking radiation by imposing reflecting boundary conditions for the quantum bulk matter in each throat. The salient features of the dynamics in a typical state are:
\begin{itemize}
    \item There exists a large number of stationary solutions of the model for a given total energy, and these can be identified with the microstates of the near-extremal black hole.
    \item In any microstate solution, the total energy splits into interior and exterior parts because a part of the gravitational charges of the hair gets locked to those of the throats while the other part decouples from the interior.
    \item For any perturbation, such as external shocks injected into the throats, any microstate relaxes to another microstate. The dynamics has features of pseudo-randomness.
    \item For large number of lattice sites, the relaxation time is independent of the initial microstate to a very good approximation and depends only on the average mass of the AdS$_2$ black holes, which is a proxy for the temperature of the full system.
    \item Most of the energy injected by the perturbations (e.g. shocks) is absorbed by the interior, i.e. the masses of the AdS$_2$ black holes, while the rest is absorbed by the hair. The fraction of the energy absorbed by the interior goes to unity in the limit when the ratio of the number of lattice sites affected by the perturbation to the total number of sites goes to zero.
\end{itemize}
These imply that the simple microstate model reproduces the absorption and relaxation properties of a semi-classical black hole.

The black hole complementarity principle emerges from the microstate model too in any typical state. This principle states that the simple postulates especially the local validity of the semi-classical description of the black hole horizon for an infalling observer implies that the information should be both inside and outside of an old black hole \cite{PhysRevD.48.3743,PhysRevD.50.2700,Harlow:2014yka,Raju:2020smc,Kibe:2021gtw}. We will argue that the same mechanisms which realize the black hole complementarity in this microstate model in a typical state also would provide a natural mechanism for the fine-tuning needed to reproduce the strange metallic behavior once the additional ingredients are added.

\begin{figure}[ht]
    \centering
    \includegraphics[width=0.6 \textwidth]{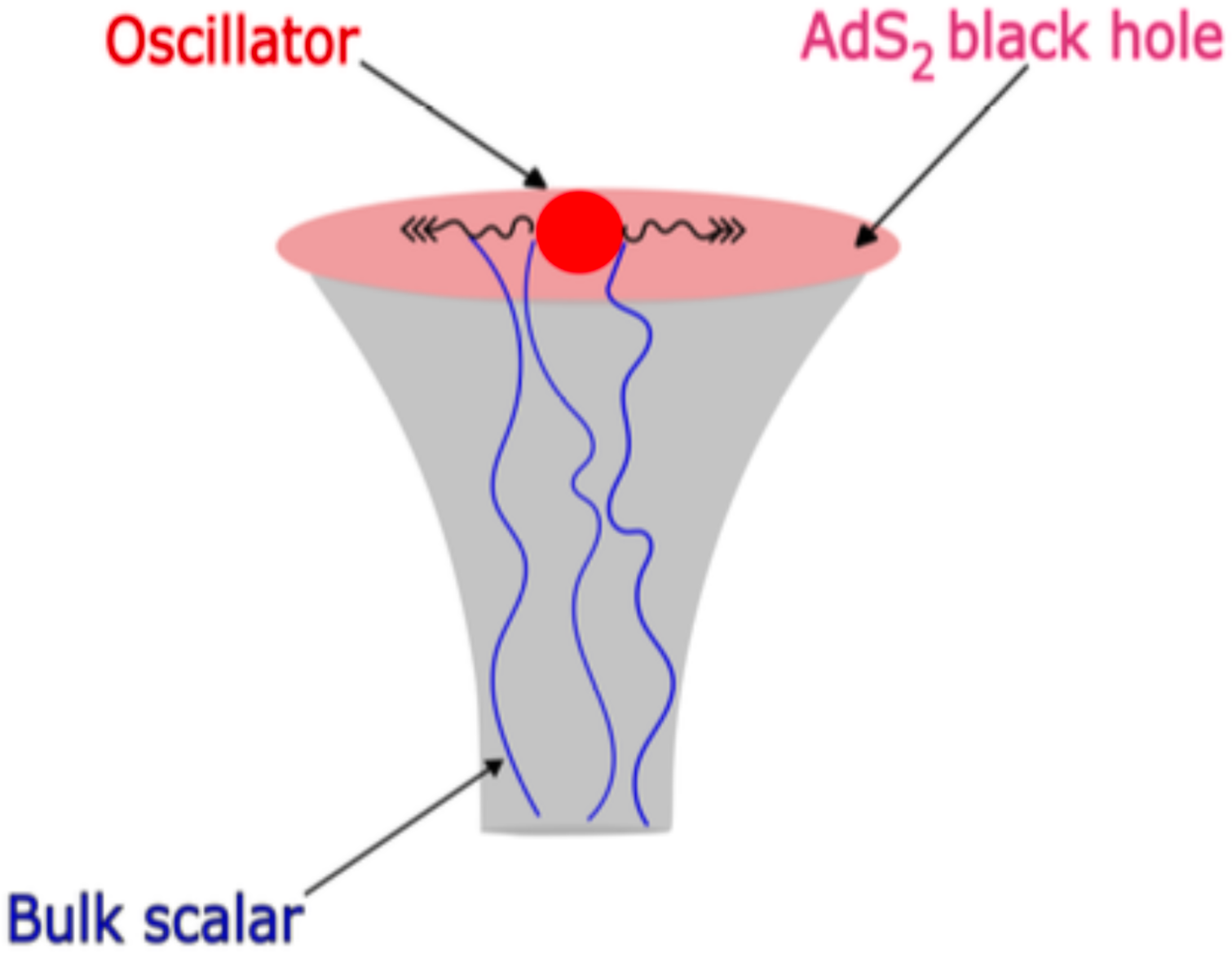}
    \caption{A simplified single throat setup of the microstate model to demonstrate information replication. Fig. from \cite{Kibe:2023ixa}.}
    \label{Fig:SingleThroat}
\end{figure}

The first puzzle of realizing the complementarity principle is how infalling information is replicated without paradoxes. A single throat setup, which arises as a simplification of the full model when the infalling scalar field is restricted to the $s$-mode, can itself illustrate this as shown in \cite{Kibe:2023ixa}. We can thus consider a quantum harmonic oscillator at the boundary coupled to a classical bulk scalar field living in the throat as illustrated in Fig. \ref{Fig:SingleThroat}. If the harmonic oscillator is classical, then it loses all its energy to the black hole and comes to its equilibrium $x = p = 0$. Although the quantum harmonic oscillator also loses most of its energy to the throat, it still retains a residual energy (proportional to $\hbar$).  Furthermore, both $\langle x \rangle$ and $\langle p \rangle$ of the oscillator also vanish at late time. Nevertheless, the final state is not the ground state but a superposition of coherent and excited coherent states which retains complete information of the initial state (this encoding is non-isometric). Crucially, the oscillator decouples from the throat and the information of the initial state of the oscillator is replicated in the (non-linear) ringdown of the throat. The decay exponents and the frequencies of the ringdown have a very narrow distribution, and remarkably, the amplitudes and decay exponents of the ringdown encode the initial state of the oscillator. 

The complementarity principle is also realized in the full model in a typical state due to a more general decoupling between various degrees of freedom. Introducing Hawking radiation in the full microstate model, it was shown that eventually, all throats decouple from each other, and the hair decouples from all the throats \cite{Kibe:2023ixa} in a typical state. Crucially, all throats homogenize, meaning that the $SL(2,R)$ charges of the throats become null (so that the throats become massless), and all components of these charges evolve towards a common value. This homogenization implies that while the information is replicated into the correlated non-linear ringdowns of the throats, the final state itself has no non-trivial remnant information (or rather, no information that can be decoded via simple measurements which are possible in the large N limit). The information is thus replicated in the algebra of time-dependent observables of the interior (but not into the final state) while a (non-isometric) residual copy is retained in the hair.\footnote{It was also shown that the quantum information mirroring \cite{Hayden:2007cs} of the infalling qubits takes place in the decoupled hair and into Hawking radiation after a scrambling time proportional to the logarithm of the entropy. The Hayden-Preskill protocol for the decoding of the information (such as time sequence and locations of shocks) was explicitly found in \cite{Kibe:2023ixa}. In consistency with the complementarity principle, this explicit decoding protocol requires no knowledge of the interior and only limited knowledge of pre-existing radiation, as can be argued from the validity of a semi-classical geometry.} We can conclude that the \textit{decoupling} of all degrees of freedom and \textit{homogenization} of the throats realize the black hole complementarity principle microscopically.

With the additional ingredients added into the microstate model as mentioned above, it would not be necessary to couple a bath to each throat as the lattice electrons of the filled bands $f_i$ can act as a bath which can equilibrate with the other degrees of freedom representing the hair (which also contains also the conducting electrons $c_i$) and the interior. The process of equilibration with the bath\footnote{Here, by equilibration we are implying that the full system is evolving towards a typical state which cannot be distinguished from a thermal state without very fine-grained measurements.} is expected to lead to a state in which all the throats decouple from each other, and also homogenize so that they have a common mass (and also same $SL(2,R)$ charges). We also expect that
\begin{itemize}
    \item the decoupling of the throats leads to a self energy of the conducting electrons, which is momentum independent since each throat gives separate contributions to the self-energy, and
    \item the self-energy contributions of each throat will also homogenize approximately (the approximation should be good enough so that information cannot be retrieved from the final state except perhaps by extremely fine-grained measurements, which are impossible in the large N limit).
\end{itemize}
Together, these mimic dynamical mean field theory \cite{RevModPhys.68.13}. The homogenization of the self-energy contributions of the throats implies that the electronic subsystem will be driven dynamically towards the strange metallic submanifold where the self-energy becomes independent of the model parameters to a very good approximation (note the self-energy function itself should be slightly modified in the full setup due to the involvement of the time-reparametrization mode). We can recall that a quasi-universal behavior can be realized only on the strange metallic manifold at least in the simple semi-holographic theory studied here, and therefore an approximate homogenization is likely to be achieved by dynamically realizing the fine tuning necessary to drive the effective coupling ($\alpha$) of the fermion in each throat to the optimal value that realizes strange metallicity. 

We can rephrase the mechanism for fine-tuning also in an effective field theoretic language. One can think of the effective description of the black hole microstates in terms of \textit{fragmented holography}, since it describes a collection of quantum dots each admitting a dual two-dimensional gravitational description. When the refined semi-holographic model is embedded in this setup, the coupling constant $\alpha$ is promoted to a field $\phi$ which assumes non-trivial (time-dependent and inhomogeneous) expectation values in typical states. Our claim is that if the effective description given by fragmented holography can reproduce the black hole complementarity principle in a typical state, then it should also be able to provide a natural dynamical mechanism in which the strange metallic submanifold is realized without the need of any external fine-tuning. 

At present, we do not have a full understanding of whether the fragmented holography setups should satisfy additional constraints to realize the black hole complementarity principle, or whether the necessary decoupling and homogenization processes are something generic due to quantum statistical reasons (analogous to how the eigenstate thermalization hypothesis \cite{DAlessio:2015qtq} is realized due to statistical reasons and leads to thermalization). The results presented in \cite{Kibe:2023ixa} do strongly suggest that fragmented holography setups in its simplest versions (with only two parameters) can realize the  black hole complementarity principle for generic values of the parameters. In the general case, the black hole complementarity principle (involving mutual decoupling and homogenization of the throats) may be realizable only for a range of parameters.  Nevertheless, even in the presence of additional constraints in the space of parameters, it is unlikely that any significant fine-tuning would be necessary to realize the black hole complementarity principle itself. So we can argue that it will be fruitful to understand effective descriptions of strange metals via the effective paradigm of fragmented holography for black hole microstates. 

In fact, it has been argued in \cite{Patel:2022gdh} that random Yukawa interactions (which is also generated by the lattice field $\phi_i$ in our enhanced microstate model) can lead to aspects of strange metallic behavior and SYK-type quantum dot description naturally emerges from disorder averaging in materials. Therefore, one can hope that it will be possible to connect the effective fragmented holographic setups to aspects of the quantum chemistry of materials although additional limits like the large $N$ limit could be necessary to enable tractability.

\section{Conclusions and Outlook}\label{Sec:Disc}

In this work, we have shown that refined semi-holographic models can give a simple description of strange metallic behavior. At the heart of this simplicity lies the feature of quasi-universal spectral function, which is realized on a co-dimension one submanifold in the space of parameters. This submanifold can be reached essentially by fine-tuning one parameter, namely the ratio of two dimensionless couplings. For tractability, we also need large $N$ type limits. Although these models can be understood as a Wilsonian effective description of non-Fermi liquids, the fine-tuning of the ratio of couplings needed to reproduce the strange metallic behavior needs a fundamental explanation. We have argued that such models can be naturally embedded into effective models of black hole microstates where the coupling that needs to be fine tuned is promoted to an effectively random field (generating random Yukawa interactions as argued to be realized in materials \cite{Patel:2022gdh}). Our arguments link the microscopic realization of the black hole complementarity principle to a dynamical mechanism for the fine-tuning needed to realize strange metallic behavior. 

From the quasi-universal behavior of the spectral function on the strange metallic submanifold, we can explain the appearance of linear-in-$T$ dc resistivity over decades of temperatures. This follows from the feature that the loop integrations over momenta receive major contribution only from the Fermi surface in refined semi-holographic models although the loop frequency integrals receive significant contribution from the high frequency tail (the quasi-universal behavior of the spectral function is realized for all frequencies near the Fermi surface). The strange metallic $T^{-3}$ scaling of the Hall conductivity is also reproduced in our model at higher temperatures. Quasi-universality implies that the conductivities in these scaling regimes are independent of the scaling exponent but do depend on the overall strength of the couplings. 

Combining these results with optical conductivity, we are able to deduce a compelling generalized Drude phenomenology for strange metallic transport which satisfies some non-trivial consistency checks. The scattering time deduced from various observables (including frequency dependent observables like optical conductivity) using Drude-like formulae agree. These Drude-like formulae can also be used to estimate the transition temperature between different regimes consistently in multiple ways. Furthermore, we obtain a Planckian scattering time and a temperature independent carrier density over a range of temperatures where linear-in-$T$ dc resistivity is exhibited. Crucially the Planckian scattering time $\tau$ can be as large $10\, T^{-1}$ when the couplings are weak but cannot be smaller than $T^{-1}$ when the couplings are strong.

Our simple model with a spherical Fermi surface is unlikely to reproduce the full spectrum of rich anomalous behavior of strange metals \cite{Phillips-Hussey} without incorporating it into a more general framework like fragmented holography involving the lattice explicitly, as discussed in the previous section. Nevertheless, it is important to investigate thermodynamic observables like specific heat in the simple model itself and incorporate these into the simplified phenomenological description. Furthermore, it is necessary to investigate if the simple model itself can reproduce anomalous magnetoresistance of strange metals (see \cite{Phillips-Hussey} for a recent review), especially the H-linear magnetoresistance at high fields \cite{Hayes_2016,PhysRevResearch.1.023011}, and see if these can also be incorporated into a phenomenological description involving a single scattering time.

Another drawback of our simple approach is that $E_F$ is the only intrinsic scale of the model and that some interesting features at the optimal ratio of coupling appear at a high scale. Particularly, the $T^{-3}$ scaling of the Hall conductivity emerges only for temperatures above $E_F$, and the slower than $\Omega^{-2}$ fall off of the real part of the optical conductivity occurs for frequencies somewhat larger than the temperature. Besides being problematic for comparison with experiments, these features do not always appear within the domain of validity of the effective approach.  However, these drawbacks can perhaps be overcome if we do not limit ourselves to a spherical Fermi surface. As discussed before, we particularly expect that the $T^{-3}$ scaling of the Hall conductivity should emerge at smaller temperatures as in the marginal Fermi liquid approach \cite{PhysRevB.68.094502} if the Fermi surface has equal amounts of positive and negative curvature. In the case of optical conductivity, the behavior at high frequency can be modified by the coupling of the charge current to the time-reparametrization mode when the refined semi-holographic model is embedded into the fully fragmented holographic setup. In fact, this can also lead to a new emergent UV energy scale. We plan to investigate these issues in the future.

One also needs a more fundamental understanding of the shape of the Fermi surface from fragmented holographic setups. Interestingly a simple variational wavefunction type approach utilizing a Gutzwiller projection applied to the BCS-wavefunction predicts preference for d-wave superconducting gap \cite{Anderson_2004}. A similar variational technique, which can be justified by the existence of generalized quasi-particles, can be adapted to our setups in fragmented holography to explain the shape of the Fermi surface in the strange metallic phase for a given lattice.\footnote{The role of the lattice in reproducing strange metallic transport in holographic models has been recently discussed in \cite{Ahn:2023ciq}.} 

The connection of our effective approach for strange metals with the fundamental physics of black hole microstates, particularly with respect to how quantum information is processed, provides a fresh interdisciplinary perspective connecting near-extremal black holes with strange metals. A natural question is whether this correspondence with black hole microstates can lead to a more detailed understanding of how the fundamental effective theory of strange metals can be subjected to robust experimental verification irrespective of material dependent complications. Perhaps this can lead to new opportunities in simulating quantum black holes as well. 

To conclude, we remark that the superconducting instability should also be investigated. The results of \cite{Doucot:2017bdm} indeed suggest that the semi-holographic model has a natural purely electronic mechanism of superconductivity based on the sharp collective modes in the particle-hole continuum at mid-infrared frequencies. We need to study the superconducting instability in the refined semi-holographic setup, where the physics of these collective modes is not cut-off dependent, adapting the Migdal-Eliashberg approach. \footnote{An interesting question is whether this is relevant also for correspondence with the physics of black holes, particularly for sufficiently large black holes (in flat space, such black holes have very small temperatures). Although there are reasons to believe that non-supersymmetric extremal black holes behave very differently from their supersymmetric counterparts, tractable models in which the generic behavior of non-supersymmetric extremal black holes can be studied are absent. This also provides a fresh motivation to study the fragmented holographic setups.} The fragmented holographic setups based on lattices should also allow us to incorporate the charge transfer super-exchange interactions between adjacent sites and explicitly study their role in the superconducting transition (see \cite{O_Mahony_2022} for some recent experimental investigations). Another related question is whether, by introducing bulk scalar fields, we can get insights into the origin of the full cuprate phase diagram including the pseudogap phase. We also plan to investigate this in the future.

\begin{acknowledgments}

We thank Sankar Das Sarma, Blaise Gout\'{e}raux, and Ren\'{e} Meyer for valuable discussions. AM acknowledges support from Fondecyt Grant 1240955. HS acknowledges support from the INSPIRE PhD Fellowship of Department of Science and Technology of India. SS acknowledges support by the US National Science Foundation under Grant DMR-1945395.
\end{acknowledgments}

\appendix
\section{The conductivities of the Fermi liquid}\label{Sec:AppendixA}
\noindent In order to derive the Fermi liquid dc and Hall conductivities, we start from the Green's function (as parametrized in \cite{Fukuyama1969})
\begin{align*}
G_{R} (\omega,\mathbf{k}) = \left(a^{-1} \omega - b^{-1}\epsilon_{k} + i \tau^{-1} \right)^{-1},
\end{align*}
where $\tau$ is the scattering time, $a$ and $b$ are renormalization factors. The specific heat particularly is renormalized by $b/a$ 
and $b$ is simply the factor renormalizing the density of states at the Fermi surface. To see the latter, we note that when $\tau$ is large, we can approximate
\begin{align}\label{Eq:FLrho1}
&\rho(0,\mathbf{k}) = - 2 {\rm Im}\,G_{R} (0,\mathbf{k}) \approx 2\pi\delta(b^{-1}\epsilon_{k} ) \nonumber\\
&= 2\pi b \delta(\epsilon_k).
\end{align}

In order to do the integral Eq.~\eqref{Eq:dcconductivity}, we proceed at follows. Firstly, at zero temperature
\begin{align}\label{zerotempfermidirac}
\frac{\partial}{\partial \omega} n_{F}(\omega) = - \delta({\omega} ).
\end{align}
Secondly, since the integral gets contribution Eq.~\eqref{Eq:dcconductivity} would get contribution only from $\omega =0$ and also from the Fermi surface when $\tau$ is large, we can approximate
\begin{align}
\rho(\omega,\mathbf{k})^2 \approx -2\pi b \delta(\epsilon_k) \times 2 \tau
\end{align}
where we have utilized \eqref{Eq:FLrho1} for the first factor, and then substituted $\omega = 0$ and $\epsilon_k =0$ in the second factor. It is then easy to see that Eq.~\eqref{Eq:dcconductivity} reduces to
\begin{align}
    \sigma_{\rm dc} = \frac{e^2}{m^2}\tau \frac{b}{4\pi}\times 2 \times m\int_0^\infty {\rm d}y\,\, y \delta (y - 2m E_F)
\end{align}
with $y = k^2$ (note we do the $\omega$ integral first). Therefore
\begin{align}
    \sigma_{\rm dc} = e^2\frac{E_F}{\pi} b\tau = \frac{n e^2}{m}b \tau,
\end{align}
since $n/m = E_F/\pi$ for the Fermi liquid. 

We can similarly proceed to evaluate  Eq.~\eqref{Eq:hallconductivity} and obtain the Hall conductivity.

We use the approximation \eqref{Eq:FLrho1} for $\rho(\omega,\mathbf{k})$. Since the integral then gets contribution from $\omega =0$ and the Fermi surface only, we can use 
\begin{align}
\frac{\partial}{\partial k_x} {\rm Re} G_R (\omega,\mathbf{k}) \approx -\,b^{-1} \tau^2 \frac{\partial \,\epsilon_k}{\partial k_x} = - \,b^{-1} \tau^2 \frac{k_x}{m}
\end{align}
Therefore,
\begin{align*}
\sigma_{\rm H} = \frac{\omega_{c} e^2 \tau^2}{2\pi^2} \int {\rm d}^2 k \,\,\delta(\epsilon_k) \left(\frac{k_x}{m}\right)^2
\end{align*}
Now
\begin{align*}
& \int {\rm d}^2 k \,\,\delta(\epsilon_k) \left(\frac{k_x}{m}\right)^2  =\frac{2}{m}  \int {\rm d}^2 k \,\,\delta(k_x^2 + k_y^2 - 2mE_F ) \,{k_x}^{2}\\
& \hspace{3cm}=\frac{2}{m}  \int_{0}^{\infty}{\rm d}k \,\,k^3\,\delta(k^2 - 2mE_F )\\
&\hspace{5cm}\times \int_{0}^{2\pi}{\rm d}\theta \,{\rm cos}^{2}\theta\\
& \hspace{3cm}=\frac{\pi}{m}  \int_{0}^{\infty}{\rm d}y \,\,y\,\delta(y - 2mE_F )\\
& \hspace{3cm}= 2\, \pi E_{F}.
\end{align*}
Therefore,
\begin{align}
\sigma_{\rm H} =  \frac{E_{F}}{\pi} e^2 \omega_{c} \tau^2 = \frac{ne^2}{m} \omega_{c} \tau^2.
\end{align}
The Hall coefficient of the Fermi liquid is
\begin{equation}
    R = \frac{\sigma_{\rm H}}{B\sigma_{\rm dc}^2} = \frac{1}{n e}\times\frac{1}{b^2}.
\end{equation}

\bibliographystyle{JHEP} 
\bibliography{cond-refs}

\end{document}